# Characterization of the Magnetism and Conformation of Single Porphyrin Molecules Adsorbed on Surfaces, and Artificial Graphene Nanoflakes

By

ZHANG Qiushi

*A Thesis Submitted to*

*The Hong Kong University of Science and Technology*

*in Fulfillment of the Requirements for*

*the MPhil Thesis Examination*

*in Physics*

*August 2017, Hong Kong*

# Authorization

I hereby declare that I am the sole author of the thesis.

I authorize the Hong Kong University of Science and Technology to lend this thesis to other institutions or individuals for the purpose of scholarly research.

I further authorize the Hong Kong University of Science and Technology to reproduce the thesis by photocopying or by other means, in total or in part, at the request of other institutions or individuals for the purpose of scholarly research.

______________________________________

**ZHANG Qiushi**

22 August 2017

# Characterization of the Magnetism and Conformation of Single Porphyrin Molecules Adsorbed on Surfaces, and Artificial Graphene Nanoflakes

**By**

**ZHANG Qiushi**

This is to certify that I have examined the above MPhil thesis and have found that it is complete and satisfactory in all respects,

and that any and all revisions required by the thesis examination committee have been made.

———————————————————————

Prof. Nian Lin, Thesis Supervisor

———————————————————————

Prof. Michael S Altman, Department Head

Department of Physics

22 August 2017

# Acknowledgements

I would like to show my great gratitude to my supervisor Prof. Nian Lin, who gave me this opportunity to study in UST and research on molecular electronics. All my study and research is conducted under full, attentive guidance and support from my supervisor. I will benefit a lot from his knowledge, enthusiasm and morality both now and in the future.

I would like to thank my former colleagues: Dr. Weihua Wang, Dr. Lei Dong and Dr. Tao Lin. They trained me from a fresh man to a skilled operator for STM. I am also grateful to my colleagues now: Guowen Kuang, Dr. Linghao Yan, Guoqing Lyu, Ran Zhang and Ziang Gao. Our discussion every day is much helpful for me.

I would like to thank Prof. Peinian Liu and his group for providing organic molecules for my study.

I would like to thank Prof. Michel A. Van Hove, Prof. Xingqiang Shi and Dr. Rui Pang for contributing great help to me in DFT calculations in my research.

I would like to thank Prof. Xuhui Huang and Dr. Xiaoyan Zheng for contributing great help to me in molecular dynamics calculations in my research.

I would like to thank Mr. T. C. Wu for contributing great help to me in tight binding calculations in my research.

I am grateful to my qualifying examination committee: Prof. Lortz Rolf W. and Prof. Ding Pan.

Thank my family and friends.

# Table of Contents

# List of Figures

distribution at two sublattices are colored with blue (sublattice A) and red (sublattice B), respectively. (h) The TB-calculated DOS of sublattice A/B and the zero-energy edge state is figured out by a red arrow. The black, red, blue and green plots representing the 1st, 2nd, 3rd and 4th curves, respectively.

42. Figure 5.5. (a) STM image of a hexagonal AGNF with the armchair edges (Set-point: -1V, 0.3nA, 48×48nm2). (b) The total dI/dV spectra over the whole flake, subtracting the contributions of coronene sites and substrate background. (Set-point: -1V, 0.3nA) (c) STS map acquired at the Dirac point: −0.30V. (d) The tight-binding calculation model in which the two sublattices are colored with blue (sublattice A) and red (sublattice B), respectively. (e) The calculated total DOS of the whole flake. (f) Tight-binding simulated spatial distribution of DOS at E=0, the size of dots is proportional to the strength.

43. Figure 5.6. (a) STM image of a deformed hexagonal AGNF with the zigzag edges (Set-point: -1V, 0.3nA, 55×55nm2). (b) The DOS of sublattice A (left) and B (right). The contributions of coronene sites and substrate background are also subtracted. The first three pLLs are figured out by red vertical lines. The zero-order pLL are figured out by red arrows. The black, red and blue plots representing the 1st, 3rd and 5th curves, respectively. (c) STS-measured energy levels of first three pLLs as a function of the square root of the orders of pLLs ( $\sqrt{n}$ ). (d) The tight-binding calculation model of this deformed AGNF, in which the two sublattices are colored with blue (sublattice A) and red (sublattice B), respectively. (e) The TB-calculated DOS of sublattice A (left) and B (right). The zero-order pLL are figured out by red arrows. The black, red, blue and green plots representing the 1st, 3rd, 5th and 7th curves, respectively. (f) TB-calculated energy levels of first three pLLs as a function of the square root of the orders of pLLs ( $\sqrt{n}$ ).

44. Figure 5.7. (a) STS map of the 0-order psedo-Landau-level, acquired at −0.26V. (b) STS map of the first-order psedo-Landau-level, acquired at −0.15V. (c) STS map of the second-order psedo-Landau-level, acquired at −0.1V (Set-point: -1V, 0.3nA, 55×55nm2). (d) Tight-binding simulated DOS map of the 0-order psedo-Landau-level at E=0. (e) Tight-binding simulated DOS map of the first-order psedo-Landau-level at E=60meV. (f) Tight-binding simulated DOS map of the second-order psedo-Landau-level at E=80eV. The distribution at two sublattices are colored with blue (sublattice A) and red (sublattice B), respectively. The size of dots is proportional to the strength.

# List of Tables

# Characterization of the Magnetism and Conformation of Single Porphyrin Molecules Adsorbed on Surfaces, and Artificial Graphene Nanoflakes

by ZHANG Qiushi

Department of Physics

The Hong Kong University of Science and Technology

# Abstract

My thesis mostly focusses on the systems of porphyrin molecules adsorbed on single-crystalline metallic surfaces. Cyclic tetrapyrrole porphyrins play key roles in many important chemical and biological processes, such as oxygen transport in heme (iron porphyrin), electron transfer and oxidation reactions in photosynthetic chlorophyll (magnesium porphyrin). Comprehensive understanding of the magnetic and conformational properties of single porphyrin molecules adsorbed on metallic substrates attracts intensive research interest. In my thesis, I have studied the structural and electronics properties of porphyrin molecules by Low-temperature scanning tunneling spectroscopy (LT-STM) and scanning tunneling spectroscopy (STS) and theoretical methods. Owing to the high resolution of LT-STM, both geometric and electronic properties at the atomic level were probed. Moreover, the experimental results were understood by comprehensive theoretical methods, i.e. density functional theory, molecular dynamics, tight-binding and plane-wave expansion calculations.

This thesis is divided into five parts, which are: (1) Introduction of the research area. (2) Principles of experimental setups and theoretical methods. (3) Switching molecular Kondo effect via supramolecular interaction. (4) Single-molecule observation of surface-anchored porphyrins in saddle, dome and ruffled conformations. (5) Manipulation and characterization of the electronic properties in artificial graphene nano-flakes.

In chapter 3, I study the switching molecular Kondo effect via supramolecular interaction. We apply supramolecular assembly to control the adsorption configuration of Co-porphyrin molecules on Au(111) and Cu(111) surfaces. By means of cryogenic STM, we reveal that the Kondo effect associated with the Co center is absent or present in different supramolecular systems. We perform first-principles calculations to obtain spin-polarized electronic structures and compute the Kondo temperatures using the Anderson impurity model. The switching behavior is traced to varied molecular adsorption heights in different supramolecular structures. These findings unravel that a competition between intermolecular interactions and molecule–substrate interactions subtly regulates the molecular Kondo effect in supramolecular systems. In chapter 4, we investigated the conformation relaxation and stabilization processes of two porphyrin derivatives (5,15-dibromophenyl-10,20-diphenylporphyrin, $Br_2TPP$, and 5,15-diphenylporphyrin, DPP) adsorbed on Au(111) and Pb(111) surfaces. We found that $Br_2TPP$ adopts either dome or saddle conformations on Au(111), but only the saddle conformation on Pb(111); whereas DPP deforms to a ruffled conformation on Au(111). We also resolved the structural transformation pathway of $Br_2TPP$ from the free-space conformations to the surface-anchored conformations. These findings provide unprecedented insights revealing the conformation adaptation process. We anticipate that our results may be useful for controlling the conformation of surface-anchored porphyrin molecules. In chapter 5, I employed the low-temperature STM manipulation to study the electronic properties of artificial graphene nano-flakes. Our major focus is the zero-energy edge state in different types of graphene nano-flakes. Due to the boundary conditions involved in the Dirac equations, the energy spectrum of the graphene system can be influenced by the edges types. This behavior is particularly true in the case of zigzag edges, which contribute unique magnetic properties to the 2D systems. Moreover, I built the hexagonal graphene nano-flakes with large deformation. The energy levels and spacial distribution of first several orders of pseudo Landau levels (pLLs) are studied in detail. To summary, the findings in my research works demonstrated that organic molecular systems exhibit interesting electronic, conformational and magnetic properties.

# Chapter 1

# Introduction

In this chapter, I will give an introduction about related fields and elaborate the motivation and significance of this study. Firstly, I will briefly the introduce development of nanoscience and on-surface self-assembly. Secondly, I will introduce the invention and development of STM. Thirdly, I will review the molecular electronics. In the fourth section, I will introduce porphyrin molecules. In the last part, I will give the overview of modification of two-dimension electron gases (2DEGs) on metallic surface.

## 1.1 Overview of nanoscience and on-surface self-assembly

'Nano' is no longer an unfamiliar word nowadays, which denotes the mathematic order of $10^{-9}$. In the field of nano-science, the nanoscale research is related to microcosmic matters, i.e. molecules, cells, atoms and nano-structures. Since early 1900s, the theory of quantum mechanics was developed following a large number of experimental and theoretical discoveries. The fundamental behaviors of matters change when they go to nanoscale [1], exhibiting new properties as Richard Feynman proposed in his famous speech entitled 'There's plenty of room at the bottom' [2]. The research of nanoscience and nanotechnologies have attracted great interests of scientists and engineers all around the world, and lot of nanoscale devices have been developed both in laboratories and also in industry for practical use [3-6]. After several decades, plenty of breakthroughs have benefited our daily lives already. To date, much of the miniaturization of computer chips has taken advantages of nanoscience and nanotechnologies, and this is expected to continue in the future [7]. For example, magnetic properties of nano materials have born a new generation of data storage [8]. Benefited from the development of nanotechnology, the fabricating methodology and energy efficiency of nanomaterial are significantly improved [9-12]. New challenges always come with technology revolution and this is also true for nanoscience. We believe that more and more breakthroughs in nano science will be made and benefit our life in the future.

Self-assembly [13-18] plays an essential role in the nanotechnology and it is one of our main research focuses for several years. The so called supramolecular chemistry makes use of self-driven processes for ordering of supramolecular or solid-state architectures from the atomic to the mesosopic scale [19-24]. The theme of this methodology is to develop complex chemical systems through self-assembly process based on non-covalent intermolecular interactions, such as hydrogen bonding, metal-organic coordination bonding, van der Waals forces, etc. [25-32].

**1.2 Development of scanning tunneling microscopy (STM)**

In 1981, Binning and Rohrer invented STM at IBM's Zürich research laboratory [33, 34] and got the Noble Prize in physics in 1986. STM is an instrument for imaging surfaces at the atomic level with a lateral resolution about 0.01 nm [35]. With this resolution, molecular structures and individual atoms can be clearly resolved. As the example shown in figure 1.2a, the atomic resolution of graphite surface was attained under ambient condition. Since the invention of STM, the attempts for improving this technique have never stopped. The majority technical improvements have been made are vacuum scanning condition, low-temperature environment, and high magnetic field. To date, urtra-high vacuum condition of ~$10^{-9}$ Pa can be achieved. The extremely low temperature of ~hundreds milli-kelvins can be reached through employing the Helium-3 cryostat, and strong magnetic field of ~10T can be applied by using superconducting coils. With these developments, a wide range of samples i.e. metals, semi-conductors and organic structures have been studied using STM [36-40]. Moreover, magnetic and electronic properties can be examined by using scanning tunneling spectroscopy (STS), such as the Kondo effect and superconducting states, see figure 1.2b and c [41,42].

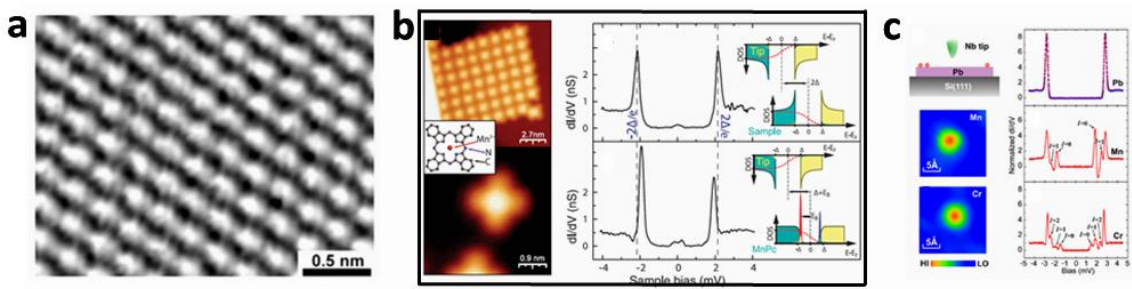

**Figure 1.1 (a) The atomic resolution STM image of graphite surface (b) Competition of superconducting phenomena and Kondo screening at the nanoscale (c) High-resolution scanning tunneling spectroscopy of magnetic impurity induced bound states in the superconducting gap of Pb thin films [41, 42].**

**1.3 Introduction to electronic properties of single molecules**

Molecular scale electronics, also called single-molecule electronics, is a branch of nanotechnology that uses single molecules, or collections of single molecules, as electronic components. Because single molecules constitute the smallest structures possible, this miniaturization is the ultimate goal for shrinking electrical circuits. [43, 44]. Essentially all electronic processes in nature, from photosynthesis to signal transduction, occur in molecular structures. As the fundamental building block, molecules play a key role in pushing forward the development of nanoscience and nanotechnology because of the advantages in size, assembly and recognition, dynamical stereochemistry and synthetic tailorability [45]. Molecular electronic devices are usually utilized as the active (switching, sensing, etc.) or passive (current rectifiers, surface passivates) elements [46]. Since the first organic molecular current rectification suggested by Aviram and Ratner, a wide range of molecular devices were demonstrated, including inter-connecters, switchers, rectifiers, transistors, nonlinear elements, dielectrics, photovoltaics, and memories [45, 52, 53].

In many examples, successful applications of molecular electronics are benefited from specific molecular properties, i.e. electronic [47], optical [48], magnetic [49], thermoelectric [50], mechanical properties [51]. These properties can be understood in the frame of the molecular orbital theory [54-59]. Molecular orbital (MO) theory is a method for determining molecular structure in which electrons are not assigned to individual bonds between atoms, but are treated as moving under the influence of the nuclei in the whole molecule. Molecular orbitals can be considered as the in-phase linear combination of individual atomic orbits. In most cases, the electronic properties are mainly determined by the molecular orbitals close to the Fermi level, especially the highest occupied molecular orbital (HOMO) and lowest unoccupied molecular orbital (LUMO) [60]. For examples, the molecular junction transportation and molecular emission spectrum are determinate by HOMO and LUMO. Thus, both theoretically and experimentally accesses of the energy scale and special distribution of molecular orbitals are firmly desirable for the research work in this field.

Molecular system adsorbed on solid substrates can be used to engineer functionalized structures or even devices on surfaces [61, 62]. Taking the advantages of advanced surface science experimental techniques, like STM, atomic-force microscopy, the structural information and electronic properties of molecular electronics can be clearly detected. Using on-surface self-assembly methodology, a large number of molecular structures can be synthesized, which provides an unique route for molecular electronics fabricating. Moreover, the structural control of molecules on surfaces also plays a key role in the molecular electronics engineering since the electronic properties of the molecules depend directly on adsorption orientations and conformations on surfaces. To date, study of molecular electronics on surface have made many breakthroughs and is attracting more and more research interests.

## 1.4 Overview of porphyrin molecules

Several major chemical classes of biologically and chemically relevant tetrapyrroles are shown in Figure 1.3 [63-69]. They are one of the most important cofactors in nature; and play a key role in regulatory effectors of many biochemical processes. Cyclic tetrapyrrole porphyrins act in many important chemical and biological processes, such as oxygen transport in heme (iron porphyrin), electron transfer and oxidation reactions in photosynthetic chlorophyll (magnesium porphyrin) [70, 71]. These properties make them the most important fine chemicals in industry; and are involved in an ever-expanding array of biochemical processes and applications ranging from use as pigments and oxidation catalysts, to emerging areas such as photodynamic cancer therapy, artificial photosynthesis, sensors, optics and nanomaterials.

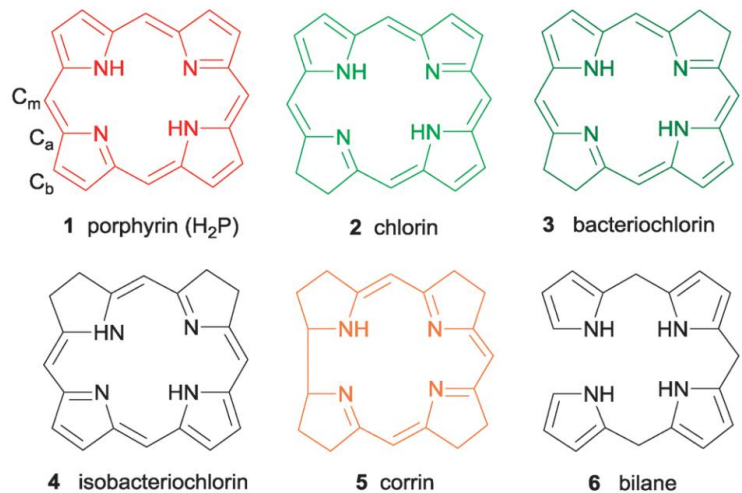

**Figure 1.2 Several major chemical classes of biologically and chemically relevant tetrapyrroles [66].**

The recent research results in the field of surface science have demonstrated that porphyrins can adapt different conformations upon adsorption on surfaces, such as saddle, dom, plannar and ruffled [72]. It is worthwhile to study these adsorption conformations since the electronic and magnetic properties are often varied for conformations. With the study of the porphyrin adsorption systems, molecular properties related to structure deformation can be controlled and engineered.

**1.5 Creating artificial graphene in two-dimension electron gases (2DEGs)**

The 2DEGs on a single crystal Cu(111) substrate are contributed by the Shockley state of Cu substrate [73-75]. The 2D free-electron on pristine Cu(111) presents a behavior with an isotropic parabolic dispersive band by the formula: $E = E_0 + \hbar^2 k^2/2m^*$, where $E_0$ is the band edge and $m^*$ is the effective mass of Cu ($0.38m_e$), as shown in Figure 1.4a [76]. In the right panel of Figure 1.4a, the two-dimension constant-energy contours graph referred to a ring of radius $k_{(E)}$ centered at the Γ corners of the Brillouin zone [77].

The 2DEGs can be modified by the electrons scattering with organic molecules. The molecular potential lattices can be achieved by supramolecular self-assembly [78, 79] and molecular manipulation [80, 81]. Making massless Dirac fermions from a modified two-dimensional electron gas was firstly proposed by C.-H. Park and S. G. Louie in 2009 [82]. As the example shown in Figure 1.4b, in the presence of an artificial penitential lattice, the 2DEGs are scattered by the hexagonal nano-potential lattices. The resulted two-dimensional massless Dirac systems are considered to be equivalent to those of the low-energy charge carriers in the real graphene layers [82]. This method can also be extended to build other artificial systems. For examples, the exotic flat band can be observed in the Lieb or Kagome lattices [83, 84], and non-Abelian SU(2) gauge field existed in the potential-induced deformed lattices [85].

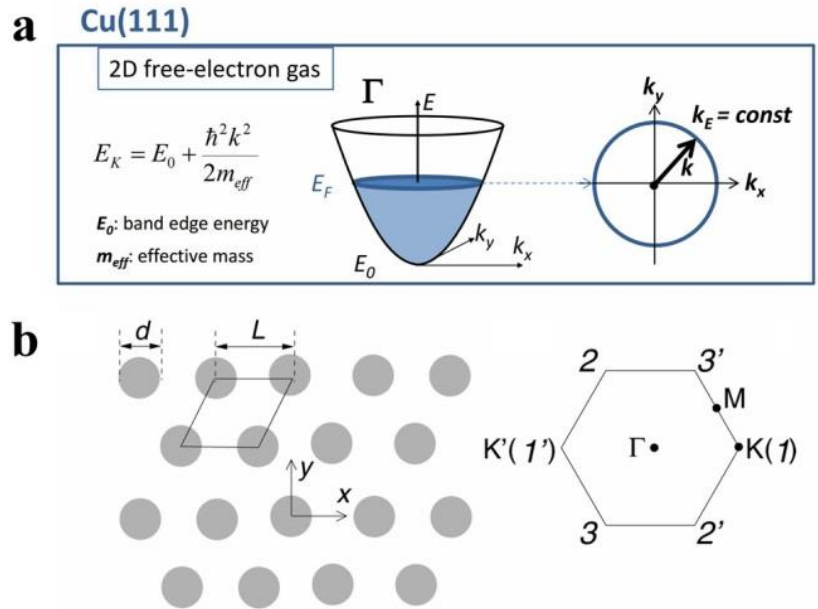

**Figure 1.3 (a) 2D free-electron gas with its characteristic parabolic dispersion and the corresponding 2D constant energy contours [77]. (b) A muffin-tin type of hexagonal periodic potential with a spatial period L. The potential is $U_0$ (>0) inside the gray disks with diameter d and zero outside. The Brillouin zone of this hexagonal lattice is on the left [82].**

# Chapter 2

# Experimental techniques and theoretical calculation methodology

In this chapter, I will introduce the experimental techniques and theoretical methods used in my research. In the experimental section, I will discuss the fundamental operation principle, general setup, scanning tunneling spectroscopy, STM working condition, sample preparation and tip manipulation. In the theoretical part, I will discuss the background of density functional theory (DFT), Gaussian09, Vienna Ab initio Simulation Package [86-88], Molecular Dynamics Simulation , Tight-binding calculation and plane wave expansion simulation.

**2.1 Experimental techniques.**

**2.1.1 The fundamental operation principle of STM**

The scanning tunneling microscope was invented in the IBM Research laboratory in 1981. After that, this tool has continuously broadened our perception about atomic scale structures and processes. STM allows one to image structures down to atomic-scale and measure physical properties of materials on a sub-nanometer scale by using a variety of different spectroscopic methods. The basic principles of STM are based on the concept of quantum tunneling. In quantum tunneling, electron can pass through an energy barrier though the electron's energy is lower than the barrier energy. Practically, when the distance between STM tip and surfaces is approached within several ångström, the tiny tunneling current can be detected by the sensitive electronics, as indicated in figure 2.1b. Moreover, the magnitude of the tunneling current decays exponentially with the distance between tip and surfaces, as a result, the features of the sub-nanometer variation can be resolved by STM.

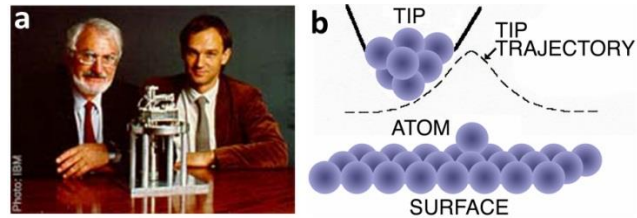

**Figure 2.1 (a) Binning and Rohrer with their Nobel prize [89] (b) As the tip moves over a surface, it "feels" atoms as increases in current [90]**

Herein, I will use a simplified 1-D quantum tunneling model as an example to introduce the theory of STM operation principle. The model describes that an electron with energy E passing a constant one-dimensional (1D) potential barrier $U_{(x)}$, which is larger than E. The electron is described by Schrödinger's equation (equation 2.1),

$$-\frac{\hbar^2}{2m}\frac{\partial^2 \psi}{\partial x^2} + U(x)\psi = E\psi \text{ and } U(x) = \begin{cases} U, & 0 \leq x \leq z \\ 0, & other \end{cases} \quad (2.1)$$

Where ℏ is reduced Planck's constant, m is the mass of the electron. Solving the Schrödinger's equation yields the relationship between the wave functions before and after the barrier, which indicates the decay of the wave after tunneling through the potential barrier, as equation (2.2) and (2.3), where z is the width of barrier:

$$\psi(z) = \psi(0)e^{-\kappa z} \quad (2.2)$$

$$\kappa = \frac{\sqrt{2m(U-E)}}{\hbar} \qquad (2.3)$$

And the probability of finding an electron after this tunneling is equation (2.4):

$$P = |\psi(0)|^2 e^{-2\kappa z} \qquad (2.4)$$

The simple model of 1D tunneling junction between a sample and a tip is shown in figure 2.2. Assuming the sample and the tip have identical work function ϕ, without bias voltage, there is no tunneling current between the sample and the tip. When a bias voltage is applied, tunneling current flows between the sample and the tip. Assuming the bias voltage is much smaller than the work function, the current can be derived as equation (2.5):

$$I \propto \sum_{E_F-eV}^{E_F} |\psi(0)|^2 e^{-2\kappa z} \quad \text{and} \quad \kappa = \frac{\sqrt{2m\Phi}}{\hbar} \qquad (2.5)$$

The local density of states at energy E and the expression of tunneling current can be further rewritten as equation (2.6) and (2.7), respectively:

$$\rho(x, E) = \frac{1}{\varepsilon} \sum_{E-\varepsilon}^{E} |\psi(x)|^2 \qquad (2.6)$$

$$I \propto V \rho_S(0, E_F) e^{-2\kappa z} \qquad (2.7)$$

As a result, the tunneling current is exponentially dependent on the separation distance between the sample and the tip, which provides STM the high resolution for imaging sample morphology.

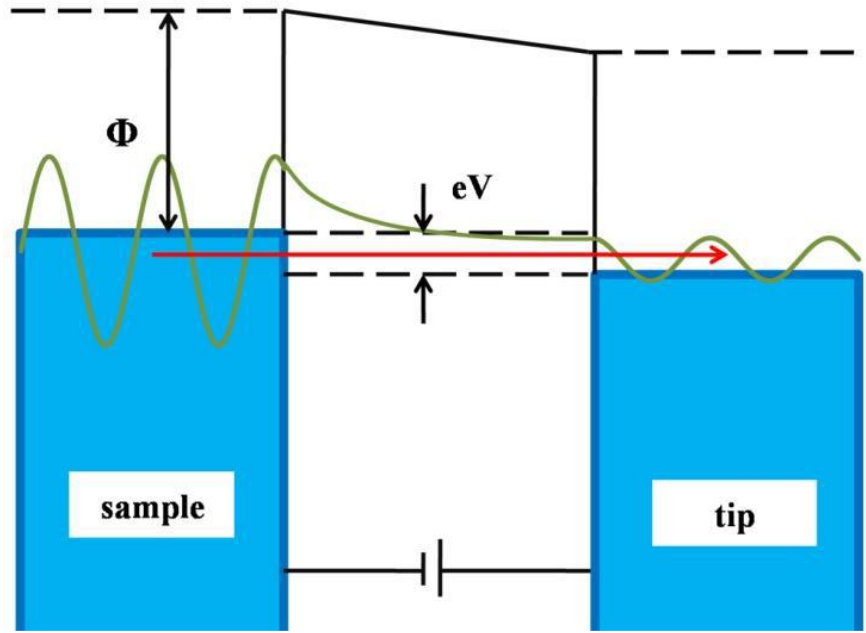

**Figure 2.2 The schematic diagram of tunneling current triggered by a bias voltage V between a sample and a tip**

### 2.1.2 The general setup of STM

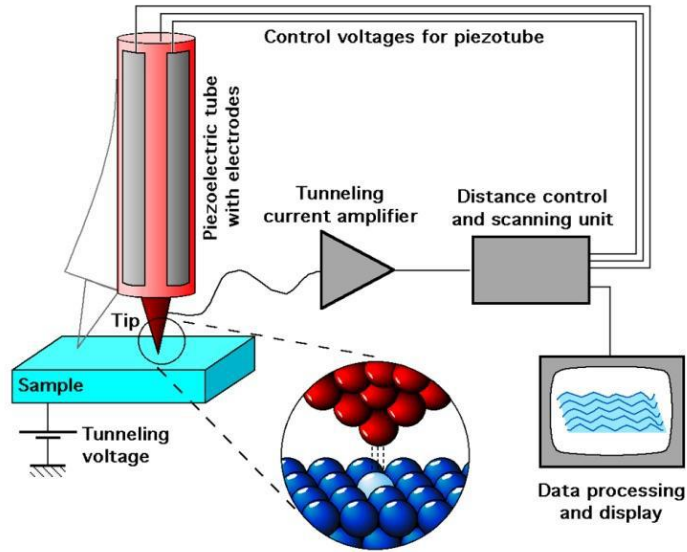

**Figure 2.3 Schematic diagram of STM setup [91]**

In figure 2.3, a schematic diagram of STM is presented. A STM experimental setup consists of three main components: (1) a very sharp metallic tip; (2) a conducting substrate; (3) a feed-back control system. STM tips are usually made from tungsten or platinum-iridium alloy where at the very end of the tip (called apex) there is one atom [92, 93]. The feedback control system controls the movement of tip. In the morphology measuring process, the tip is approached close to the sample surface within a separation of few angstroms by a coarse motion controller. And in this small separation, a tunneling current can be triggered by a bias voltage between the tip and the sample. Once the tunneling current is established, the tip is moved across the surface and the variation of the tunneling current is recorded, which represents the morphology of the sample and can be further converted into a STM image.

In general, there are two variables can be detected in STM, magnitude of tunneling current and tip height. If the tip is moved across the sample in the x-y plane, the changes in surface height and density of states cause changes in current. These changes are mapped in images. There are two main working modes of STM of constant current mode and constant height mode, as illustrated in Figure 2.4. Constant current mode is commonly used to obtain high-quality STM images. In this mode, the STM tip is approaching or retracting as moving across the surface while the tunneling current is kept as a constant. In other words, the gap between the sample and the tip is kept constant during measurement assuming the work functions and the density of states of the different sites on the sample are identical. In order to maintain the constant tunneling current, the vertical displacement of the tip is varied simultaneously during measurement controlled by the feedback control unit. In constant height mode, the vertical displacement of the tip is kept constant and the morphology of the surface is obtained by recording the changing of the tunneling current. Constant height mode enables higher scanning speed compared to constant current mode, because it does not require varying the vertical displacement of the tip. As a result, this mode allows studying fast dynamic process in real time. However, the quality of the image or other information detected is not as accurate as the constant current mode.

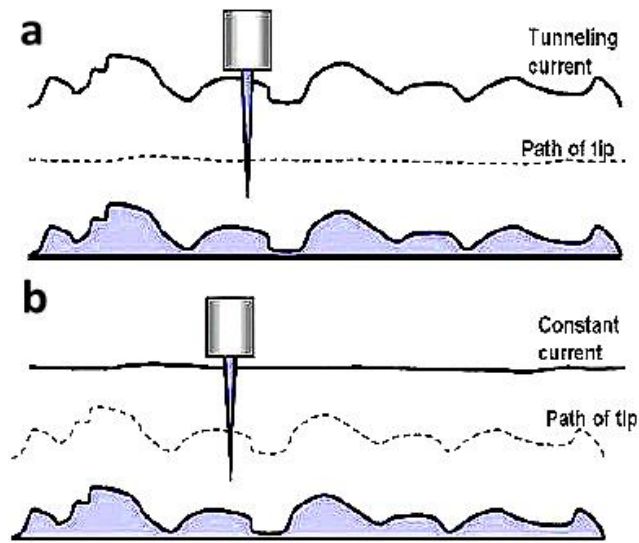

**Figure 2.4 Comparison of (a) constant-height and (b) constant-current mode for STM [94].**

### 2.1.3 Scanning tunneling spectroscopy (STS)

For better understanding of the data acquired by STM, different theoretical models have been developed since the invention of STM. The Tersoff-Hamann theory [95, 96], based on the Bardeen's tunneling theory [97], is one of the most famous due to its simplicity and good agreements with experimental findings. In the Tersoff-Hamann theory, solving the Bardeen's matrix element can be significantly simplified by a series of assumptions. The assumptions including: (1) the tip density of state (DOS) is spherically symmetric (s-wave symmetric wave function); (2) the tunneling matrix element does not depend on energy level; (3) The tip DOS is a constant within the energy range of interest; and (4) the sample DOS varies in energy not larger than kT. According to [95-97], the expression of tunneling current can be derived as equation (2.8) and the simplified expression based on the assumption is equation (2.9), where f is the Fermi distribution function, $\rho_S$ and $\rho_T$ are the density of states (DOS) for the sample and tip, respectively, M is the tunneling matrix.

$$I = \frac{4\pi e}{\hbar} \int_{-\infty}^{+\infty} [f(E_F - eV + \varepsilon) - f(E_F + \varepsilon)] \rho_S(E_F - eV + \varepsilon) \rho_T(E_F + \varepsilon) |M|^2 d\varepsilon \quad (2.8)$$

$$I \propto \int_0^{eV} \rho_S(E_F + \varepsilon) d\varepsilon \quad (2.9)$$

Then the differential tunneling conductivity can be approximately derived as equation (2.10):

$$\frac{dI}{dV} \propto \rho_S(E_F + eV) \quad (2.10)$$

Based on the equation (2.10), we can find that the value of dI/dV is directly proportional to the local density of electron (either occupied or unoccupied). Thus, point STS is commonly used to detect local density of states (LDOS) at a particular place of sample experimentally. The point STS is carried out with following steps: (1) placing the tip above an interested place of the sample; (2) with the height of the tip fixed, measuring the tunneling current as a function of the bias voltage between the tip and the sample, i.e. recording I-V curve; (3) numerical differentiating the I-V curve or alternatively using lock-in technique to obtain dI/dV.

## 2.1.4 The working conditions of STM

STM is used to measure nano-scale objects, for example, organic molecular structures adsorbed on conductive surfaces. Most of the single molecules and many molecular structures are not stable enough to scan at room temperature, so low-temperature experimental condition is desired. To uncover the features of molecules, the experimental environment should be clean enough, in another word, ultra-high vacuum condition is essential. In our experiment, the STM system is working under ultra-high vacuum (UHV) condition (~ $10^{-11}$ mbar) in order to get rid of influences from atmosphere, e.g. oxidation, etc. To achieve this UHV condition, a series of pumps including mechanical pumps, turbo molecular pumps, ion pumps and titanium sublimation pumps are used [99, 100].

To achieve low temperature experimental condition, liquid nitrogen and liquid helium are used, as shown in figure 2.5a. The scanning system of STM is surrounded by a Dewar system, which consists of two flasks, the inside and outer ones, as shown in figure 2.5b. The gap between the two flasks is filled with liquid nitrogen or helium, which prevents heat transfer by conduction or convection [98]. When the system is filled by liquid nitrogen, the scanning system can be cooled under to 77K. If using liquid helium, as shown in figure 2.5c, to fill the inner flask, the whole system can be further cool down to 5K.

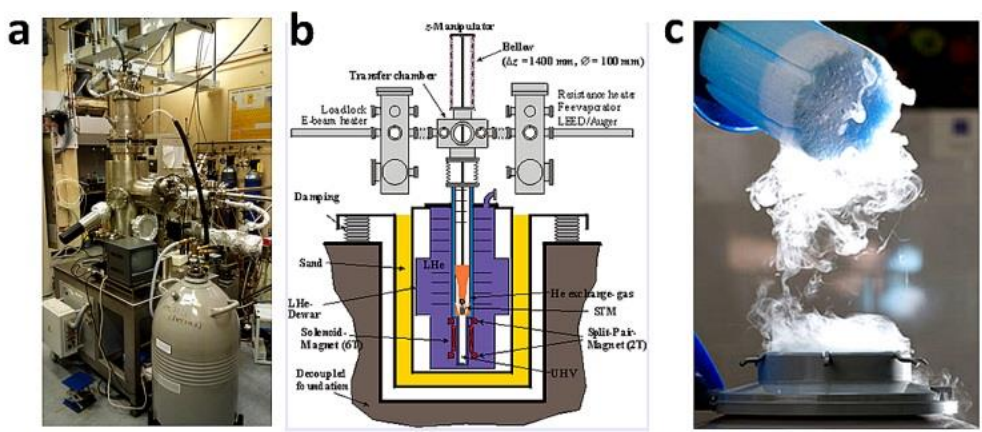

**Figure 2.5 (a) Filling liquid nitrogen into the Dewar flask (b) Structrual scheme of STM Dewar flask (c) Image of liquid helium [98].**

- 14 -

**2.1.5 Tip manipulation**

STM tip manipulation is an useful experimental method for placing molecules/atoms into desired arrangements. The mechanism of tip manipulation is as follow: when the distance between tip and molecules/atoms is reduced within several angstroms, the electronic force or van der Waals force is dominant and the molecules/atoms are attracted by the STM tip. Two main types of manipulations are illustrated in Figure 2.6: lateral manipulation and vertical manipulation. Lateral manipulation of an molecules/atoms involves three steps, as presented in figure 2.6a [93]: (1) step A: the tip is placed at the top of a single molecules/atoms and then approached to the molecules/atoms in order to increase the interactions between the tip and the molecules/atoms; (2) step B: the molecules/atoms is pulled by the tip to the desired position; (3) step C: the tip is gradually retracted to the normal imaging position and as a result, leaves the molecules/atoms at the new position. Vertical manipulation involves transferring the molecules/atoms between the tip and the substrate. Typical approaches to realize vertical manipulation, as presented in figure 2.6b, are applying particular bias voltage between the tip and the sample, making mechanical contact, etc. The process can be explained by a double potential well model as shown in the right panel of Figure 2.6b [100]. During normal imaging, the energy curve of the atom can be modeled as two potential wells which are separated by an energy barrier. The two wells correspond to two possible positions of the atom: on the tip apex and on the substrate. When a particular voltage is applied, the shape of the double potential well changes and the energy barrier decreases. In this case, the atom can be transferred from the substrate to the tip. When the tip and the atom has a mechanical contact, these two potential wells overlap into one well and the atom can be transferred to the tip. In figure 2.6c, a STM image of artificial molecular graphene is shown, which was built by tip manipulation.

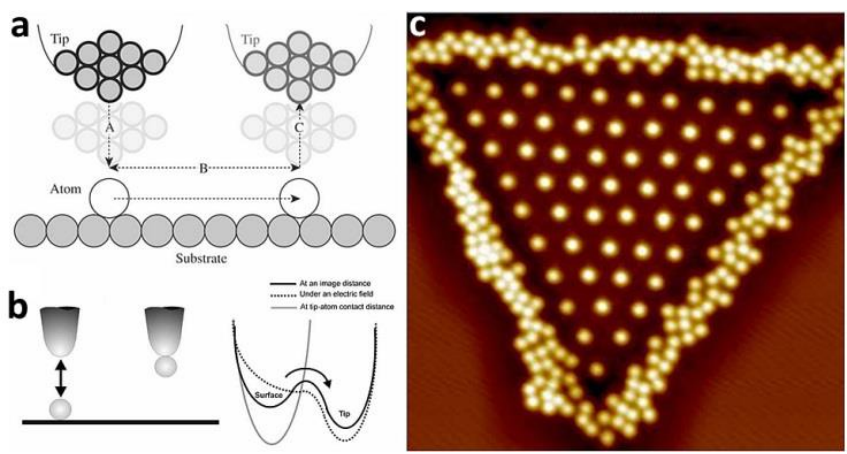

**Figure 2.6** Schematic illustration of (a) lateral manipulation [97] and (b) vertical manipulation [99] (c) Molecular artificial graphene in triangle shape.

### 2.1.6 Sample preparation

The sample preparation, as shown in Figure 2.7, consists of three parts: substrate cleaning, organic molecules/metal atoms evaporation and sample annealing. The single crystals of Cu(111), Pb(111) and Au(111) were used as substrates in my experiments. The cleaning procedure of sample is composed of several repeated circles of sputtering and annealing processes. An Ar sputtering ion gun was used for sputtering [101]. In detail, Ar gas was leaked into an ion gun through a leak valve and then the high voltage (usually about 1.0~1.5 kV) applied by ion gun allow Ar molecules to be ionized into $Ar^+$ and to be accelerated to the sample, bombarding the sample. After the sputtering process, the surface of substrate is pretty rough, an annealing process is essential. The sample is annealed by an electron-beam heating system or a resistor to 800 K to 1000 K for Cu(111), Au(111), and 500K for Pb(111).

After cleaning the substrate, organic molecules or metal atoms were deposited onto the substrate. An organic molecular beam epitaxy (OMBE) [101, 102] is used to deposit organic molecules. In an OMBE, organic molecules in ultra-pure form are heated incrucibles until they begin to slowly sublime. The term "beam" means that evaporated atoms do not interact with each other or vacuum chamber gases until they reach the substrate, due to the long mean free paths of the elements. In deposition of metal atoms, electron beam, which was emitted from a filament is accelerated by high voltage to the metal material, elevating the temperature of the metal material to sublime it. Another approach of metal evaporation is to wrap the metal wire around a filamenting and thermally evaporatethe metal materials. Table 2.1 lists the molecules used in my thesis work.

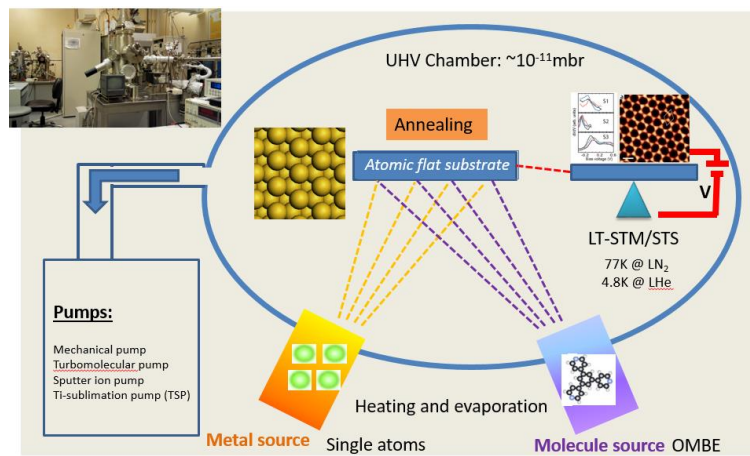

**Figure 2.7 Schematic illustrations of sample preparation system [103].**

**Table 2.1 The name and the skeletal formula of molecules used in this thesis.**

| Name | Chemical Structure | Sublimation Temperature |
|---|---|---|
| **Br2-TPP** | 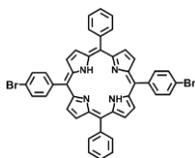 | 325°C |
| **Br2-DPP** | 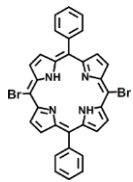 | 285°C |
| **L-TPyB** | 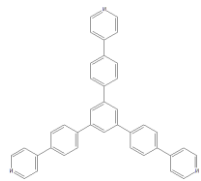 | 390°C |
| **TPyB** | 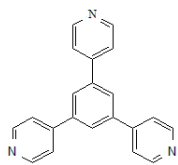 | 340°C |
| **Co-TPyP** | 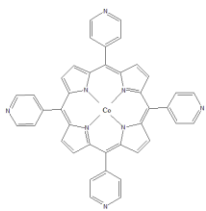 | 330°C |

## 2.2 Theoretical calculation methodologies.

### 2.2.1 Density functional theory

Density functional theory (DFT) is a computational quantum mechanical method to investigate the electronic structure (principally the ground state) of many-body systems, in particular molecules and the condensed phases. DFT method has been widely applied in the field of solid-state physics. DFT theory was established according to the Thomas–Fermi model and the Hohenberg–Kohn theorems [104], which assumes that the ground state properties of a many-electron system are uniquely determined by an electron density that depends on only 3 spatial coordinates. The properties of a many-electron system can be obtained by using the spatially dependent electron density. Hence the name density functional theory comes from the use of functional of the electron density. DFT is among the most popular and versatile methods available in condensed-matter physics, computational physics, and computational chemistry [103]. Compared with traditional Hartree–Fock theory, the computational cost is lower for DFT and the accuracy for solving the complex many-electron wave function is relatively good. Since 1970s, the accuracy of DFT method was greatly improved with a series of approximations, i.e. the exchange and correlation interactions developed in 1990s.

Practically, several computational softwares are commonly used to conduct DFT calculations, such as, Gaussian, VASP and Quantum Espresso. In the next section 09, I will briefly introduce some of these softwares.

### 2.2.2 Gaussian software

To date, many computational softwares of DFT have been developed and applied in research. Gaussian is one of the most widely-used calculation tools. Gaussian 09 is the latest version of the Gaussian series. Gaussian 09 can calculate the energies, molecular structures, vibrational frequencies and molecular orbital properties of molecules and reactions in a wide variety of chemical environments. Gaussian 09 provides the most advanced modeling capabilities available today, and it includes many new features and enhancements which significantly expand the range of problems and systems which can be studied. The main advantages of Gaussian series can be summarized as below [105]:

1. Accurate, reliable and complete models without cutting corners.
2. Applicable to the full range of chemical conditions and problem sizes and across the entire periodic table.
3. State-of-the-art performance in single CPU, multiprocessor/multicore and cluster/network computing environments.
4. Setting up calculations is simple and straightforward, and even complex techniques are fully automated. The flexible, easy-to-use options give you complete control over calculation details when needed.

For the latest version of Gaussian, a large number of DFT functionals are available, including Exchange Functionals, Correlation Functionals, Specifying Actual Functionals, Correlation Functional Variations, Standalone Functionals, Hybrid Functionals, Functionals including dispersion, Long range corrected functionals. Practically, the functionals like B3LYP and CCSD series can provide best accuracy for general systems, m06 enables the long-range van der Waals attraction and PM6 is a semi-empirical method to calculate large systems [106, 107]. Gaussian 09 provides many types of basis sets, such as 3-21G and 6-31G, and adding polarization and diffuse functions is also available.

### 2.2.3 VASP software

Vienna Ab initio Simulation Package (VASP) is a computer program for atomic scale materials modeling, e.g. electronic structure calculations and quantum-mechanical molecular dynamics. VASP computes an approximate solution of the many-body Schrödinger equation, either within density functional theory (DFT), solving the Kohn-Sham equations, or within the Hartree-Fock (HF) approximation, solving the Roothaan equations. Hybrid functionals that mix the Hartree-Fock approach with density functional theory are implemented as well. Furthermore, Green's functions methods (GW quasiparticles, and ACFDT-RPA) and many-body perturbation theory (2nd-order Møller-Plesset) are available in VASP. In VASP, quantities like the one-electron orbitals, the electronic charge density and the local potential are expressed in plane wave basis sets. The interactions between the electrons and ions are described using norm-conserving or ultrasoft pseudopotentials, or the projector-augmented-wave method. To determine the electronic groundstate, VASP makes use of efficient iterative matrix diagonalisation techniques, like the residual minimisation method with direct inversion of the iterative subspace (RMM-DIIS) or blocked Davidson algorithms. These are coupled to highly efficient Broyden and Pulay density mixing schemes to speed up the self-consistency cycle [109]. Functionals, like LDA, GGAs, meta-GGAs and Hartree-Fock, Hartree-Fock/DFT hybrids are available in VASP. Projector augmented-wave (PAW) together with accurate cut-off energy are used to describe the potential for all elements. Generally the PAW potentials are more accurate than the ultra-soft pseudopotentials. There are two reasons for this: first, the radial cutoffs (core radii) are smaller than the radii used for the US pseudopotentials, and second the PAW potentials reconstruct the exact valence wavefunction with all nodes in the core region. Since the core radii of the PAW potentials are smaller, the required energy cutoffs and basis sets are also somewhat larger. If such a high precision is not required, the older US-PP can be used [110, 111]. There are many fields can be calculated using VASP: Dynamics and relaxation, Magnetism, Linear response to electric fields, Linear response to ionic displacements, Optical properties, Berry phases, Green's function methods and Many-body perturbation theory [112-117].

**2.2.4 Introduction to Molecular Dynamics Simulation**

Modern simulation methods are usually mixed quantum and classical mechanics, which greatly assist experimental understanding in many ways like molecular interaction, structural deformation and chemical reaction processes [118, 119]. The molecular dynamics(MD) method was firstly developed in 1960s [120] and became one of the most commonly used simulation tools in the fields of theoretical physics, organic chemistry, biochemistry, biophysics and material science [121].

MD simulation method is used to solve the time-dependent properties of molecular systems. Atomic or molecular structures are commonly allowed to relax and interact for a long-enough fixed period of time, which reveals the dynamical evolution of the whole system. In some most modern versions, a large number of simulation trajectories are used to determinate the motion of particles. During the simulation trajectories, many experimental conditions, like annealing, cooling, electrical dipoles and chemical solution environment, can be simulated the forces and potential energy between each particle are calculated by using interatomic potentials or molecular mechanics force fields. MD is able to provide accurate information on the fluctuations and structure-to-function relationships of real physical systems. Information about the dynamic properties of simulated systems is duly gathered, which can be used to to analyze both electronic and conformational of single molecules and molecular ensembles [122, 123].

# Chapter 3

# Switching Molecular Kondo Effect via Supramolecular Interaction

## 3.1 Molecular Kondo effect

The Kondo effect is a many-body problem, which was first explained by Japanese theorist Jun Kondo in 1964 [126-129]. The Kondo effect is arising from the exchange interaction between the localized spins of magnetic impurities and the conducting electrons in the host metal. It occurs if the temperature is below the Kondo temperature with an interesting behavior that the electrical resistance increases as the temperature is lowered. The Kondo effect is predicted to be used in many future applications such as spintronics and quantum information processing [126-129]. Scanning tunneling microscopy/spectroscopy (STM/STS) is an ideal method to study the Kondo resonance in atomistic scale. The resonance is observed as a Fano-shaped resonance near $E_F$ in tunneling conductance spectra $dI/dV$ [130-132]. The shape of resonance can either be a peak or a dip, which is determined by the coupling strength between the STM tip and the detecting target [133].

The mechanism of the Kondo resonance can be understood as an exchange processes: flipping the spin of the impurity from spin up to spin down, or vice versa, while simultaneously creating a spin excitation in the Fermi Sea [126-129]. As indicated in figure 3.1, firstly, an electron is taken from the localized impurity state and put into an unoccupied energy state at the surface of the Fermi sea. Although, the energy needed for such a process is large, between about 1 and 10 electronvolts for magnetic impurities. In quantum mechanics, however, the Heisenberg uncertainty principle allows such a configuration to exist for a very short time – around h/|εo|. Within this timescale, another electron must tunnel from the Fermi sea back towards the impurity. However, the spin of this electron may point in the opposite direction. This spin exchange qualitatively changes the energy spectrum of the system. When many such processes are taken together, one finds that a new state – known as the Kondo resonance – is generated with exactly the same energy as the Fermi level. Such a resonance is very effective at scattering electrons with energies close to the Fermi level. Since the same electrons are responsible for the low-temperature conductivity of a metal, the strong scattering contributes greatly to the resistance [129].

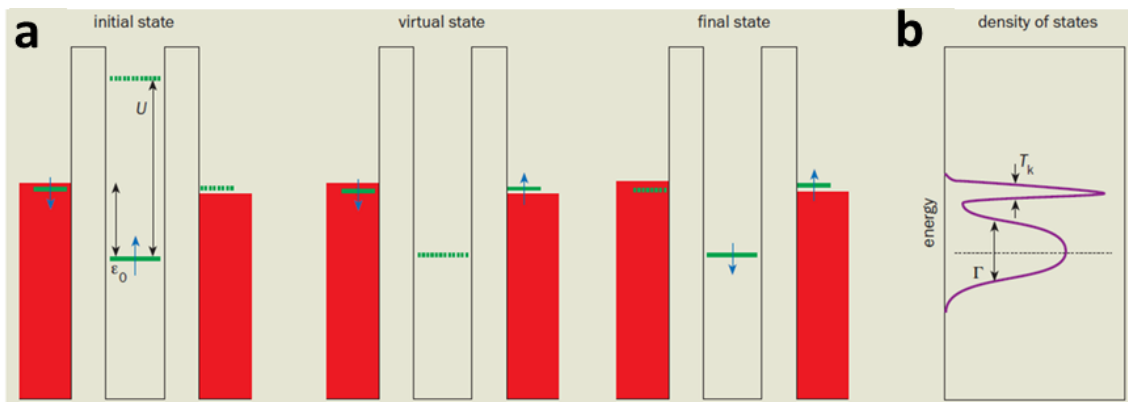

**Figure 3.1 [129] (a) The mechanism of the exchanging process, which is the cause of Kondo effect. (d) The Kondo resonance state at Fermi level, $T_K$ is the Kondo temperature.**

Recent advances in nanotechnology have enabled the observation and understanding of the Kondo effect down to a single atomic or molecular level. Due to the potential applications in spintronics and quantum storage, the so-called molecular Kondo effect attracted more and more research interests and most of the works are focusing on the understanding and controlling the Kondo resonance [133-140]. The strength of the Kondo resonance is usually measured by the Kondo temperature $T_K$ in unit of Kelvin, which can be obtained experimentally by fitting by the Fano-shape resonance dI/dV spectrum [130, 131]. To understand the strength of Kondo resonance qualitatively, a simplified model was commonly used in previous works, as presented in equation (3.1), where ρ and J are the density of states at the Fermi energy and the exchange coupling between the spin of the adsorbed molecule and that of the host, respectively [141-143].

$$k_B T_K \propto e^{-1/\rho J} \tag{3.1}$$

In order to calculate the Kondo temperature quantitatively, the Anderson single-impurity model [148, 149] was also applied in recent works [144-147]. The Haldane expression to calculate the Kondo temperature based on the Anderson single-impurity model is shown in equation (3.2).

$$k_B T_K = \frac{1}{2}\sqrt{\Gamma U} e^{\pi E_d(E_d + U)/\Gamma U} \tag{3.2}$$

The physical meaning of each parameter is presented in figure 3.2, where $\Gamma$ is the broadness of the impurity state, $E_d$ is the energy level of the impurity state with respect to the Fermi level ($E_d < 0$), and U is the on-site Coulomb repulsion energy [144-147]. According to equation 3.2, $T_K$ monotonically depends on $\Gamma$, and a smaller $\Gamma$ leads to a lower Kondo temperature. TK dependence on U is not monotonic: exponential decays when U is small (U < 2|$E_d$|) and near linearly increases when U is large (U > 2 |$E_d$|). Generally speaking, the magnitude of U is proportional to the value of magnetic moment and $\Gamma$ represents the molecule-substrate coupling strength, a strong molecule substrate coupling enlarges $\Gamma$ and reduces U. These parameters can be extracted from the DFT+U-calculated electronic structure [147, 150].

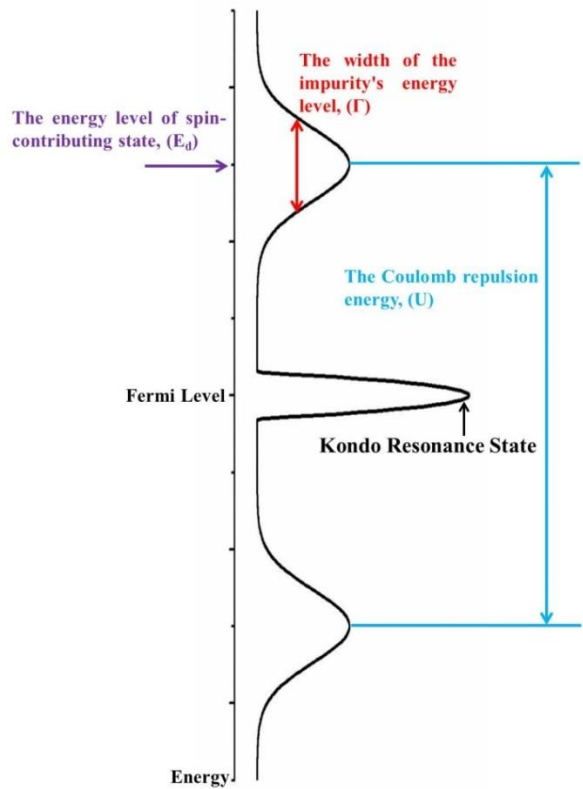

**Figure 3.2 The physical meaning of each parameter inside the Haldane expression based on the Anderson single-impurity model.**

In the qualitative model or the Anderson one, the strength of molecular Kondo effect is definitely determined by two key factors: the density of spin impurities and the molecule-substrate coupling strength. Adjusting these two key factors is able to control and engineer the molecular Kondo effect. Recently, many experimental and theoretical approaches have been reported to control the Kondo effect. Figure 3.3 illustrates some example approaches: deforming the molecular structure [148], changing the neighboring molecules [142], NO molecule absorption on the impurity site [151] and the quantum size effects [141].

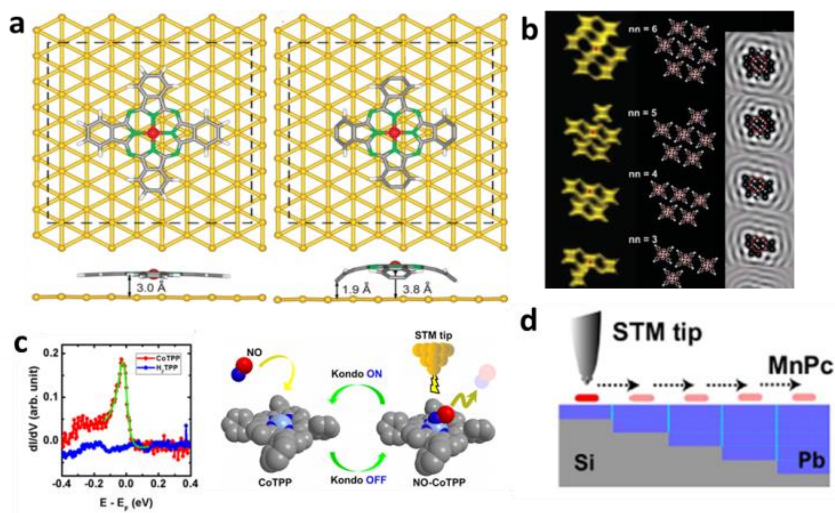

**Figure 3.3 Some example approaches to control the molecular Kondo effect. (a) deforming the molecular structure (b) changing the neighboring molecules (c) NO molecule absorption on the impurity site (d) the quantum size effects.**

### 3.2 Introduction

Recently, the Kondo effect manifested in magnetic organic molecules adsorbed on a metallic substrate has been extensively studied owing to their potential applications in molecular spintronics and magnetic storage.[152-156] Numerous experimental studies showed that the Kondo temperature $T_K$, a parameter describing the screening strength, ranges from millikelvin to hundreds of kelvin.[157-171] This large variation is attributed to distinctive molecule–substrate interaction and molecular spin configuration.[159-162] Such phenomena have triggered great efforts to manipulate the molecular Kondo effect, for example, by cleaving periphery atoms from the molecules, [162] attaching gas molecule(s) at the molecular spin center,[163, 164] modifying molecular conformation or structure,[165, 166] adsorbing the molecules at different substrate lattice sites,[157, 158] coupling magnetically with the neighboring molecular spin centers,[167, 168] or modulating the substrate electron density.[169-171] These studies primarily address individual molecules one by one. In this Article, we demonstrate that supramolecular interactions are able to effectively switch the molecular Kondo effect, which opens a route to control the Kondo effect in ensembles of molecules collectively.

We select Co-tetra-pyridyl-porphyrin (Co-TPyP) molecules adsorbed on Cu(111) and Au(111) surfaces as a model system, taking the advantage of single Co-TPyP molecules exhibiting a Kondo resonance on Cu(111) but no Kondo resonance on Au(111). We first demonstrate how to switch on and switch off the molecular Kondo effect on Au(111) and Cu(111), respectively, by organizing the molecules in specific supramolecular structures. Next, we present the spin-polarized electronic structures obtained from the ground-state spin-polarized density-functional theory (SP-DFT + U) calculations and use the Haldane relation derived from the single-orbital Anderson impurity model to compute the Kondo temperature.[172, 173] The computed values amounts to 85% of the experimental ones. On the basis of this good agreement, we propose a mechanism for the supramolecular-induced switching: The competition between the intermolecular interactions and the molecule–substrate interactions varies the molecular adsorption configuration, which effectively modulates the three determinative parameters in the Anderson model, namely, broadness, energy level, and on-site Coulomb repulsion energy of the spin state. [172, 173]

### 3.3 Experimental and simulation method

All samples were prepared in an ultrahigh vacuum system with a base pressure of $8\times10^{-10}$ mbar. Single-crystalline Au(111) and Cu(111) substrates were cleaned by $Ar^+$ sputtering and annealing cycles. Co-TPyP and molecules 1 and 2 were deposited onto the substrates that were held at room temperature. The evaporation temperatures were 340, 390, and 225°C for Co-TPyP, 1, and 2, respectively. Fe was evaporated using an e-beam evaporator. All STM and STS data were acquired at 5 K in a constant current mode. STS spectra were measured with the set point of V = −1 V and I = 300 pA, using lock-in technique with a modulation of 4 mV (rms) and a frequency of 1.5 kHz.

DFT calculations were performed on Au(111) and Cu(111)–(8 × 8) substrates for single molecules and an Au(111)–(6 × 6) substrate for the molecular array. Three atomic layers were considered in each slab model with the bottom two layers fixed. Structural optimizations adapted gamma-point-only K sampling. A 4 × 4 × 1 K-point sampling and a Gaussian broadening of 0.02 eV were used. The plane-wave cutoff energy was 400 eV. An optimized version of the van der Waals (optB88-vdW) density functional was applied in the VASP package.[191, 192]

**3.4 Results and discussion**

Panels a and b of Figure 3.4 show STM topographs of a single Co-TPyP adsorbed on Au(111) and Cu(111), respectively. On Au(111), the molecule features an elongated central protrusion and an apparent height of 2.1 Å. On Cu(111), the molecule features a saddle shape and a higher apparent height of 2.5 Å. Figure 3.4c shows the dI/dV spectra measured at the two species. On Au(111), Co-TPyP does not show a zero-bias anomaly, indicating absence of detectable Kondo effect at the experimental condition of 5 K. In contrast, on Cu(111), Co-TPyP exhibits a zero-bias dip that can be fitted with a Fano function[174]

$$\frac{dI}{dV} \propto \frac{[q+(V-V_0)/\Delta]^2}{1+[(V-V_0)/\Delta]^2} + G_{EL} \quad (3.3)$$

where q is the line-shape factor, $V_0$ is the offset of the resonance from the Fermi level, $G_{EL}$ is the background dI/dV signal,[175] and $\Delta = k_B T_K$. We conducted site-dependent measurements at the single molecules and found that the dip feature is localized at the center of the molecules, indicating the Kondo resonance is associated with the Co center. We surveyed ~30 single molecules and found that the fitted $T_K$ ranges from 142 to 152 K. The mechanism of the absence and presence of the molecule Kondo effect on the two surfaces will be discussed later. The two surfaces offer us playgrounds to explore on/off switching of the molecular Kondo effect.

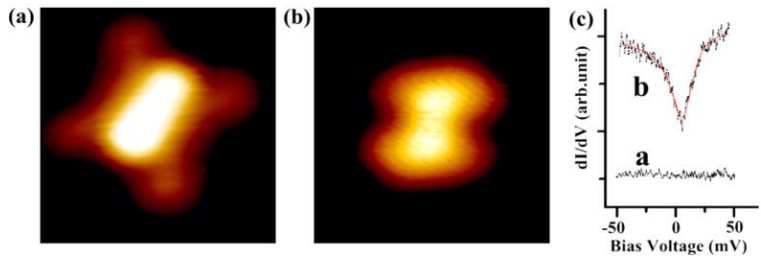

**Figure 3.4. STM topographs (2 × 2 nm², +0.8 V, 0.3 nA) of a single Co-TPyP adsorbed on (a) Au(111) and (b) Cu(111). (c) The dI/dV spectra measured at the two molecules shown in (a) and (b) (−1.0 V and 0.3 nA). The red curve represents the Fano fitting.**

On Au(111), after an annealing treatment at 100 °C, Co-TPyP molecules formed close-packed molecular arrays, as shown in Figure 3.5a. Figure 3.5d presents a structural model. The interaction that stabilizes the molecular arrays is intermolecular hydrogen bonds.[176] In the molecular arrays, the Co-TPyP displays a spherical central protrusion with an apparent height of 2.8 Å, substantially higher than the single molecules adsorbed on Au(111) (2.1 Å). The top curve in Figure 3.5g shows that the Co-TPyP in the molecular array exhibits a pronounced zero-bias resonance peak. It is worthwhile to note that the zero-bias anomaly of Co-TPyP is a dip on Cu(111) but a peak on Au(111). The line shape of the Kondo resonance reflects the nature of the tunneling process taking place in the STM junction.[174] Here we attribute the peak (dip) shape to a stronger (weaker) coupling of the Co spin states with the tip as compared with the surface-tip coupling.[177, 178] We fitted the peaks measured at ∼30 molecules using eq. 3.1 and obtained $T_K$ ranging from 100 to 118 K.

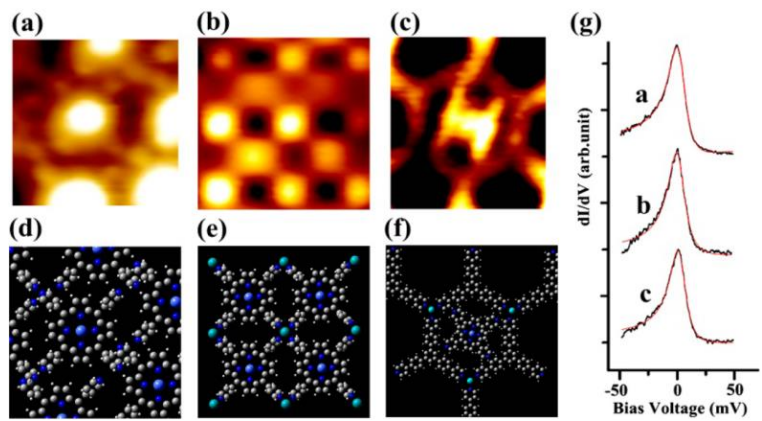

**Figure 3.5. Co-TPyP supramolecular structures formed on Au(111). STM topographs of (a) (3.5 × 3.5 nm², +0.8 V, 0.3 nA) a molecular array, (b) (2.5 × 2.5 nm², −1 V, 0.3 nA) a coordination network self-assembled with Fe (brighter protrusions), and (c) (3.5 × 3.5 nm², −1 V, 0.3 nA) a Co-TPyP trapped inside a 2D pore of the network 1. (d–f) Structural models of (a–c), respectively. (g) The dI/dV**

**spectra measured at Co-TPyP in the three supramolecular structures (−1.0 V and 0.3 nA). The red curves represent the Fano fitting.**

We applied another type of supramolecular interaction to organize Co-TPyP: Co-TPyP molecules and Fe atoms were deposited onto Au(111) and the sample was annealed at 150 °C. This process generated a metal–organic structure in which the Co-TPyP molecules are coordinated with Fe atoms.[179] Panels b and e of Figure 3.5 show a STM topograph and a structural model, respectively, of this supramolecular structure. The molecules in this structure also feature a larger apparent height of 2.4 Å. The middle curve in Figure 3.5g shows that the molecules in this supramolecular structure exhibit a pronounced zero-bias resonance peak. The fitted $T_K$ obtained from ∼30 molecules ranges from 91 to 98 K. The third approach to switch on the molecular Kondo effect is to prepattern the Au(111) surface with supramolecular networks. 1,3,5-Tris(4-(pyridin-4-yl)phenyl) benzene molecules and Fe atoms self-assembled into a network structure, denoted as network 1, on Au(111).[180] This structure contains two-dimensional (2D) hexagonal pores with an inner diameter of 2.6 nm. Subsequently, deposited Co-TPyP molecules were found trapped inside the supramolecular pores, as shown in Figure 3.5c. A structural model is drawn in Figure 3.5f, revealing that the size of the trapped Co-TPyP molecule is comparable to the pore size. Therefore, we expect the trapped molecules experience lateral interaction exerted by the supramolecular pores. This is evidenced by that fact that the apparent height of the trapped molecules is 2.4 Å, higher than that of the molecules adsorbed on bare Au(111) (2.1 Å). The bottom curve in Figure 3.5g is a tunneling spectrum measured on a trapped molecule, featuring a pronounced zero-bias resonance peak. The fitted $T_K$ obtained from ∼20 trapped molecules ranges from 93 to 107 K. In summary, the absent molecular Kondo effect in the single molecules adsorbed on bare Au(111) is switched on in all three supramolecular structures.

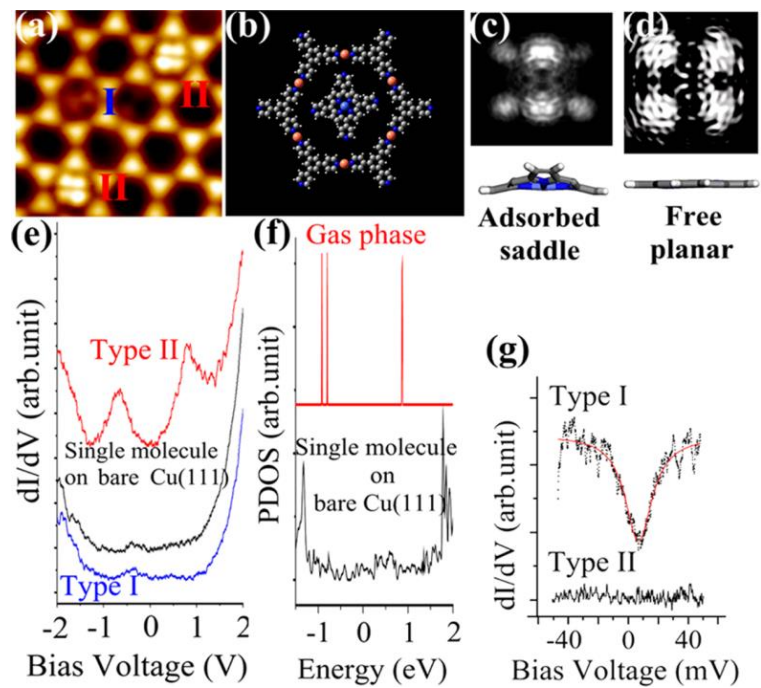

**Figure 3.6.** (a) STM topograph (9 × 9 nm$^2$, +0.8 V and 0.3 nA) showing single Co-TPyP molecules trapped inside the 2D pores of the network 2 on Cu(111), where I and II denote two types of adsorption, respectively. (b) Structural model of a Co-TPyP molecule inside a 2D pore. (c and d) DFT-calculated DOS contour maps at +0.8 eV of a single Co-TPyP adsorbed on bare Cu(111) and a gas-phase Co-TPyP. Side views of the porphyrin macrocyclic core are shown below. (e) The dI/dV spectra measured at a type-I and a type-II molecules, and a single molecule adsorbed on bare Cu(111). (f) DFT-calculated PDOS of a gas-phase Co-TPyP (top) and a single Co-TPyP adsorbed on bare Cu(111) (bottom). (g) The dI/dV spectra measured at the type-I and type-II molecules (−1.0 V and 0.3 nA).

We constructed a different type of supramolecular networks on Cu(111) to switch off the molecular Kondo effect on Cu(111). 1,3,5-Tris(pyridyl)benzene molecules were evaporated on Cu(111). After an annealing at 100 °C, the molecules coordinated with Cu adatoms to form networks, denoted as network 2.[181] The network 2 contains hexagonal 2D pores with an inner diameter of 2.7 nm. Panels a and b of Figures 3.6 present a STM topograph and a structural model of the network 2, respectively. When Co-TPyP molecules were deposited onto the sample, they were trapped inside the pores in two different configurations, denoted as type-I and type-II, respectively, as marked in Figure 3.6a. The type-I molecule dispalys a saddle shape and its apparent height is 2.3 Å, comparable to a single molecule adsorbed on bare Cu(111) (cf. Figure 3.4b). Figure 3.6e presents dI/dV spectra that were acquired in a bias window of [−2.0, 2.0 V] to resolve the frontier molecular orbitals. One can see that the type-I molecule and the single molecule adsorbed on bare Cu(111) give nearly identical dI/dV spectra with gradually increasing intensity beyond ±1.5 V without any peak-like features. Figure 3.6c shows a DFT-simulated DOS contour map of a molecule adsorbed on bare Cu(111), displaying 2-fold symmetry. The calculated PDOS, as presented at the bottom part of Figure 3.6f, is peakless within ±1.5 V due to hybridization of the molecular states with the substrate. Overall the theoretical results of a single molecule on bare Cu(111) reproduce the signatures of the type-I molecule fairly well. We thus infer the type-I molecules adsorb on the surface without being affected by the 2D pore. This picture is corroborated by the fact that the type-I molecules exhibit a dip-like zero-bias resonance as shown in Figure 3.6g, which resembles the single molecules adsorbed on bare Cu(111) (cf. Figure 3.4c).

The type-II molecules, however, behave very differently. First, they feature four lobes and a large height of 3.5 Å. Second, their dI/dV spectra show two pronounced peaks at +0.83 and −0.60 V (Figure 3.6e top). Third, the absence of a zero-bias anomaly shown in Figure 3.6g indicates that the Kondo effect is switched off. The greater adsorption height and the well-resolved molecular states hint that the type-II molecules are elevated and decoupled from the substrate. We compare the type-II molecules with a theoretically calculated gas-phase molecule. Figure 3.6d shows that the porphyrin core of the gas-phase molecule is planar and the simulated DOS contour map displays a four-lobe shape, matching the four-lobe STM of the type-II molecules. The calculated PDOS of the gas-phase molecule, as plotted in the top part of Figure 3.6f, nicely reproduces the two peaks resolved in the dI/dV spectrum. The good agreements between the gas-phase molecule and the type-II molecules imply that the type-II molecules are elevated from the substrate by the 2D pores and they are nearly free from the substrate coupling.

As summarized in Table 3.1, the supramolecular structures switch on the molecular Kondo effect on Au(111) but switch off the molecular Kondo effect on Cu(111). To understand the underlying mechanism, we performed SP-DFT calculations for a gas-phase molecule, a single Co-TPyP adsorbed on Au(111), an array of Co-TPyP adsorbed on Au(111), and a single Co-TPyP adsorbed on Cu(111). A Hubbard-like term $U_{eff}$ was employed to describe the metal complexes. An effective value of 6.2 eV was chosen so that the DFT + U calculated electronic structure matches the results obtained from the GW calculation reported in ref 206. The calculated frontier molecular orbitals with this $U_{eff}$ reproduce nicely the experimentally resolved molecular states (cf. panels e and f in Figure 3.6. Figure3.7a–c presents the optimized structures and Table 3.1 lists the calculated heights (defined as the distance between the Co and the substrate top layer). The calculated heights of the Co-TPyP on Au(111), Co-TPyP in the molecular array on Au(111), and single Co-TPyP on Cu(111) are 2.3, 2.8, and 2.6 Å, respectively. These values are consistent with the trend of the experimental apparent heights also shown in Table 3.1. Compare the molecular conformation of the single molecule and the molecule in the molecular array, one can see in the molecular array the pyridyl groups rotate out of the porphyrin core plane by ~45° (cf. side view in Figure 3.7b), In contrast, the pyridyl groups of the single molecule lie nearly coplanar with the porphyrin core (cf. side view in Figure 3.7a). Presumably, the intermolecular interactions in the molecular array impose a force to the pyridyl groups to cause their ~45° rotation. Under such a molecular conformation, the porphyrin core is lifted from the surface, as illustrated in Figure 3.7b (side view). Table 3.1 also lists the apparent molecular heights in other supramolecular systems. On Au(111), the molecules in the three supramolecular structures are higher than the single molecule. We argue that a same mechanism is at work here, that is, the intermolecular interactions in the supramolecular systems apply lateral force to lift the molecules. On Cu(111), the supramolecular structures do not change the height of the type-I molecules but significantly lift the type-II molecules. We ascribe this difference to different molecular adsorption position with respect to the supramolecular pore: the type-I molecule is at the pore center and interacts weakly with the pore boundary, whereas the type-II molecule nears the pore boundary and thus interacts strongly with the pore boundary. Hereby, we propose that the lateral intermolecular interactions may readily alter the molecular conformation and, consequently, the adsorption height.

**Table 3.1. Seven Adsorption Configurations of Co-TPyP, $T_K$ Fitted from the Experimental Zero-Bias Resonance, and Experimental Apparent Height (±0.1 Å, Imaged at −1.0 V) and DFT Calculated Height**

| substrate | configuration | $T_K$(K) | apparent height/DFT calculated height (Å) |
|---|---|---|---|
| Au(111) | Single molecule | N/A | 2.1/2.3 |
| Au(111) | In 2D pore 1 | 93-106 | 2.4 |
| Au(111) | Molecular array | 100-118 | 2.8/2.8 |
| Au(111) | Coordination network | 91-98 | 2.4 |
| Cu(111) | Single molecule | 142-152 | 2.5/2.6 |
| Cu(111) | Type-I in 2D pore 2 | 107-268 | 2.3 |
| Cu(111) | Type-II in 2D pore 2 | N/A | 3.5 |

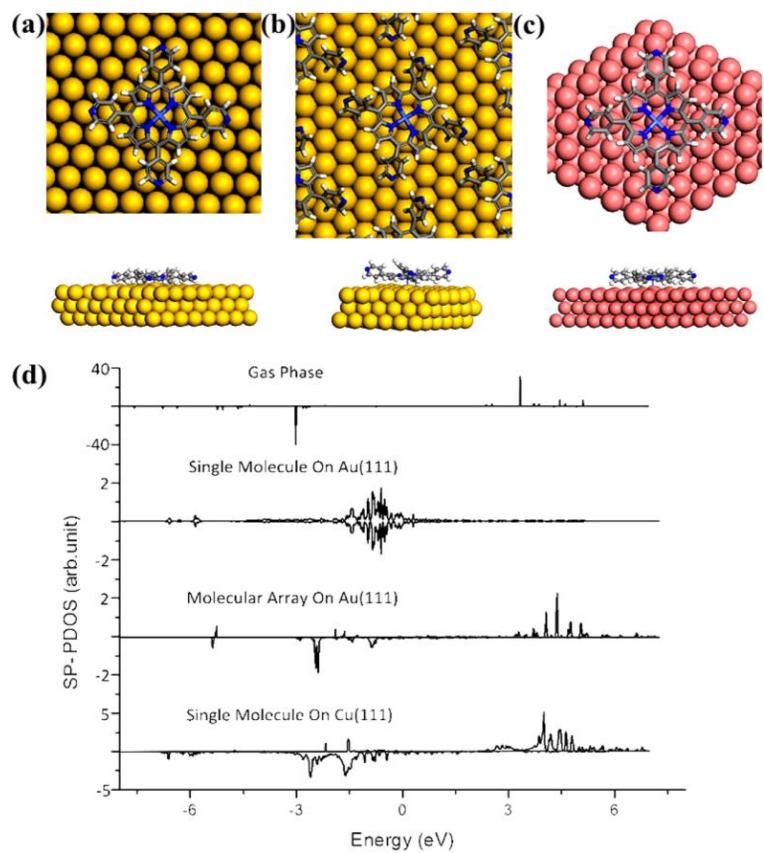

**Figure 3.7. DFT optimized structures of (a) a single Co-TPyP on Au(111), (b) an array of Co-TPyP on Au(111), and (c) a single Co-TPyP on Cu(111). (d) Spin-polarized Co $3d_{z^2}$ density of states of the gas-phase Co-TPyP, single Co-TPyP, and Co-TPyP molecular array on Au(111), and single Co-TPyP on Cu(111).**

The calculated spin-polarized projection densities of states (SP-PDOS) of Co $3d_{z^2}$ orbital are plotted in Figure 3.7d. We focus our discussion on the $3d_{z^2}$ orbital because other Co 3d orbitals are either fully occupied or empty. Note that the $3d_{z^2}$ orbital is known to hybridize to the substrate strongly than the other orbitals thanks to its shape and orientation.[160] In the gas-phase Co-TPyP, the Co atom is in a 3d7 state with a half-filled $3d_{z^2}$ orbital. The $3d_{z^2}$ orbitals of the three adsorbed molecules are split and broadened due to crystal field splitting and Co-substrate coupling. In addition, the coupling with the surbstrate shifts the orbitals with respect to the Fermi level. Despite these effects, the Co atoms in the single Co-TPyP on Cu(111) and the molecular array on Au(111) retain their 3d7 configuration with a half-filled $3d_{z^2}$ orbital. The most dramatic change occurs in the single Co-TPyP on Au(111) where the spin-up channel shifts below the Fermi level. As a result, the Co is in a 3d8 state and the net spin is diminished. This change can be attributed to a significant charge transfer (0.9 electrons, see Table 3.2) from the Au substrate to Co. Note that Perera et al. reported that the Co–Br2TPP molecules feature a planar conformation as deposited at a 120 K Cu(111) in which there is a charge transfer of 0.5 e between the Co center and the substrate.[161] Therefore, Co is in a 3d8 configuration with a nearly zero magnetic moment and the Kondo resonance is distributed at the porphyrin molecular lobes.

Table 3.2. Energy Level ($E_d$) and Width ($\Gamma$) of the Spin-Down Co $3d_{z^2}$ Orbital, the Coulomb Repulsion Energy U between the Spin-Up and Spin-Down Channels, Charge Transfer, a Magnetic Moment Derived from SP-DFT + U Calculation, and the Kondo Temperature ($T_K$) Computed Using eq. 3.2

| configuration | $E_d$ (eV) | $\Gamma$ (eV) | U (eV) | charge transfer* (e) | magnetic moment ($\mu_B$) | $T_K$ (K) |
|---|---|---|---|---|---|---|
| Gas Phase | −3.052 | ~0 | 6.317 | N/A | 1.07 | ~0 |
| Single/Au(111) | −0.723 | 1.02 | N/A | 0.9 | 0.01 | N/A |
| Array/Au(111) | −1.811 | 0.81 | 6.052 | 0.33 | 0.64 | 94 |
| Single/Cu(111) | −1.784 | 0.85 | 6.147 | 0.28 | 0.78 | 123 |

a Asterisk (*) indicates from the substrate to Co.

The most-widely used method to compute Kondo temperature is a combination of DFT and numerical renormalization group (NRG) solution of the Anderson impurity model.[177-186] An alternative is to use Haldane expression,

$$k_B T_K = \frac{1}{2}\sqrt{\Gamma U} e^{\frac{\pi E_d(E_d+U)}{\Gamma U}} \quad (3.2)$$

where $\Gamma$ is the broadness of the impurity state, $E_d$ is the energy level of the impurity state with respect to the Fermi level ($E_d < 0$), and U is the on-site Coulomb repulsion energy.[172,173] According to eq 3.2, $T_K$ monotonically depends on $\Gamma$, and a smaller $\Gamma$ leads to a lower Kondo temperature. $T_K$ dependence on U is not monotonic: exponential decays when U is small (U < 2|$E_d$|) and near linearly increases when U is large (U > 2 |$E_d$|). Generally speaking, a strong molecule–substrate coupling enlarges $\Gamma$ and reduces U. These parameters can be extracted from the DFT-calculated electronic structure.[187,188] It is worthwhile to note that using DFT calculated electronic structures to compute Kondo temperature often fails because of inaccurate molecular levels, and DFT + U methods are a common way to solve the problem.[189]

We quantified $\Gamma$, $E_d$, and U from the SP-PDOS as follows: $\Gamma$ equals the standard deviation of the spin-down states,[188] that is, larger splitting and broadening results in a larger $\Gamma$; $E_d$ is the weighted average level (WAL) of the spin-down states with respect to the Fermi level; U is the energy difference between the WAL of the empty spin-up states and the WAL of the filled spin-down states. Table 3.2 summarizes $\Gamma$, $E_d$, and U of the four configurations. The gas-phase molecule has a deep $E_d$, a near zero $\Gamma$, and the largest U. Upon adsorption, $\Gamma$ increases to between 0.8 and 1.0 eV, $E_d$ shifts closer to the Fermi level, and U is reduced. In the single molecule on Au(111), the spin-up channel coincides with the spin-down channel and both are filled. So eq. 3.2 is not applicable. The magnetic moments of the four configurations are also shown in Table 3.2. The adsorbed molecules exhibit a smaller magnetic moment than the gas-phase molecule (1.07 $\mu_B$). As the extreme case, the single molecules on Au(111) show a magnetic moment of just 0.01 $\mu_B$, apparently due to the 0.9 e charge transfer. One can see from Table 3.2 that broadening of $\Gamma$, up-shifting of $E_d$, reduction of U, and magnetic moment are in accordance with the trend of the charge transfer which speaks about molecule–substrate coupling strength.

Knowing $\Gamma$, $E_d$, and U, we can compute $T_K$ using eq 3.2. As shown in Table 3.2, the gas-phase Co-TPyP represents an extreme case of weak coupling where $\Gamma$ approaches zero and U is unreduced. As a result, the Kondo effect is absent. The single Co-TPyP adsorbed on Au(111) represents another extreme case where eq. 3.2 is not applicable as the magnetic moment approaches zero. As shown in Table 3.1, the type-II molecule in a 2D pore on Cu(111) is situated farthest from the surface while the single molecule on Au(111) is located closest to the substrate. The Kondo resonance is not observed in either case, but eq 3.2 explains different mechanisms are at work. The former case renders a very small $\Gamma$ because of the weak molecule–substrate coupling, whereas the latter case is associated with very strong molecule–substrate coupling which completely quenches the magnetic moment. We argue that Kondo effect cannot be experimentally detected unless the molecule–substrate coupling is neither too weak nor too strong, but falls in an appropriate regime that provides appreciable $\Gamma$, $E_d$, and U. Co-TPyP's in the other five configurations are subjected to intermediate molecule–substrate couplings, which give rise to appreciable $\Gamma$, $E_d$, and U, and hence detectable $T_K$. Table 3.2 shows that the single Co-TPyP adsorbed on Cu(111) and Co-TPyP in the molecular array adsorbed on Au(111) manifest Kondo resonance with $T_K$ at 123 and 94 K, respectively. These values are approximately 85% of the average $T_K$ of the two configurations, 147 and 110 K, obtained experimentally.

DFT calculation for the other four supramolecular structures is beyond our computational capability. On Au(111), the increased molecular height in the three supramolecular structures as compared with the single molecules suggests the intermolecular hydrogen bonds, the metal-molecular coordination bonds, and the supramolecular pores lift the Co-TPyP molecules. The Co $3dz^2$ orbital in the three configurations is expected to behave similarly and the Kondo effect is switched on. On Cu(111), the type-II molecules are elevated by the supramolecular pore so high that they are nearly free from the substrate coupling. Consequently, $\Gamma$ is greatly reduced, leading to a very low $T_K$ which switches off the Kondo effect at 5 K. We propose this scenario also explains the absence of Kondo resonance at Co-TPyP on Ag(111).[190]

**3.6 Conclusion**

In conclusion, through comparative investigation of seven configurations of Co-TPyP molecules that are either free from or subjected to intermolecular interactions in the different supramolecular structures on Cu(111) and Au(111), we have found that the intermolecular interactions can effectively switch off/on the Kondo effect in ensembles of molecules. The switching mechanism is rationalized in the frame of the Anderson impurity model, and the Kondo temperatures computed based on the DFT + U calculated electronic structures agree satisfactorily with the experimental values. These findings highlight a cooperative character of the molecular Kondo effect in the supramolecular systems.

# Chapter 4

# Single-Molecule Investigations of Conformation Adaptation of Porphyrins on Surfaces

## 4.1 Introduction

Adsorption of organic molecules at transition-metal surfaces is important because these systems show promise as components in (opto)-electronic devices [61,62, 193-197, 202-204]. Depending on the strength of the interaction between the molecule and the surface, the binding is typically classified as either physisorption or chemisorption [198-201]. In the case of relatively weak overlap of electron orbitals between the adsorbate and the substrate surface, the ubiquitous van der Waals (vdW) interactions together with the covalent or ionic bonding are frequently the majority force that binds the molecule to the surface [197]. By definition, the strength of the adsorption process is measured by the magnitude of adsorption energy, which is the sum of molecule-substrate bonding energy and deformation energy of gas-phase molecule. Generally, larger adsorption energy can lead to stronger molecule-substrate bonds and obvious structural deformation [205-209]. The adsorption energy for the system of organic molecules adsorbed on metallic substrates is ranging from hundreds of meV to several eV and the deformation energy involved is from tens of meV to hundreds of meV [210-215]. Taking the simplest benzene rings adsorbed on ordinary Au(111) and Pt(111) surfaces as an example, the interaction between the π orbital of benzene and d orbits of substrates leads to an adsorption height 0.2~0.3nm and adsorption energy 0.8 (Au) and 1.6 eV (Pt) [197]. Detailed computational and experimental adsorption energy and heights are listed in table 4.1 referred from [197]. With such large adsorption energy, all of the hydrogen atoms attached to the carbon ring are lifted up, which is a common type of deformation upon adsorption, as shown in figure 4.1.

**Table 4.1 Computational and experimental adsorption energy and heights of the Bz/Pt(111) system and Bz/Au(111) system [197].**

| System | Method | $E_{ad}$ [eV] | Height$_{ad}$ [Å] |
|---|---|---|---|
| Bz/Pt(111) | PBE+vdW | 1.96 | 2.08 |
| Bz/Au(111) | PBE+vdW | 0.74 | 3.05 |

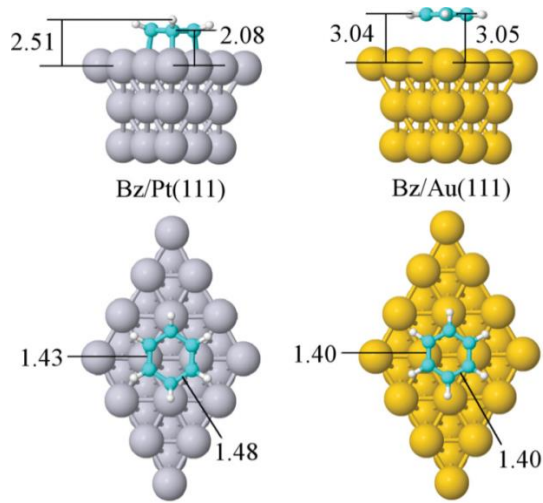

**Figure 4.1 Adsorption structures of the Bz/Pt(111) system and Bz/Au(111) system, both at the so-called bri-30° adsorption site [197].**

Importantly, the fundamental behaviors of molecules will also change, as the electronic and magnetic properties of a molecule are subject to the coupling of the molecule with its environment [216, 217]. One of the widely-studied systems is porphyrin molecules [63-71]. As shown in figure 4.2, the porphyrins are commonly observed as ruf, sad, dom and wav conformers due to different environmental conditions [72]. However, the gas-phase porphyrin is considered to be a planar conformation [218-221]. That is to say, the change of environmental condition, especially the adsorption process, can significantly alert the structural conformation and in order to engineer the molecular properties [72]. For example, the Fe(II) in a gas-phase FeP or Fe-phthalocyanine (FePc) is at an intermediate spin state of $S = 1$ [222-225]. However, the spin state will change from $S = 1$ to 2 if the molecule adsorbed on a substrate, either metallic or semiconductor. The underlying mechanisms multiple mechanisms, including (1) deformation of the macrocyclic ring, (2) Fe − Au(111) interaction, and (3) weakened Fe-N bonds, when the molecule is adsorbed on the surface [226-232]. Taking the advantage of this, the molecular structures can be sensitively controlled by changing environmental conditions and as a result, the relative electronic and magnetic properties can be engineered simultaneously.

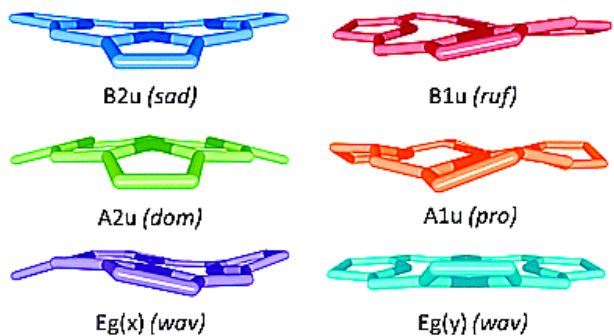

**Figure 4.2 [72] Different common conformation types of porphyrin.**

The structural flexibility of porphyrins allows their macrocyclic core to be distorted into non-planar conformations [234-244]. For example, outside a protein microenvironment, the heme macrocycle is nearly planar, while within the microenvironment it can be distorted into non-planar conformations, which is thought to be critical for the functions of porphyrin-containing proteins [245]. Similarly, non-planarity in synthetic porphyrins has been found to be associated with their redox potential, axial ligand affinity, spin state, excited state lifetimes and rate constants for metallation insertion [245]. This conformation-function relationship has given rise to the concept of "conformational control" [246-250], that is, the macrocycle conformation is tailored to achieve desired physicochemical properties. Various methods have been developed to control the conformation of the porphyrin macrocycle, including substituting bulky groups at the peripheral sites to introduce steric repulsion, metallating the macrocycle with metal centers and anchoring axial ligands at metal-porphyrins [245, 251-252].

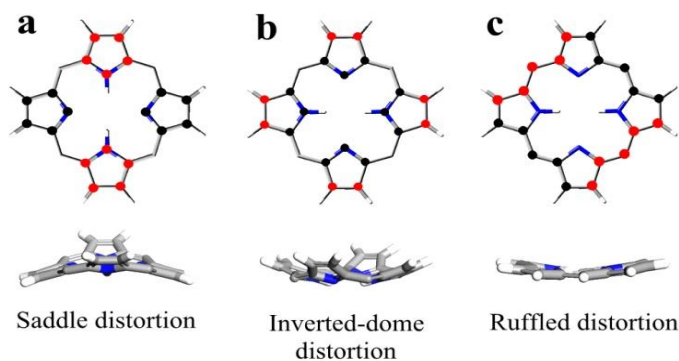

**Scheme 4.1. Non-planar conformations of the macrocylic core of porphyrins identified in this study: (a) saddle (*sad*), (b) inverted-dome (*dom\**) and (c) ruffled (*ruf*). Black and red dots indicate displacements on opposite sides of the mean plane.**

Recently, surface–anchored porphyrins have been extensively studied in the light of potential applications in single-molecule electronics and magnetism, photovoltaic, sensing and heterogeneous catalysis [253, 254]. As adsorbed on a surface, the molecule-substrate interaction restricts the structural flexibility, and the macrocycle core often adopts specific conformations, some rarely happen in free space [255-257]. In this regard, surface adsorption provides a route to control conformation, and hence functionality, of porphyrins. However, the dynamic pathway of the conformation transformation when porphyrins are adsorbed on surface is largely elusive. In this study, we used complementary theoretical and experimental methods to unravel the dynamic process of conformational change as well as the thermodynamic equilibrium structures of two porphyrin derivatives adsorbed on a Au(111) and a Pb(111) surfaces.

In free space, the porphyrin macrocyclic cores exhibit six out-of-plane normal deformations: saddle (*sad*), dome (*dom*), ruffled (*ruf*), waves (wav-x and wav-y), and propeller (*pro*) [245-250]. Each of these conformations corresponds to a lowest-frequency normal mode of a symmetry type. Scheme 4.1 depicts three of them that were found to be stabilized on the surfaces in our study: the *sad* features pyrrole rings tilted alternately up and down, the *dom* or inverted dome (*dom\**) features four pyrrole rings tilted in the same direction, and the *ruf* features meso carbons displaced alternately up and down [258]. We found that the free-space molecules undergo a multiple-step conformational transformation on the surface before reaching the equilibrium states. We also demonstrated that the conformation of the surface-adsorbed porphyrin depends on the choice of surface or the peripheral groups.

**4.2 Results and discussion**

The first molecule we studied is 5,15-dibromophenyl-10,20-diphenylporphyrin (Br$_2$TPP). We used DFT to analyze Br$_2$TPP adsorbed on a three-layer Au(111) slab. Initially, we set the molecule in six conformations (planar, *dom*, *dom\**, *pro*, *ruf* and *sad*) and relaxed these structures at different adsorption sites and heights on top of the Au(111) slab. The calculations revealed that, except *ruf*, the rest five conformers could converge to energetically stable configurations. Fig. 4.3 shows the optimized structures of the five conformers. The adsorbed molecule retains their initial conformational characteristics. Table 4.2 lists the corresponding adsorption height (the porhyrin macrocycle plane to the top layer Au atoms), adsorption energy and distortion energy. The adsorption energy is defined as $E_{adsorption} = E_{Au} + E_m - E_{total}$, where $E_{total}$ is the energy of the molecule with the Au slab, $E_{Au}$ is the energy of the Au slab, and $E_m$ is the energy of a free-space Br$_2$TPP molecule at its ground state, which is in a planar conformation. The distortion energy is defined as $E_{distortion} = E_{conformer} - E_m$, where $E_{conformer}$ is the energy of the same structures as shown in Fig. 4.3 but without the Au(111) slab. $E_{distortion}$ thus quantifies the cost of distorting a planar molecule to a specific conformation in free space. It can be viewed as an energy barrier for a ground-state molecule in free space to reach a specific conformation adsorbed on the surface. As listed in Table 4.2, amongst the five structures, the *sad* conformer is adsorbed with the highest adsorption energy (2.16 eV), also with the highest distortion energy (0.49 eV). Therefore, this structure is energetically most favored, but kinetically slow to reach. The second most stable structure is the *dom\** conformer (1.35 eV) with lower distortion energy (0.41 eV). The *pro* conformer features intermediate adsorption energy (1.21 eV) and distortion energy (0.38 eV). The *dom* and planar conformers are weakly adsorbed (0.73 eV and 0.91 eV), but the *dom* has a much high distortion energy. Overall the *dom* is an un-favored structure both energetically and kinetically.

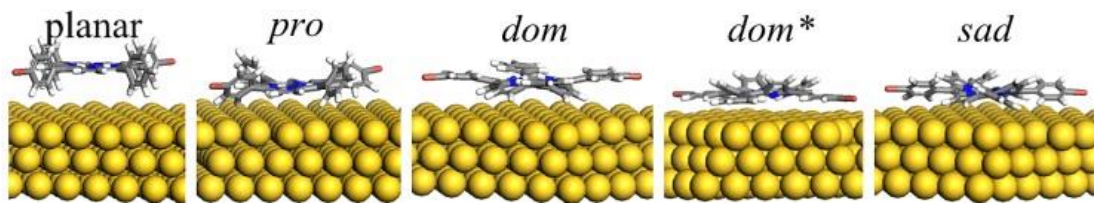

**Figure 4.3. DFT-optimized structures of five conformers stabilized on Au(111).**

**Table 4.2. DFT-optimized height, adsorption energy and distortion energy of Br$_2$TPP of five converged structures on Au(111).**

| Conformation | planar | *pro* | *dom* | *dom\** | *sad* |
|---|---|---|---|---|---|
| Adsorption Height (Å) | 3.95 | 3.80 | 3.80 | 3.50 | 3.75 |
| Adsorption Energy (eV) | 0.91 | 1.21 | 0.73 | 1.35 | 2.16 |
| Distortion Energy (eV) | 0.08 | 0.38 | 0.45 | 0.41 | 0.49 |

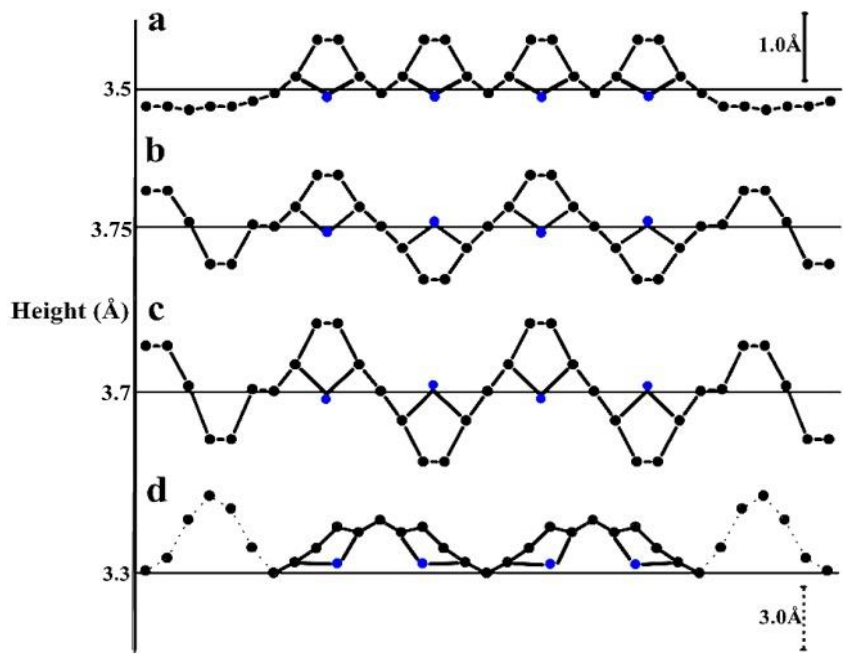

**Figure 4.4. Vertical heights of the porphyrin atoms with respect to the top-layer surface atom plane (only two phenyl groups are shown on the two sides). (a)** *dom\** **Br$_2$TPP on Au(111). (b)** *sad* **Br$_2$TPP on Au(111). (c)** *sad* **Br$_2$TPP on Pb(111). (d)** *ruf* **DPP on Au(111). Blue: N, black: C.**

In Fig. 4.4, we plot the DFT-calculated vertical heights of the atoms with respect to the top atom plane of the substrate of the two stable conformers of *dom\** and *sad* on Au(111). In the *dom\** conformer (Fig. 4.4(a)), all four pyrrole moieties bend upward in the same manner, while the peripheral phenyl groups (two are plotted on two sides) lie nearly parallel to the substrate. The *sad* conformer (Fig. 4.4(b)) features pyrrole moieties displaced alternately up and down. The peripheral phenyl groups are tilted with respect to the surface.

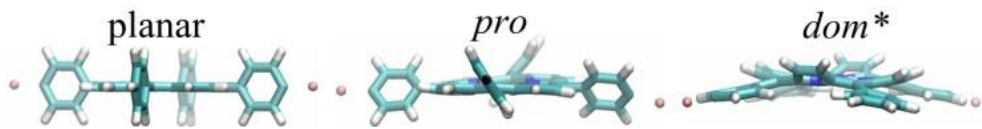

**Figure 4.5. Three conformations of Br$_2$TPP in free space found in the 550 K MD trajectories.**

All conformers are at a height more than 3.5 Å above the top atom layer of the surface, implying the molecule does not form strong chemical bonds, but interact weakly, with the substrate atoms. Presumably, the conformers undergo dynamic structural relaxation at finite temperatures. The DFT calculations only provide the ground-state energy and structure, i.e., at zero temperature. We applied molecular dynamics (MD) simulation to obtain the dynamic conformations at the experimental-relevant temperatures. We first simulated the conformation dynamics for a free-space molecule. Consider the planar is the lowest-energy conformation in free space, we set it as the initial conformation in free space and sampled the structures at 550 K, comparable to the temperature of evaporating Br$_2$TPP in the experiments. The simulations revealed that the molecule was transformed alternatively among planar, *pro*, and *dom* (*dom\**) conformations, as depicted in Fig. 4.5, in two independent 500 ns MD trajectories. The *sad* conformation was not observed in the MD trajectories. These results corroborate the DFT-calculated distortion energy of these conformers shown in Table 4.2.

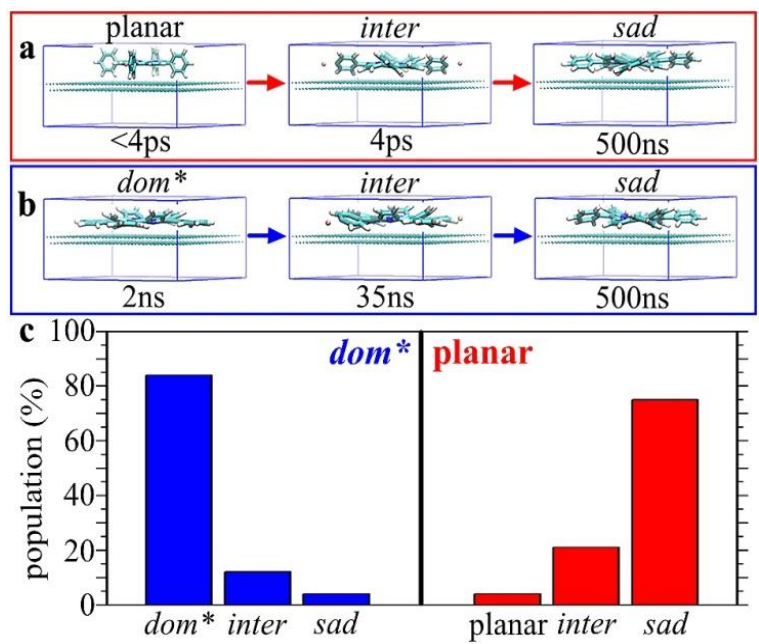

**Figure 4.6. Snapshots of MD simulations for Br$_2$TPP to adopt the (a) planar and (b) inverted-dome (*dom\**) conformations adsorbed on a Au(111) substrate. (c) The populations of the final conformation of Br$_2$TPP obtained from 100 paralleled temperature-annealing MD simulations initialed from the *dom\** (left panel) and planar conformations (right panel), respectively. *inter* stands for the intermediate structures.**

Next, we used MD simulations to resolve the structural relaxation of the planar and *dom\** conformers when they are adsorbed on a Au(111) substrate. As the snapshots shown in Fig. 4.6a and b, the planar conformer quickly transformed to a *sad* conformation within 4 ps at 250K (the substrate temperature when the molecules were deposited). In sharp contrast, at the same temperature, the *dom\** conformer retained its conformation for 2000 ps before transforming to an intermediate conformation, and finally reached the *sad* conformation after 35 ns (see Movies in SI). Transitions among different conformations (*sad*, *dom\** and planar) were simulated using metadynamics path-collective-variables simulations (SI), revealing that the *dom\** conformation is metastable with a small free energy barrier (~10 kJ/mol) preventing its transition towards the *sad* conformation (Fig. S4.5a), while the planar-to-*sad* transition is barrier-free (Fig. S4.5b). We also simulated the cooling processes of $Br_2TPP$ on Au(111) (SI). Fig. 4.6c presents the statistical analysis of the final conformations obtained from 100 paralleled cooling MD simulations of each case. While the system cools down from 250K to 5K (the STM imaging temperature), the molecules in the *dom\** conformation are very stable with over 80% of them still keep their conformations unchanged. However, under the same condition, more than 70% of the planar ones change into the *sad* conformation. We also carried out the simulations at higher temperatures (300, 350, and 450 K) and found that the life time of the *dom\** is shortened (Table S1) while the *sad* conformer is preserved at the same temperatures in the whole trajectories.

The DFT and the MD results thus unravel a unified picture of how the molecule adapts the surface adsorption: In free space, the planar conformation is energetically most favored and the *sad* is the least favored, while at 550 K the planar conformation could alternatively transform among the planar, the *pro* and the *dom* conformations but not to the *sad*. Upon adsorption on a Au(111) surface, however, the *sad* becomes the energetically most stable conformation, followed by the *dom\** and *pro*. At 250 K, the planar conformer quickly transforms to the *sad* conformation, while the *dom\** has a longer life time before transforming to the *sad*.

Experimentally, we deposited $Br_2TPP$ molecules on a clean Au(111) substrate held at 250 K and transferred the sample to a scanning tunneling microscope (STM) stage cooled at 5 K. Fig. 4.7a is a STM topograph showing at lower molecular coverage, the molecules organized into dimers; at higher coverage, the molecules formed long chains (Fig. S4.1). Careful inspection of the STM data revealed two distinctive configurations of the molecules (Fig. 4.7b, two upper panels): configuration I (conf-I) features a groove along the central axis of the molecule with two central protrusions on each side, which resembles the previously reported saddle-shape conformation [259, 260]; configuration II (conf-II) appears flat without internal corrugations. Surface-adsorbed porphyrins often exhibit distinctive topographic and electronic characteristics originated from various mechanisms, for example, deprotonation or metalation of the macrocyclic core [261-264], proton tautomerization [265-267], adatom attachment [268], or adsorption-site dependent conformational variation [269]. We have conducted series control experiments and concluded that these effects cannot explain the two configurations observed here (SI).

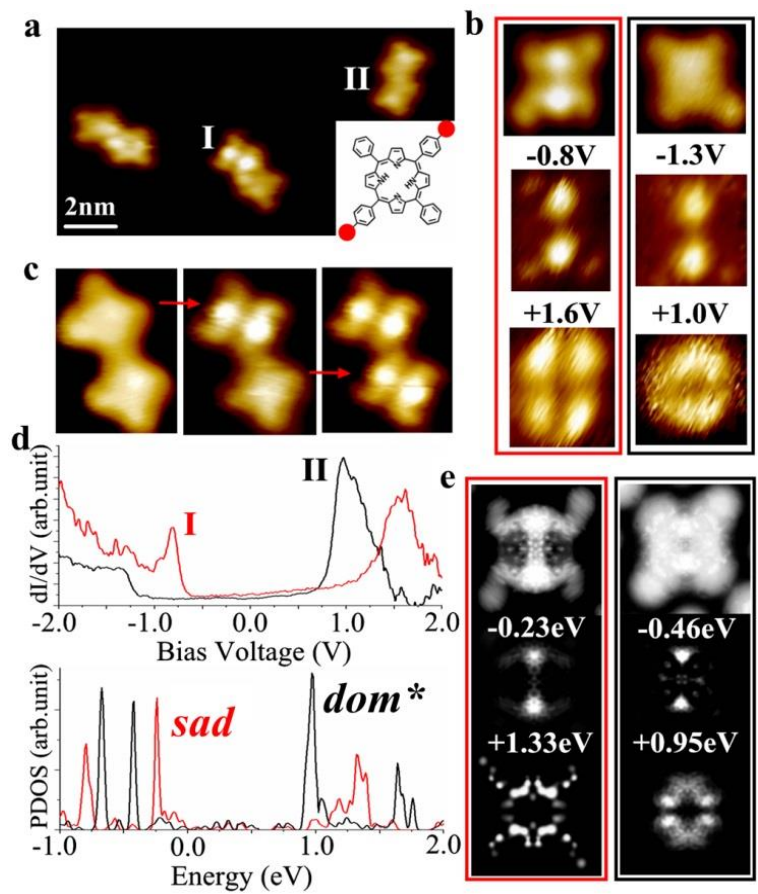

**Figure 4.7**. **(a) STM topograph of Br$_2$TPP molecules on Au(111). Inset: Chemical structure of Br$_2$TPP (red dots indicate Br). (b) Upper panels: STM topographs of a conf-I molecule (left) and a conf-II molecule (right). Middle and lower panels: STS**

maps of a conf-I molecule (left) and a conf-II molecule (right) acquired at the indicated voltage. (c) STM topographs showing conversion of a dimer from conf-II to conf-I. (d) Upper panel: dI/dV spectra of molecules with each configuration. Lower panel: DFT-calculated projected densities of states (PDOS) of the two conformers. (e) Simulated STM images and constant-height PDOS contour maps of the *sad* (left) and *dom\** (right) conformers at the indicated energy. (STM image conditions: -1 V and 0.3 nA. Scale bars, 2 nm.)

The theoretical analysis presented above suggest that the *sad* and *dom\** conformers are the two most likely structures. Here we compare the experimental-resolved topographic and electronic features of conf-I and conf-II with the *sad* and *dom\** conformers. Fig. 4.7e (two upper panels) shows the simulated images of the two conformers. The simulated image of the *sad* conformer (red frame) reproduces the two-protrusion feature of the conf-I molecules in the STM topographs (Fig. 4.7b, upper panels), while the simulated image of the *dom\** conformer (black frame) features a square shape with a dent in its center, which is absent in the STM image of the conf-II molecule. We attribute the absence of the central dent to tip convolution effects in STM data acquisition.

Fig. 4.7d (upper panel) presents the differential tunneling (dI/dV) spectra acquired at the molecules of the two configurations. The conf-I spectrum (in red) has a broad peak at 1.6 V and a sharp, half-height peak at -0.8 V. The conf-II spectrum (in black) has a strong peak at 1.0 V and a shallow step at -1.3 V. The DFT-calculated projected densities of states (PDOS) of the two conformers are plotted in Fig. 4.7d (lower panel). The HOMO and LUMO levels of the *dom\** conformer are downshifted relative to those of the *sad* conformer, which is consistent with the shift in the dI/dV peaks of conf-II relative to the peaks of conf-I. STS maps of molecules in the two configurations at the characteristic energies are shown in Fig. 4.7b (middle and lower panels). The occupied states of both configurations display two lobes symmetrically positioned with respect to the central axis. The empty states of the two configurations display quite different spatial distributions: conf-I shows four lobes separated by two orthogonal central axes, whereas conf-II shows two half-moon shapes divided by a groove along the central axis. These STS maps can be compared with the DFT-simulated constant height contour maps of HOMO and LUMO as shown in Fig. 4.7e (middle and lower panels). The HOMO maps of the two conformers are nearly identical, with both showing two protrusions. In contrast, the LUMO maps of the two conformers are quite different: the *sad* conformer displays a four-fold pattern, while the *dom\** conformer displays a two-fold pattern. The HOMO and LUMO contour maps of *sad* match well the occupied and empty states of the conf-I resolved by STS (Fig. 4.7b), while the HOMO and LUMO contour maps of *dom\** match the occupied and empty states of conf-II molecules. Taken together, comparison of the experimentally-resolved topographic and electronic data with the DFT results supports the assignment of the conf-II (conf-I) to the *dom\** (*sad*) conformer.

We found that annealing the sample to 450 K raised the population of conf-I molecules from 25% to 75% (SI), indicating that thermal excitation may convert *dom\** conformers to *sad*. We also used the STM tip to convert *dom\** conformers to *sad*. An example of a sequential conversion of a covalently-linked dimer, which was formed by 450 K annealing, is shown in Fig. 4.7c: in the first step, a 2-V voltage pulse was applied when the tip was positioned on top of the upper molecule; after acquiring an STM image the second 2-V pulse was applied to the lower molecule. The conversion did not displace or rotate the molecule, indicating that the conversion involves a subtle conformation change. Such tip-induced conformation conversion was reported before [270]. It is worthwhile to note that the molecules cannot be changed from conf-I to conf-II using tip-applied voltage pulse. Both thermal- and electron-excited conversions manifest that the *dom\** conformers can be transformed to *sad* but not vice versa, which agrees with the DFT-calculated energies of the two structures and the MD-simulated temperature dependent structural transformation.

Here we discuss the effects of surface and periphery groups to the conformation adaptation. To investigate the surface effects, we used Pb(111) as the substrate. We deposited $Br_2TPP$ on Pb(111) following the same procedure as on Au(111). Fig. 4.8a is a STM topograph showing that the $Br_2TPP$ molecules formed close-packed islands on Pb(111). The molecules show bright protrusions at their four corners and a dark depression in the center. Thorough inspection revealed that all molecules exhibit the same features, evidencing that the $Br_2TPP$ molecules adopt a single conformation on Pb(111). We applied DFT to analyze a *sad* and a *dom\** conformers of $Br_2TPP$ on Pb(111). The adsorption energy is 1.89 eV for the *sad* conformer and 0.72 eV for the *dom\** conformer. On both surfaces, the *sad* conformers are adsorbed strongly (2.16 eV on Au(111) and 1.89 eV on Pb(111)) as compared with the *dom\** conformers (1.35 eV on Au(111) and 0.72 eV on Pb(111)). Nevertheless, Pb(111) renders weaker molecule-substrate interaction for both conformers. Presumably, the *dom\** conformers undergo faster structural transformation as deposited on Pb(111) and relax to the *sad* conformation in very short time. This explains the experimental observation that only one type of molecules were present on Pb(111). In other word, the weaker molecule-substrate interaction makes Pb(111) more selective in adapting the molecular conformation.

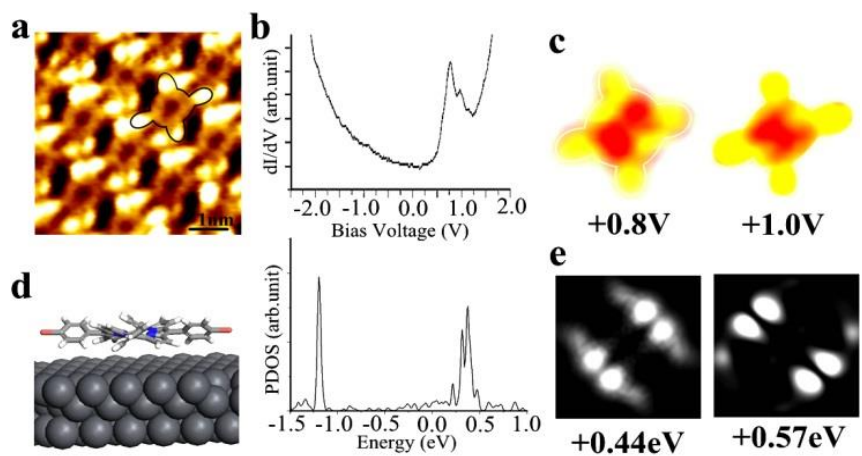

Figure 4.8. (a) STM topograph of a close-packed island of Br$_2$TPP on Pb(111) (image conditions: 5.5 × 5.5 nm$^2$, -1 V and 0.1 nA). (b) Upper panel: molecular dI/dV spectrum. Lower panel: PDOS of a *sad* conformer. (c) STS maps at the indicated bias voltage. (d) DFT-optimized structure of a *sad* conformer. (e) Constant-height PDOS contour maps of a *sad* conformer at the indicated energies.

The intra-molecular features resolved in Fig. 4.8a appear differently from those shown in Fig. 4.7b despite both are of the *sad* conformation. To understand this difference, we inspected the electronic structures of Br$_2$TPP on Pb(111). A dI/dV spectrum (Fig. 4.8b) shows two split peaks at 0.8 and 1.0 V, which contrasts the single broad peak shown in Fig. 4.7d. The spatial distribution of the two peaks is depicted in the STS maps shown in Fig. 4.8c. Both maps feature a depression line dividing the molecule into two halves (Note that the dividing lines of the two maps are perpendicular with each other.), which is different from the four-lobe map shown in Fig. 4.7b. The calculated

PDOS (Fig. 4.8b, lower panel) nicely reproduces the two-peak features resolved in the dI/dV spectrum, showing a peak at 0.44 eV and one at 0.57 eV. The PDOS contour maps at the two peak energies are displayed in Fig. 4.8e, which are comparable with the STS maps (Fig. 4.8c).

To trace down the origin of the differences exhibited by the *sad* conformer on Au(111) and Pb(111), we compare the vertical heights of the atoms of the *sad* conformers adsorbed on Au(111) and on Pb(111), as presented in Fig. 4.4b and 4.4c, respectively. The atoms of both conformers exhibit the same pattern of distortion. However, the displacements from the mean molecular plane on Pb(111) are ~25% larger than those on Au(111), indicating that the *sad* conformer is more distorted on Pb(111). This larger distortion of the macrocyclic core lifts the degeneracy of two LUMOs, which are 90° rotated with respect to each other in a planar porphyrin [271, 272], to result in the split double peaks and the 90° rotated STS maps.

Last, to study how peripheral groups affect the conformation adaptation, we designed a molecule bearing two phenyl moieties at positions 5 and 15 of the porphyrin (5,15-diphenylporphyrin, DPP, chemical structure shown in Fig. 4.9a). STM images revealed that as adsorbed on Au(111), the DPP molecules appear an elliptical shape, with two bright ends that can be assigned to the two phenyl moieties, and a depression in the porphyrin core (Fig. 4.9b). All molecules in the STM image show the same features (Fig. S4.3). So we conclude DPP molecules adopted a single conformation on Au(111).

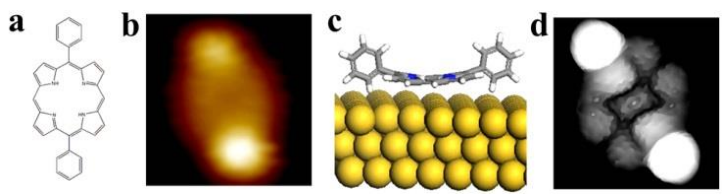

**Figure 4.9. (a) Chemical structure of DPP (b) STM topograph of a single DPP molecule adsorbed on Au(111) (2 × 2 nm$^2$, -1 V and 0.3 nA). (c) DFT-optimized structure and (d) DFT-simulated image of the *ruf* conformer.**

Considering that the *ruf* conformation matches the two-fold symmetry of the molecule, we carried out DFT analysis on a *ruf* and a *sad* conformers on Au(111). The adsorption energies are 1.75 eV for *ruf* and 0.82 for *sad*. Therefore it is expected that at room temperature, the *ruf* conformers are stably adsorbed on Au(111). Fig. 4.9c shows the DFT-optimized structure of the *ruf* conformer. The two phenyl moieties stand upright, and the macrocyclic core shows a ruffled distortion. The *ruf* conformation is evident in Fig. 4.4d which shows that the displacement of the two pyrrole moieties on the left (with respect to a central atom) is a mirror reflection of the displacement of the two pyrrole moieties on the right and both sides bend upward. The two phenyl groups erect on the surface with an outstanding height (note that the phenyl atoms are plotted in a different scale). The DFT-simulated map of the *ruf* conformer (Fig. 4.9d) shows two bright protrusions corresponding to the erecting phenyl groups and a ruffled macrocylic core with a depression. This map agrees fairly well with the STM image (Fig. 4.9b). In contrast, the simulated *sad* conformer appears very different from the STM data (see Fig. S4.4). Interestingly, the *ruf* conformation is unstable for $Br_2TPP$ but becomes stable for DPP; whereas the most stable *sad* conformation for $Br_2TPP$ becomes less stable for DPP. This comparison exemplifies that removing/attaching periphery side groups may drastically change the energy landscape of specific conformers adsorbed on a surface. Three main effects of periphery side groups are at work: (1) steric repulsion with the macrocyclic core, (2) interaction with the substrate, and (3) raising the macrocyclic core from the substrate. The conformation of the macrocylic core is determined by a subtle balance of the three effects. Hence we propose that periphery substitution is an effective mean to control the conformation of the porphyrins adsorbed on surfaces.

## 4.3 Conclusion

In summary, we have investigated the dynamic conformation adaptation processes of porphyrin molecules as being adsorbed on surfaces. Our results evidence that the macrocyclic core of the surfaces-adsorbed porphyrin molecules can be distorted to the conformations (dome or ruffled) that are not present or less-stable in free space, highlighting the prominent roles of the molecule-substrate interaction in determining the porphyrin conformation. The atomic modeling allows us to resolve the detail pathway of the surface-induced structural transformation. Moreover, our comparative studies of choosing different surfaces and molecules with different periphery groups shed lights on the effects of these parameters.

## 4.4 Synthesis of the molecules

The synthesis of **5,15-bis-(4-bromophenyl)-10,20-diphenylporphyrin** ($Br_2$TPP) and **5,15-dibromo-10,20-diphenylporphyrin** ($Br_2$DPP) has been described in our earlier works [273, 274].

## 4.5 Experimental and computational methods

**Experiment.** All the experiments were conducted in an ultrahigh vacuum system with a base pressure of $8\times10^{-10}$ mbar. Single-crystalline Au(111) and Pb(111) substrates were cleaned by $Ar^+$ sputtering and annealing cycles. $Br_2$TPP and $Br_2$DPP were deposited using an organic molecular evaporator onto the substrates, which were held at 250 K. The evaporation temperatures were 325 °C and 285 °C for $Br_2$TPP and $Br_2$DPP respectively. After annealing at 450 K, $Br_2$DPP molecules did not form covalently-linked chains but closely packed as molecular islands (shown in Fig. S3). Presumably, the Br groups were replaced by H to form DPP. All STM and STS data were acquired at 4.9 K in a constant current mode. The set point of V = -1 V and I = 300 pA was applied to measure STS spectra, using lock-in technique with a modulation of 10 mV (rms) and a frequency of 1.5 kHz.

**DFT calculation.** DFT calculations were performed in the VASP package. The plane-wave cutoff energy was 400 eV. For the substrates Au (111) and Pb (111), a (8×8) system with three atomic layers was employed. Among the three layers, only the top layer was relaxed with the bottom two layers fixed. Structural optimizations adapted gamma-point-only and 4 × 4 × 1 K−point sampling. A Gaussian broadening of 0.02 eV and an optimized version of the van der Waals (optB88-vdW) density functional were used.

**Computational methods of molecular dynamics (MD) simulation**

**Derivation of force field parameters.** To obtain the dynamic conformations of $Br_2$TPP both in free space and on the Au(111) substrate, we performed all-atom MD simulations by using GROMACS (version 4.5.4) software [275]. The atom types and force field parameters of $Br_2$TPP were taken from the General Amber Force Field (GAFF) [276] except the partial charge of each atom. To obtain the partial charges, we first optimized the geometric structures of $Br_2$TPP, then calculated the electrostatic potential, and finally we applied the Restrained Electrostatic Potential (RESP) method [277, 278] to derive the partial charges. Here, all the quantum mechanics calculations were performed by the Density Functional Theory (DFT) at B3LYP/6-31G* level using the Gaussian 09 software [279]. The charge for Au was set to zero. The Lennard-Jones parameters for Au were assigned as $\sigma_{AuAu} = 0.32$ nm and $\varepsilon_{AuAu} = 0.65$ kJ/mol, respectively [280].

**MD simulation set-up.** We performed all-atom MD simulations for $Br_2TPP$ on Au(111) surface to obtain the corresponding conformations. A crystalline substrate of atomic gold was setup with the Au(111) facet exposed, with surface dimensions of 6.05 nm $\times$ 5.99 nm. The substrate was two atomic layers thick in the direction normal to the substrate. The positions of all Au atoms were kept frozen during the MD simulation. Then, we chose the representative planar, dome and saddle conformations of $Br_2TPP$ and placed each of them on the Au(111) surface, respectively, as the initial conformations for MD simulations. The distance between the center-of-mass of $Br_2TPP$ molecule and Au(111) substrate were set to 3.8 Å, consistent with previous DFT results [281] Next, we placed both the $Br_2TPP$ and the Au(111) substrate in a rectangular box, with size 6.05 nm $\times$ 5.99 nm $\times$ 3 nm. As a result, there was a 20 Å vacuum layer above the $Br_2TPP$ molecule to avoid interactions due to periodic images. Then, we performed energy minimization by steepest descent algorithm, followed by two independent paralleled 300 ns MD simulations for each system at 250 K, 300 K, 350 K and 450 K, respectively (see Table S4.1). During MD simulations, the $Br_2TPP$ on Au(111) substrate was fully relaxed. The conformations were stored at a time interval of 10 ps. We then collected 60,000 conformations of each system at each temperature, respectively. The trajectories of 250 K simulations are shown in the three movies (M1_planar, M2_dom and M3_sad). Each movie runs for 40 ns and each frame elapses 10 ps if not indicated otherwise.

For the system in free space, we chose the DFT optimized structure of $Br_2TPP$ in vacuum (a planar structure) as our initial conformation. We first performed energy minimization using steepest descent algorithm. Then, we performed two independent 500 ns long production MD simulations at 550 K, respectively. The conformations were stored at a time interval of 10 ps. In total, we collected 100,000 conformations for analysis.

| Table 4.3. The conformational change time for Br$_2$TPP on Au(111) substrate from the *dom\** to the *sad* conformations via an intermediate state obtained from MD simulations at different temperatures. Temperature (K) | 250 | 300 | 350 | 450 |
|---|---|---|---|---|
| Changed to intermediate (ps) | 1950 | 250 | 10 | 2 |
| Changed to *sad* (ps) | 35200 | 4450 | 850 | 4 |

**Set-up of temperature annealing simulations for Br$_2$TPP on the Au(111) substrate.** To illustrate the stability and the distinct populations of different conformations (planar and dome conformations) of Br$_2$TPP on the Au(111) substrate, we performed large number of annealing simulations to mimic the experimental cooling process. First, we chose the planar and dome conformations obtained from DFT calculations as the initial conformations and placed each of them on the Au(111) substrate. Here, the box size and the distance between Br$_2$TPP molecule and Au(111) substrate are the same as the above setup for Br2TPP on Au(111) substrate. Then, we performed energy minimization by steepest descent algorithm. Next, we performed temperature-annealing MD simulations with temperature linearly reduced from 250 K to 5 K within 3 ns followed by 2 ns MD simulation at final temperature for further relaxation of the system. From each initial conformation (i.e. planar and dome conformation), we initiated 100 independent annealing simulations with different initial velocities. Finally, we collected the final conformation of each MD simulation for further analysis.

Here, all the production MD simulations were performed under NVT ensemble. The time step was 2 fs. Weak couplings to the external heat bath were applied based on the V-rescale [282] schemes. Periodic boundary conditions were applied in all three directions to minimize the edge effects in a finite system. In addition, the long-range electrostatic interactions were evaluated by the Particle Mesh Ewald (PME) method [283, 284]. The cutoff distances for both short-range electrostatic interactions and the Van der Waals interactions were chosen as 1.0 nm.

**4.6 Self-assembly of Br$_2$TPP on Au(111).**

Br$_2$TPP molecules form linear chains as deposited on Au(111) without annealing treatment. Fig. S4.1(a) shows a STM topograph. The molecules aggregate as chains along the herringbones of the reconstructed Au(111) surface or form closely-packed islands. The neighboring molecules' distance indicates the molecules are packed with weak inter-molecular interactions. STM data reveal that the majority molecules (80%) are in conf-II, as shown in a representative image of Fig. S4.1(a), which was acquired under a specific tip condition while the conf-II molecules appeared brighter. After annealing the sample to 450 K, the 75% molecules are in conf-I (Fig. S4.1(b)), indicating that thermal excitation may convert the conf-II molecules to conf-I.

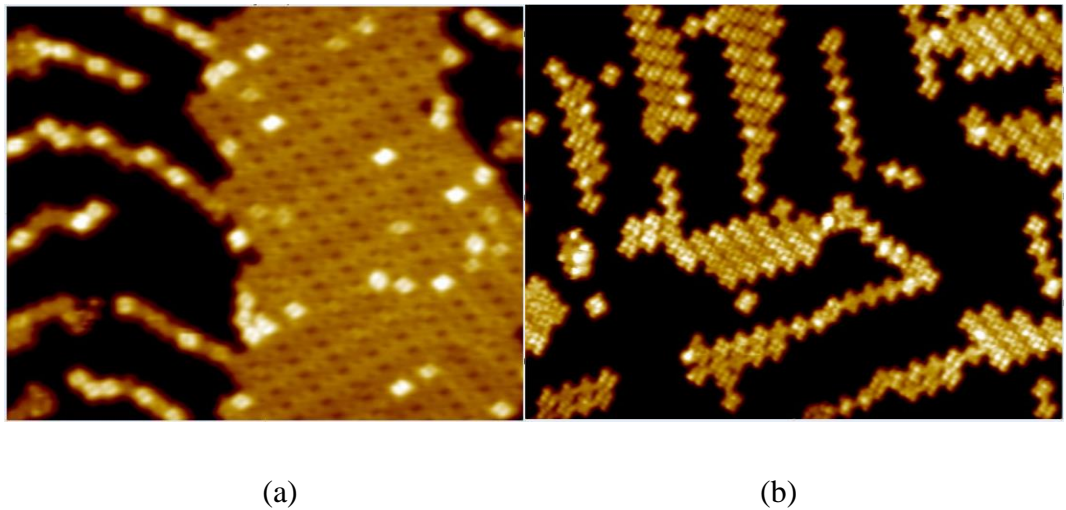

(a)                            (b)

**Figure 4.10. STM topographs (40×30 nm$^2$) of Br$_2$TPP at high molecular coverage on Au(111). (a) As-prepared sample showing the molecules form chains and closely-packed islands where 20% molecules are in conf-I and 80% molecules are in conf-II. (b) After 450 K annealing. 75% molecules are in conf-I.**

**4.7 Possible mechanisms of the two configurations.**

There are several mechanisms that lead to distinctive topographic and electronic characteristics of the porphyrin molecules adsorbed on surface, including deprotonation of the macrocyclic core, metalation of the macrocyclic core, proton tautomerization, adatom attachment, and adsorption-site dependent conformational changes. In our experiments, since both conf-I and conf-II are present for as-deposited molecules, we exclude that the two configurations are associated with deprotonation of the porphyrin core. Metallation of the porphyrin core can also be excluded since there are no foreign metal atoms deposited on the surface. The sequential conversions shown in Fig. 4.7c evidence that the two configurations are not originated from specific adsorption sites. Their electronic characteristics (dI/dV) and non-reversible conversion are in contradiction with tautomerization switching. It was reported that TPP can bond to a gold adatom on Au(111) in Ref. 12. Indeed we also observed such species in our experiment when we artificially created Au atoms on the surface. However, the conf-II molecules exhibit the topographic and electronic characteristics that are different from the Au bonded molecules. Furthermore, we conducted a manipulation experiment as shown in Fig. S4.2. A conf-II molecule (Fig. S4.2(a)) was moved to a new site by the tip at 2 V. It was converted to con-I, but there was no Au adatom left on the surface (Fig. S4.2(b)).

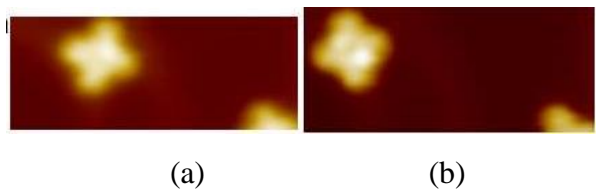

(a)　　　　　　(b)

**Figure 4.11. STM topographs (6.8×2.8 nm$^2$) showing tip-manipulation of a single molecule. (a) Before the manipulation the molecule is conf-II. (b) After the manipulation, the molecule is changed to conf-I while no Au adatom observed on the surface.**

**4.8 Self-assembly of DPP on Au(111).**

Besides few individual ones, the majority of DPP molecules form closely-packed islands, as exemplified in Fig. S4.3. The black trace in Fig. S4.3 highlights a DPP molecule. It shows two brighter tips and a depressed center. This shape is identical to the single molecules. One can see in Fig. S4.3 that all the molecules display such an appearance, implying that the DPP molecules take only one type of conformation as adsorbed on Au(111).

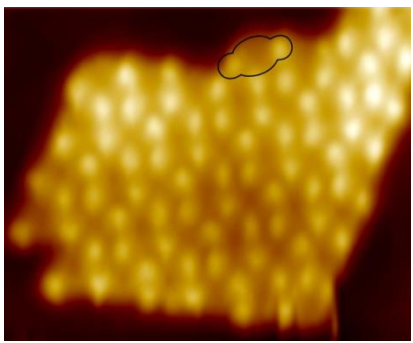

**Figure 4.12. STM topograph (10×14 nm$^2$) showing a closely-packed island assembled out of DPP molecules. The black trace highlights a single molecule.**

**4.9 DFT-calculated *sad* conformer of DPP.**

The DFT optimized structure of a DPP in *sad* conformation is shown in Fig. S4.4(a). As the molecule possesses two peripheral phenyl groups with two-fold symmetry, the saddle-shaped porphyrin core is highly distorted. This structure is 4.0 eV higher in total energy than the *ruf* conformer. The simulated image of the sad conformer is shown in Fig. S4.4(b). It obviously contradicts against the STM topograph shown in Fig. 4.9(b). So we conclude DPP does not take a *sad* conformation but adopts a *ruf* conformation as adsorbed on Au(111).

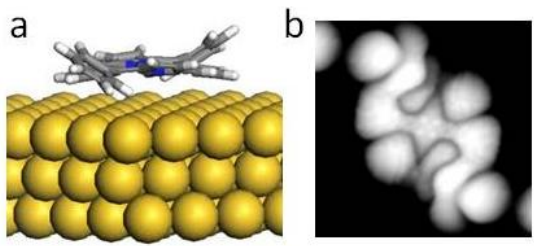

**Figure 4.13. DFT-optimized DPP in *sad* conformation. (a) Structure and (b) Simulated image.**

## 4.10 Computational details of metadynamics simulations

To give more insights into the transitions among different conformations (saddle, dome and planar) for Br2TPP on Au(111) surface, we performed additional metadynamics simulations. Our metadynamics simulations was performed using the path-collective-variables (PCV) defined on the transition paths sampled in our unbiased MD simulations [285], including the dome-to-saddle and planar-to-saddle transitions. We found that both transitions are down-hill, i.e. the free energy of dome and planar conformations are both significantly higher than the saddle conformation (~170 kJ/mol and ~350 kJ/mol respectively). The difference is that while the dome conformation is metastable with a small free energy barrier (~10 kJ/mol) preventing its transition towards the saddle structure (orange block in Figure S4.5a), the planar-to-saddle transition is barrier-free (orange line in Figure S4.5b). In addition, we also report the snapshot of the transition state for the dome-to-saddle transition: the face-to-face two hydrogen-containing pyrrole moieties were parallel to the substrate. This indicates that the dome-to-saddle transition occurred by rotating the two hydrogen-containing pyrrole moieties simultaneously in the inverted direction.

The metadynamics simulations were performed by GROMACS (version 4.5.4) [275] and the PLUMED plugin (version 2.1.1) [286]. Because we already observed the dome-to-saddle and planar-to-saddle transitions in our unbiased molecular dynamics (MD) simulations (see Figure 4.5 in main text), we defined two PCVs using two trajectories each containing the dome-to-saddle and planar-to-saddle transition respectively. We extracted the conformations from the unbiased MD trajectory based on the RMSD metric, computed using all carbon and nitrogen atoms of the molecule. For dome-to-saddle, we extracted 27 conformations as the path with a RMSD = 0.0193 nm between every pair of neighbor nodes. For planar to saddle, we extracted 19 conformations as the path with a RMSD = 0.0187 nm between every pair of neighbor nodes.

Then, the PCV $s(X)$ and $z(X)$ were defined on the two paths and used to describe the position of any configuration relative to them:

$$s = \frac{\sum_{i=1}^{N} i \exp\left(-\lambda \left(R[X - X_i]\right)^2\right)}{\sum_{i=1}^{N} \exp\left(-\lambda \left(R[X - X_i]\right)^2\right)} \tag{4.1}$$

$$z = -\frac{1}{\lambda} \ln\left[\sum_{i=1}^{N} \exp\left(-\lambda \left(R[X - X_i]\right)^2\right)\right] \tag{4.2}$$

Where $i$ represents the index of the nodes on the path (from 1 to $N$), represents the RMSD between any conformation X and this node. For any conformation $X$, $s(X)$ and $z(X)$ indicate the progress along the path and average deviation from this path respectively. $\lambda$ is comparable to the inverse of mean square displacement between successive frames. Here, for both path, $\lambda$ was set 6200 nm$^{-2}$.

One dimensional metadynamics simulation was then performed on $s(X)$ for both paths, with wall potential at $z(X) = 0.0004$ nm$^2$. The height and width of the Gaussian hills were set to 0.25 kJ/mol and 0.25, respectively, with a deposition interval of 1 ps. For dome-to-saddle paths, we performed 95 ns long time metadynamics simulation, For the dome-to-saddle transition, we performed 95 ns metadynamics simulation based on PCV. For convergence check, we plotted the one dimensional free energy profiles at five time points: 35 ns, 50 ns, 65 ns, 85 ns and 95 ns respectively (Figure S4.5a). It is obvious that the five free energy profiles are quite similar, indicating convergence. The convergence of the 55 ns metadynamics simulations for the planar-to-saddle transition were also illustrated in Figure S4.5b (see free energy profile at 30 ns, 35 ns, 40 ns, 45 ns, 50 ns and 55ns, respectively).

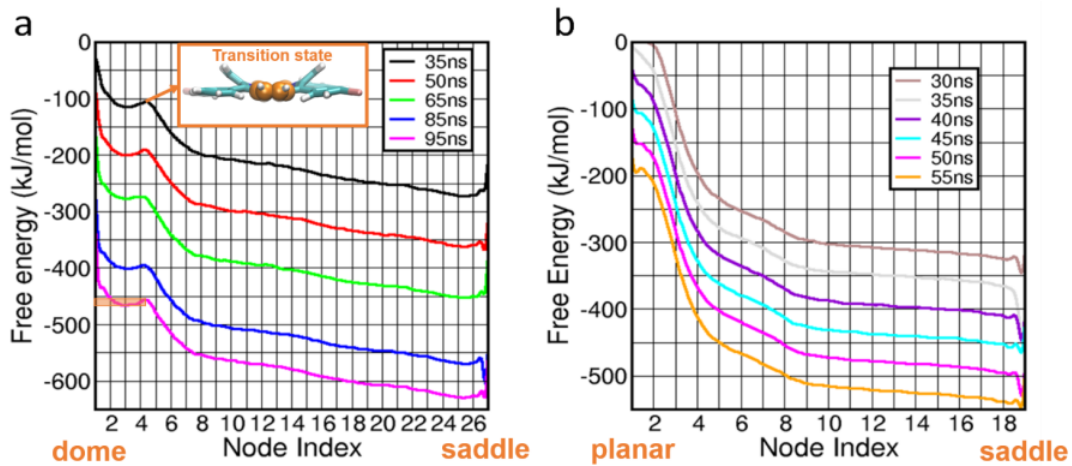

**Figure 4.14.** Free energy evolution of the metadynamics simulations along the path-collective variable representing (a) the dome-to-saddle transition and (b) the planar-to-saddle transition.

# Chapter 5

# Manipulation and Characterization of Artificial Graphene Nanoflakes

## 5.1 Introduction

Graphene has attracted intensive research interests in the past decade [287-290]. Following the successes in experimental fabrication, a rich variety of electronic and chemical properties of graphene were uncovered [37, 291-294]. Nanostructured graphene systems, suce as nano-ribbons (GNR) [295] and nanoflakes (GNF) [296] are predicted to exhibit remarkable electronic properties. Compared with nanoribbons, the nanoflakes present several advanced electronic properties. First, GNFs can behave as quantum dots [297]. Due to the small spin-orbit and hyperfine coupling in carbon, GNFs hold the promise of extremely long spin relaxation and decoherence time, which can be used to build the solid –state spin qubits [298]. Secondly, GNFs have a large number of configurational degrees of freedom associated with different types of shapes and edges. This structure freedom provideS a range of opportunities for engineering GNFs for future applications [300]. According to the theoretical works [299], the electronic properties, such as the density of states (DOS) and the edge states, may differ in GNFs of different shapes. For example, the numbers of edges can be different for varied shapes of GNFs. Thus, the distribution of two sublattices can also be different between the GNFs with odd or even numbers of edges. As a result, the electronic properties induced by edge effect, such as the zero-energy edge state, are more flexible inside the GNFs [300]. However, in contrast to the numerous theoretical works, the experimental studies of GNFs are relatively rare. Although some GNFs have been successfully manufactured by the soft-landing mass spectrometry [292], it is still very challenging to produce GNFs with controllable shapes, sizes and edge types. An alternative to study GNFs physics is to use artificial systems. Experimentally, this can be achieved in cold atoms [301] and artificial molecular graphene system [81]. In this work, we will employ the artificial molecular graphene system. The mechanism of making massless Dirac fermions from a modified two-dimensional electron gas was firstly proposed by C.-H. Park and S. G. Louie in 2009 [82]. By applying a penitential lattice on a metallic surface, the 2DEGs of the surface are scattered by lattices to result in a two-dimensional massless Dirac systems, which resembles the real graphene [81, 82]. This proposal has been achieved by manipulating CO atoms on a Cu(111) substrate [301] and coronene molecules on Cu(111) [81]. The advantages of artificial graphene systems are that both the lattice constant and hopping parameters are tunable. With this advanced experimental methodology, one can study the electronic behaviors of various GNFs, such as making the triangle, hexagonal and inhomogeneous strained ones [301].

In this work, we employed low-temperature STM manipulation to make several types of GNFs. Our major focus is the zero-energy edge state in different types of GNFs. There are two fundamental types of edge in GNFs, zigzag and arm-chair ones [302, 303]. Due to the boundary conditions involved in the Dirac equations, the energy spectrum of the graphene system can be influenced by the edges types. And as a result, the unique electronic properties of graphene nano-structures can be tuned. This behavior is particularly true in the case of zigzag edges. The zigzag edges feature a half-filled flat band at the Fermi level, which contributes a zero-energy state located at the edges. Using the STM-manipulation technique, we have made triangular and hexagonal GNFs with the two types of edges. The comparisons of these GNFs provide us a platform to study the electronic behaviors of the zero-energy state. We have found that the edge state can only be observed in the nano-flakes with the zigzag edges. Furthermore, we built a hexagonal GNF with structural deformation. This deformed artificial graphene system can behave like a real graphene flake under elastic strain. The elastic inhomogeneous strain changes the hopping amplitude between neighboring carbon atoms and breaks sublattice symmetry. As a result, an effective vector potential is created [304], which generates a large pseudo-magnetic field of tens of tesla in the graphene flake without breaking time reversal symmetry [37]. In such a condition, the electrons in the artificially deformed graphene form the pseudo Landau levels. We have studied the distribution of first several orders of pseudo Landau levels (pLLs) in real space. The $0^{th}$-order pLL is highly sublattice dependent. In the real space, that state is located at the center region of sublattice A, but the edges of sublattice B.

## 5.2 Experimental and theoretical methods

Artificial graphene nano-flakes (AGNFs) were constructed on Cu(111) using coronene molecules following the procedure described before. The coronene molecules were arranged in a triangular lattice with a lattice constant of 3 nm. The coronene molecules provide repulsive potentials to the 2DEG of the Cu(111) surface state electrons, which results in a linear dispersion simulating the electronic structures of graphene. Additional coronene molecules were packed closely to form walls surrounding the triangular lattice to make AGNFs. The orientation of the walls with respect to the triangular lattice defines the edge morphology of the AGNFs, i.e. zigzag or arm-chair edges. Scanning tunneling spectroscopy was performed to acquire spatially-resolved density of states of the AGNFs.

We carried out the tight-binding (TB) calculations with a nearest-neighboring hopping parameter of 65meV using the Python package *pybinding*. The hopping parameter was obtained with the formula: $t = v_F \frac{2\hbar}{\sqrt{3}d}$, where d is the lattice constant (3 nm) and $v_F$ is the Fermi velocity (2.6 × 10$^5$ m/s, the Fermi velocity of AGNFs system is calculated by the formula: $v_F = \frac{2\pi\hbar}{3m^*d}$). The local density of states is obtained through the Green function, which is derived from the Kernel Polynomial Method with Gaussian broadening of 2.3 meV. The density of states is calculated from the eigenvalues with the same Gaussian broadening. The theoretical introduction and calculated results of 2DEG in the repulsive potential pillars are discussed in Supporting Information section.

## 5.3 Results and discussion

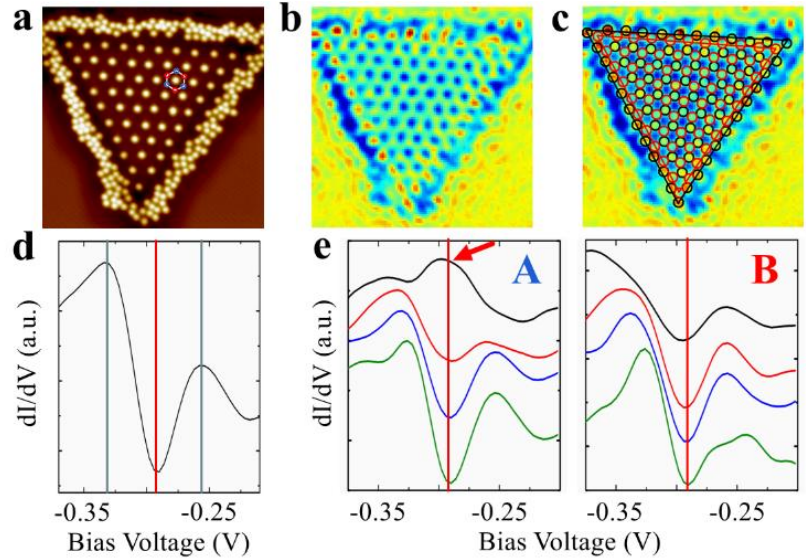

**Figure 5.1. (a) STM image of a triangular AGNF with three zig-zag edges. (Set-point: -1V, 0.3nA, 40×40nm²) Blue (red) denotes sublattice-A (B) atoms and the hexagonal frame represents a six-member ring. (b) STS map acquired at the Dirac point: −0.295 V, the brighter color representing stronger feedback (Set-point: -1V, 0.3nA, 40×40nm²). (c) The sketch of each sublattice site, black (red) circles represent sublattice-A (B) atoms. (d) The total dI/dV spectra over the whole flake, subtracting the contributions of coronene sites and substrate background. (Set-point: -1V, 0.3nA) The energy level of the Dirac point is indicated by the red line. (e) The DOS of sublattice A (left) and B (right). The contributions of coronene sites and substrate background are also subtracted. The zero-energy edge state is figured out by a red arrow. The black, red, blue and green plots representing the 1st, 2nd, 3rd and 4th layers of "atoms", respectively.**

An equal-lateral triangular shape AGNF with zigzag edges is shown in Figure 1(a). The color circles overlaid highlight the artificial "carbon atoms", in which blue (red) denotes sublattice-A (B) atoms and the hexagonal frame represents a six-member ring. Figure 1(d) shows the STS averaged over the whole area of this AGNF, subtracting the contributions of the coronene molecules and the substrate background. The STS displays a V-shape which features the density of states (DOS) of the massless Dirac fermions. The valley of the V at -0.29 V corresponds to the Dirac point. The spatial distribution of the DOS at the Dirac point is plotted in Figure 1(b), showing that the DOS is highly localized at the three edges of the triangular AGNF. We also analyzed the sublattice-resolved DOS. As sketched in Figure 1(c), black (red) circles represent sublattice-A (B) atoms. The zigzag edges (the black triangle) consist of sublattice-A atoms only. One can see that the high-density spots coincident with the sublattice-A sites along the three edges. These spots, indicated with the black circles along the three sides of the black triangle in Figure 1(c), are equivalent sites of the AGNF. Similarly, the red circles along the three sides of the red triangle are equivalent sites of the sublattice-B. We plot the equivalent-site averaged DOS of the both sublattices in Figure 1 (e). The top curves in the right and left panels are the average DOS of the outmost triangles of the two sublattices (defined as 1st), respectively. The 2nd, 3rd and 4th layers correspond to the DOS at the triangles of step-by-step reduced size. The DOSs in Figure 1(e) manifest two features: (1) the Dirac point state only exists at the sublattice-A sites; (2) it decays rapidly from the edges towards the interior. Like our previously work on artificial graphene ribbons, the triangular AGNF with zigzag edges also exhibits an edge state at the Dirac point.

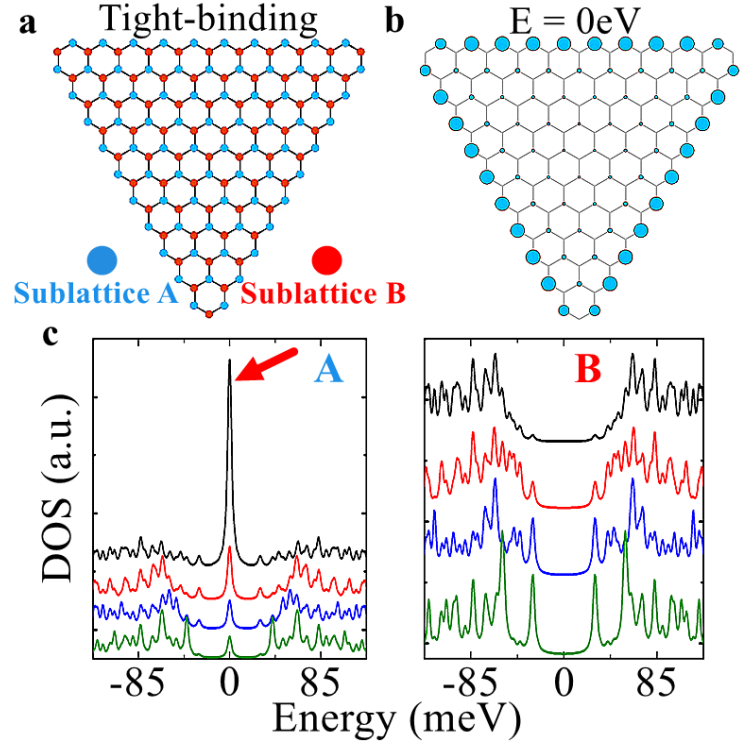

**Figure 5.2.** (a) The tight-binding calculation model in which the two sublattices are colored with blue (sublattice A) and red (sublattice B), respectively. (b) Tight-binding simulated spatial distribution of DOS at E=0, the size of dots is proportional to the strength. The distribution at two sublattices are colored with blue (sublattice A) and red (sublattice B), respectively. (c) The calculated DOS of sublattice A (left) and B (right). The zero-energy edge state is figured out by a red arrow. The black, red, blue and green plots representing the 1st, 2nd, 3rd and 4th layers, respectively.

Figure 2(a) is the TB calculation model in which the two sublattices are colored with blue (sublattice A) and red (sublattice B), respectively. Due to the electron-hole symmetry, the Dirac point is at E=0. Figure 2(b) displays the spatial distribution of DOS at E=0, showing that only sublattice-A atoms exhibit appreciable DOS and the three edges feature highest DOS. Figure 2(c) shows the DOS of the two sublattices with the outmost triangles defined as 1st, featuring that the zero-energy state is absent in the sublattice B, but only presents in the sublattice A and decays from the edges to the interior. Qualitatively, the TB calculation captures the major characteristics of the edge state of the triangular AGNF. Overall, the TB calculation confirms the edge state observed experimentally.

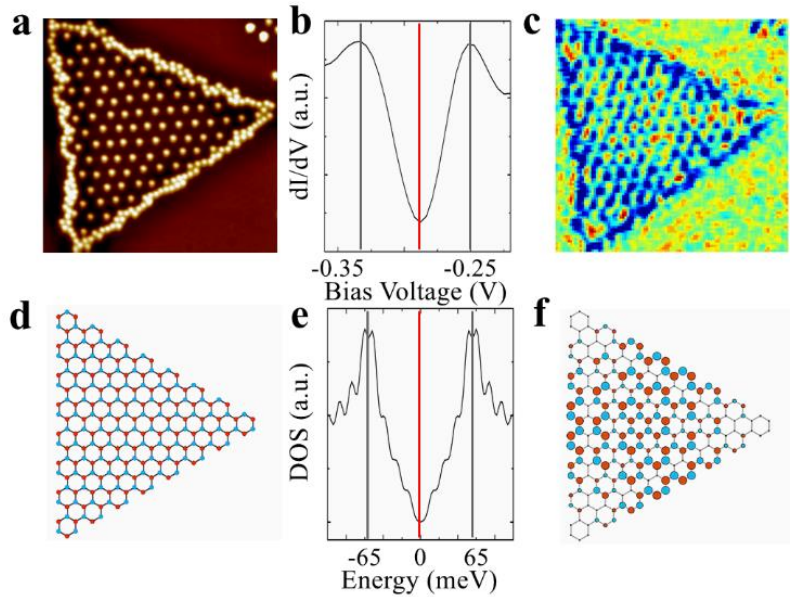

**Figure 5.3.** (a) STM image of a triangular AGNF with three armchair edges (Set-point: -1V, 0.3nA, 46×46nm$^2$). (b) The total dI/dV spectra over the whole flake, subtracting the contributions of coronene sites and substrate background. (Set-point: -1V, 0.3nA) (c) STS map acquired at the Dirac point: −0.29V. (d) The tight-binding calculation model in which the two sublattices are colored with blue (sublattice A) and red (sublattice B), respectively. (e) The calculated total DOS of the whole flake. (f) Tight-binding simulated spatial distribution of DOS at E=0, the size of dots is proportional to the strength.

We also built a triangular AGNF with armchair edges. The STM image is shown in Figure 3(a). Figure 3(b) shows the total DOS over the whole area of this AGNF. The spectrum also displays a clear V-shape with the Dirac point located at -0.29V. Figure 3(c) shows the spatial distribution of the DOS at the Dirac point, in which we can observe that the density is spread over the armchair AGNF without any edge-state featrue. Theoretically, we employ the TB calculation to simulate the electronic properties of this armchair sample. The corresponded TB model is presented in Figure 3(d), the two sublattices are identified by different colors. In Figure 3(f), the spatial distribution plot of DOS at E=0, we can see that the DOS is spread over the whole flake. Moreover, the DOS distribution is identical for the two different sublattices. Compared with the zigzag AGNF, the armchair one is featured as the two sublattices being equivalently distributed at each edge. The zero-energy state is obviously missing in the armchair AGNF. Thus, we confirm the missing of the zero-energy edge state in this armchair AGNF.

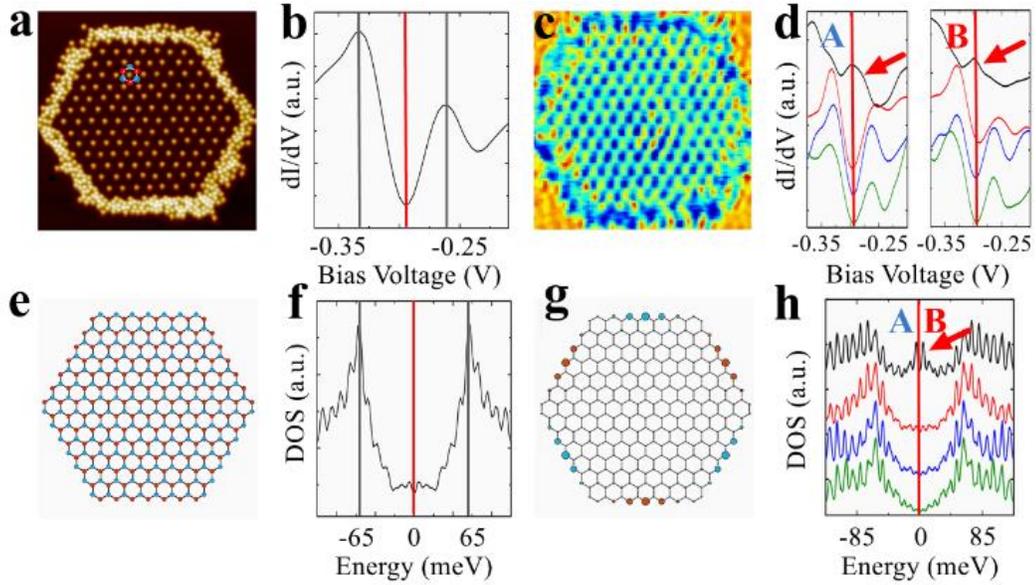

**Figure 5.4.** (a) STM image of a hexagonal AGNF with six zig-zag edges. (Set-point: -1V, 0.3nA, 50×50nm²) Blue (red) denotes sublattice-A (B) atoms and the hexagonal frame represents a six-member ring. (b) The total dI/dV spectra over the whole flake, subtracting the contributions of coronene sites and substrate background. (Set-point: -1V, 0.3nA) (c) STS map acquired at the Dirac point: −0.29V, the brighter color representing stronger feedback (Set-point: -1V, 0.3nA, 50×50nm²). (d) The DOS of sublattice A (left) and B (right). The contributions of coronene sites and substrate background are also subtracted. The zero-energy edge state is figured out by a red arrow. The black, red, blue and green plots representing the 1st, 2nd, 3rd and 4th layers, respectively. (e) The tight-binding calculation model in which the two sublattices are colored with blue (sublattice A) and red (sublattice B), respectively. (f) The TB-calculated total DOS plot over the whole AGNF. (g) Tight-binding simulated spatial distribution of DOS at E=0, the size of dots is proportional to the strength. The distribution at two sublattices are colored with blue (sublattice A) and red (sublattice B), respectively. (h) The TB-calculated DOS of sublattice A/B and the zero-energy edge state is figured out by a red arrow. The black, red, blue and green plots representing the 1st, 2nd, 3rd and 4th layers, respectively.

We also studied the hexagonal AGNFs with different types of edges. The STM image of a hexagonal AGNF with six zigzag edges is shown in Figure 4(a) with two sublattices sites identified by different colors. One important thing to note is that the two sublattices A and B are equivalnet in this hexagonal AGNF. Among the six edegs, three of them are sublattice A, where the other three are B. The total DOS also has a V-shape, as presented in Figure 4(b) showing that Dirac point is located at -0.294V. The distribution of the DOS in real space at the Dirac energy is shown in Figure 4(c). As can be seen, the DOS also display a strong concentration at the edges. One may also find some distribution of DOS at the flake center, which is caused by the defects. We plot the DOS of sublattices A (left) and B (right) in Figure 4(d). The top curves in both panels are the average DOS of the edges of two sublattices (defined as 1st). The 2nd, 3rd and 4th curves correspond to the DOS at the lines of step-by-step close to the center. The DOSs in Figure 4(d) manifest two features: (1) the Dirac point state (figured out by a red arrow) exists at both sublattices A and B sites; (2) it decays rapidly from the edges towards the interior. In Figure 4(e), we present the corresponded TB model calculation with the two sublattices are also indicated as blue (sublattice A) or red (sublattice B) dots. We can clearly identify the equivalent distribution feature of the two sublattices. Panel (f) plots out the total DOS calculated by TB. We can find a resonance peak induced by the zero-energy state at the Dirac energy (E=0). We also map out the TB-calculated spatial distribution of DOS at the Dirac energy, as shown in Figure 4 (g). It is clear that the zero-energy state is mainly confined at the edges. Moreover, we can find that the distribution of DOS is exactly identical for both sublattices. The TB-calculated DOS plot is demonstrated in Figure 4(h). Again, the edges are defined as 1st curve, followed by the 2nd, 3rd and 4th lines of each sublattice, showing that the zero-energy state are present at the edges of both the sublattices. Qualitatively, the TB calculation captures the edge-confined characteristics and identically-distributed feature at different sublattices of the zero-energy state. Compared with the triangular case, similar V-shape total DOS and zero-energy edge state can be found in the zigzag AGNF. However, owing to the geometry symmetry, the two sublattices are equivalently located here. Thus, we can observe the zero-enery edge state at the edges of both types of sublattices, while it only existing at one sublattice in the triangular case.

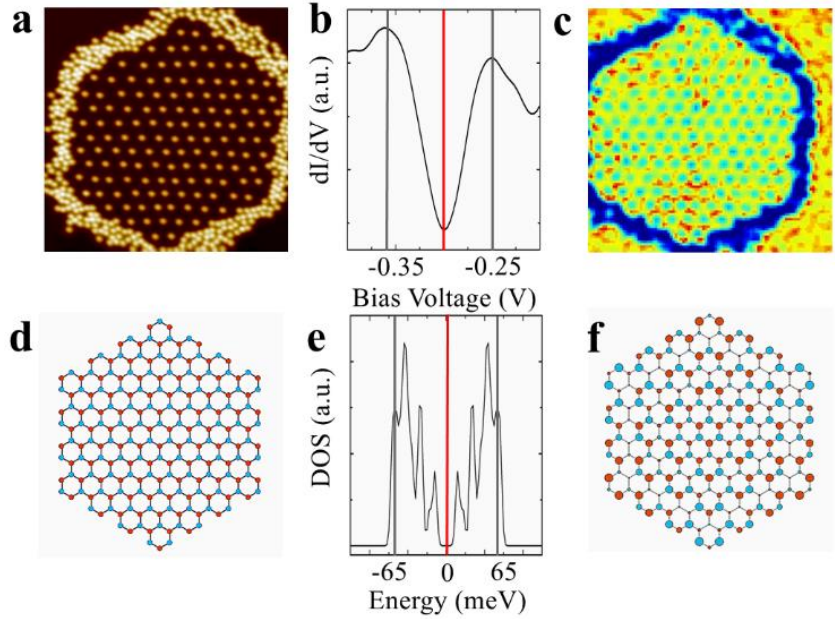

**Figure 5.5. (a) STM image of a hexagonal AGNF with six armchair edges (Set-point: -1V, 0.3nA, 48×48nm$^2$). (b) The total dI/dV spectra over the whole flake, subtracting the contributions of coronene sites and substrate background. (Set-point: -1V, 0.3nA) (c) STS map acquired at the Dirac point: −0.30V. (d) The tight-binding calculation model in which the two sublattices are colored with blue (sublattice A) and red (sublattice B), respectively. (e) The calculated total DOS of the whole flake. (f) Tight-binding simulated spatial distribution of DOS at E=0, the size of dots is proportional to the strength.**

We also built a hexagonal AGNF with the armchair edges, as shown in the STM image in Figure 5(a). The V-shape total DOS is plotted out in Figure 5(b), showing the Dirac point is located at -0.3V. The spatial distribution of DOS at the Dirac energy plotted in panel (c) shows that DOS spread over the whole flake without any edge-state feature. The corresponding TB-calculation model is shown in Figure 5(d). The TB-calculated total DOS is presented in panel (e). The total DOS nicely reproduces the V-shape curve without any zero-energy resonance peak. We can find the state at E=0 spreads out in the entire flake, as shown in Figure 5(f). Overall, the TB calculation reproduces the experimental observations, and confirms that there is no edge state existed in the armchair case. As the conclusion made in the triangular cases, we can conclude that only the zigzag edge lead to the zero-energy edge state in the hexagonal AGNFs.

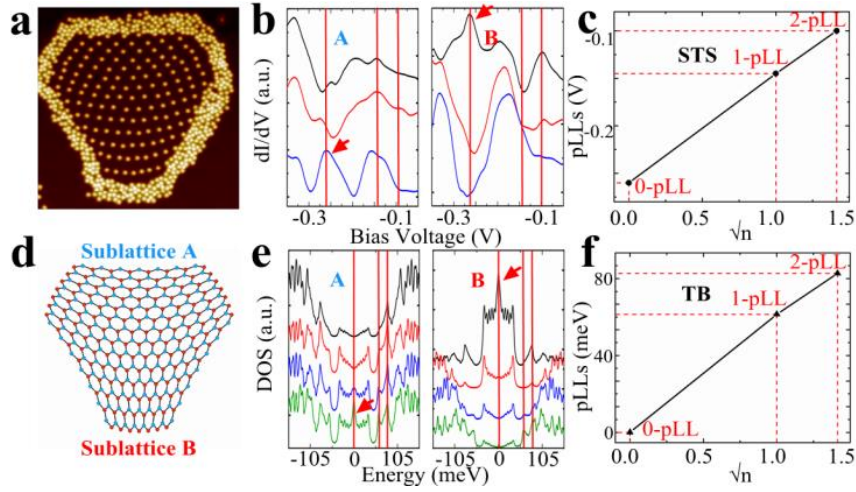

**Figure 5.6. (a) STM image of a deformed hexagonal AGNF with six zigzag edges (Set-point: -1V, 0.3nA, 55×55nm$^2$). (b) The DOS of sublattice A (left) and B (right). The contributions of coronene sites and substrate background are also subtracted. The first three pLLs are figured out by red vertical lines. The zero-order pLL are figured out by red arrows. The black, red and blue plots representing the 1st, 3rd and 5th layers, respectively. (c) STS-measured energy levels of first three pLLs as a function of the square root of the orders of pLLs ($\sqrt{n}$). (d) The tight-binding calculation model of this deformed AGNF, in which the two sublattices are colored with blue (sublattice A) and red (sublattice B), respectively. (e) The TB-calculated DOS of sublattice A (left) and B (right). The zero-order pLL are figured out by red arrows. The black, red, blue and green plots representing the 1st, 3rd, 5th and 7th layers, respectively. (f) TB-calculated energy levels of first three pLLs as a function of the square root of the orders of pLLs ($\sqrt{n}$).**

Here in the last part, we will discuss the electronic properties of the AGNFs under inhomogeneous structural deformation. This deformed artificial graphene system can equivalently behave like the real graphene flakes under elastic strain [304]. Theoretically, the elastic strain can significantly change the hopping amplitude of the carbon atoms, and then an effective vector potential (gauge field) is induced that shifts the Dirac point. This strain-induced gauge field can further give rise to a large pseudo–magnetic field distributed over the whole flake. In such a situation, the charge carriers in graphene can be regarded as circulating under the influence of an external out-of-plane magnetic field.

As the STM image shown in Figure 6(a), the pattern is deformed from a hexagonal zigzag AGNF. The deformation can be described by the following displacement equations (place each carbon site into the radial coordination): $u_r = qr^2 \sin(3\theta)$; $u_\theta = qr^2 \cos(3\theta)$, where q is a parameter denoting the strength of deformation. Here, the value of q is $0.001 Å^{-1}$ and the lattice constant (d) is 2.7nm. Based on the theoretical calculation in [80], the strength of induced pseudo-magnetic field ($B_s$) for the artificial graphene system is: $B_s = \frac{16\pi q}{3d}$ (given in the unit of $\hbar/e =1$). The induced pseudo-magnetic field of our sample is about 45T, which is much stronger compared with real external magnetic field. Under such a strong pseudo–magnetic field, a number of pseudo-Landau-levels (pLLs) have been observed inside the deformed AGNF. Figure 6(b) plots the DOS of the both sublattices from edges to center. The top curves in the right and left panels are the average DOS of the edges of the two sublattices (defined as 1st), respectively. The 3rd (red) and 5th (blue) curves correspond to the DOS at equivalent sites closer to the center. These pLLs are regarded as resonance peaks in the DOS. After considering the both sublattices, we can locate the energy levels of first three pLLs, which are indicated by the red vertical lines. The zero-order pLL is located at -0.26V with the first and second ones are at -0.145V and -0.1V, respectively. Figure 6(c) shows the STS-measured energy levels of first three pLLs as a function of the square root of the orders of pLLs ($\sqrt{n}$). The energy difference between the $pLL_n$ and $pLL_0$ ($dE_{(Ln-L0)}$) is proportional to $\sqrt{n}$ as the function: $dE_{(Ln-L0)} \sim v_F \times \sqrt{nB}$. Thus, the obtained linear curve in panel (c) can corroborates our determination of the first three pLLs. The DOS in panel (b) shows that the $pLL_0$ has strong sublattice dependence. To specify, the $pLL_0$ is located at the center of sublattice A, but at the edges of sublattice B. The reason is: Under the pseudo magnetic field, the time-reversal symmetry is not broken, but sublattice symmetry is broken in [80].

Figure 6(d) presents the TB-calculation model of this deformed AGNF. The TB-calculated DOS for the both sublattices are presented in Figure 6(e). We determinate the

- 76 -

energy levels of first three pLLs. As the red vertical lines identified, the $pLL_0$ is located at E=0, $pLL_1$ and $pLL_2$ are at 60meV and 81meV, respectively. Clear difference in DOS between the two sublattices can be observed. The distribution of $pLL_0$ is located at the center or edges of sublattice A or B, respectively. In Figure 6(f), we illustrate the TB-calculated energy levels of first three pLLs as a function of the square root of the orders of pLLs ($\sqrt{n}$). The linear relation can also provide strong supporting of the determination of pLLs. Both the DOS and corresponding pLLs are quantitatively comparable between the experimental and TB-calculation results. In conclusion, we have observed pLLs in this deformed AGNF and the determinate energy levels of first three pLLs. Moreover, the distribution of $pLL_0$ is sublattice dependent, which is due to the broken of sublattices symmetry.

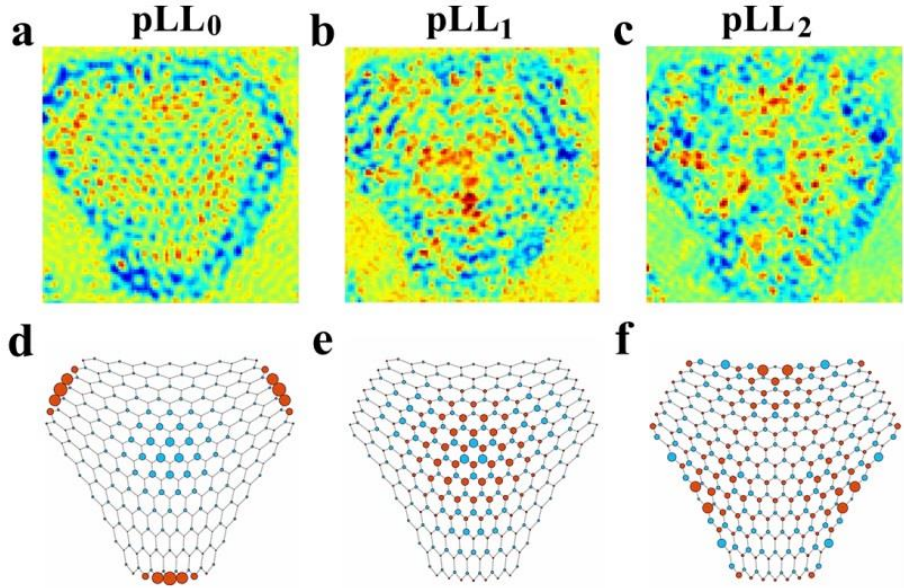

**Figure 5.7. (a) STS map of the 0-order psedo-Landau-level, acquired at −0.26V. (b) STS map of the first-order psedo-Landau-level, acquired at −0.15V. (c) STS map of the second-order psedo-Landau-level, acquired at −0.1V (Set-point: -1V, 0.3nA, 55×55nm$^2$). (d) Tight-binding simulated DOS map of the 0-order psedo-Landau-level at E=0. (e) Tight-binding simulated DOS map of the first-order psedo-Landau-level at E=60meV. (f) Tight-binding simulated DOS map of the second-order psedo-Landau-level at E=80eV. The distribution at two sublattices are colored with blue**

**(sublattice A) and red (sublattice B), respectively. The size of dots is proportional to the strength.**

# Chapter 6

# Summary and Future Outlook

In this thesis, I have used low-temperature scanning tunneling microscopy, spectroscopy and density functional calculation to investigate on-surface synthesis and characterization of single molecular properties adsorbed on metallic surfaces utilizing. In chapter 3, the work of Switching Molecular Kondo Effect via Supramolecular Interaction is presented. At first, we applied supramolecular assembly to control the adsorption configuration of Co-porphyrin molecules on Au(111) and Cu(111) surfaces. We found that the Kondo effect associated with the Co center is absent or present in different supramolecular systems is detected. We performed first-principles calculations to obtain spin-polarized electronic structures and compute the Kondo temperatures using the Anderson impurity model. The switching behavior is traced to varied molecular adsorption heights in different supramolecular structures. These findings unravel that a competition between intermolecular interactions and molecule–substrate interactions subtly regulates the molecular Kondo effect in supramolecular systems. In chapter 4, the work on Single-Molecule Observation of Surface-Anchored Porphyrins in Saddle, Dome and Ruffled Conformations is presented. We demonstrated at the single-molecule level that anchoring porphyrin molecules on a surface can distort the macrocyclic core into dome, saddle or ruffled conformations. The distortions arise from competition among substrate-molecule interactions, metallation of the macrocyclic core, and steric hindrance from peripheral groups. These results may help achieve conformational control for surface-anchored porphyrin molecules.

In chapter 5, I focused on the designing of artificial graphene nanoflakes. We employed low-temperature STM manipulation to study the electronic properties of the artificial graphene nanoflakes. Our major focus is the zero-energy edge state in different types of graphene nano-flakes. We also built an artificial graphene nano-flakes with large deformation and studied the distribution of pseudo Landau levels (pLLs) in real space. The $0^{th}$-order pLL was found to have highly sublattice dependent. In the real space, the states are located at the center region of sublattice A, but the edges of sublattice B. In conclusion, a lot of interesting electronic, conformational and magnetic properties are demonstrated within organic molecular systems. These findings can be further applied in the research of related applications[1–13].

For the future outlook of my research works, I will mainly continue my study on the properties of both porphyrin molecules and artificial graphene nanoflakes. Firstly, the magnetic properties of porphyrin molecules are still open to further research works. For example, the magnetic moment contributed by the center metalated atom can be influenced by the some factors in experiments. As a result, it is very important to find some controllable methods to finely tune the magnetic moment of porphyrin molecules, which can greatly assist future application. Additionally, expect for magnetic atom, we can also apply other types of metal atom to metalate porphyrin molecules, such as heavy metal atoms. Because of the tremendous atomic size and strong spin-orbital coupling of heavy metal atom, it is reasonable to estimate that not only the geometry, but also the electronic structure of heavy-metal-metalated porphyrin molecules will be adjusted.

Secondly, I am eager to design some other types of artificial graphene nanoflakes with special electronic properties. Up to now, I have successfully conducted the research of various types of regular nanoflakes. For the future research, as discussed in my thesis defense, I will try to study the nanoflakes under the broken of sublattice symmetry. To achieve this, I can adapt another type of molecular potential, with a triangular shape, to replace the original round one. With this new type of molecular potential, the symmetry between sublattice A and B will be thus largely broken. As a result, the density of electronic state will be different among different sublattice sites in the nanoflake. Furthermore, more works can also be done followed the deformed nanoflake. Although the topological state in this deformed nanoflake is a trivial one, there are still some novel electronic properties, if some specific defects are introduced into the nanoflake. Combine theoretical calculation and further experimental study, more and more properties of artifiacial graphene nanoflakes will be gradually uncovered.

# Appendix

## Appendix 1. Density functional theory

As discussed above, the DFT theory can be considered as solving the many-electron time-independent Schrödinger equation in order to obtain the ground state. Bedsides, the theorems can be further simplified by the Born–Oppenheimer approximation, which assumes that the nuclei of the treated molecules or clusters are seen as fixed generating a static external potential V in which the electrons are moving and the wave function of a many-body system can be decomposed into its electronic and nuclear components. The simplified expression of Schrödinger equation is equation (2.11), where, for the N-electron system, H is the Hamiltonian, E is the total energy, T is the kinetic energy, V is the potential energy from the external field due to positively charged nuclei, and U is the electron-electron interaction energy.

$$\hat{H}\Psi = \left[\hat{T} + \hat{V} + \hat{U}\right]\Psi = \left[\sum_i^N \left(-\frac{\hbar^2}{2m_i}\nabla_i^2\right) + \sum_i^N V(\vec{r}_i) + \sum_{i<j}^N U(\vec{r}_i, \vec{r}_j)\right]\Psi = E\Psi \quad (2.11)$$

However, even with this simplification, it seems still impossible to strictly solve the Schrödinger equation of a many-body system. In the late 1920s, soon after the derivation of Schrödinger equation, D. R. Hartree introduced a procedure, which he called the self-consistent field method to calculate the approximate wave functions and energies of atoms and ions. Lately, this method was complimented by Slater and Fock. This half empirical method was named as Hartree-Fork (HF) method thereinafter. The HF method was successfully been used in many systems after the invention of computer in the late 1950s. Soon, it was found that the accuracy of this approach was not acceptable in many cases, and the calculation for larger many-body systems using this method was a big challenge because of exceptional high demanding of computational resources. In 1965, Kohn and Sham developed a new method using the functional of electron density (named density functional theory (DFT)). In the next few years, DFT had been greatly developed and widely used in many fields. In 1998, the Nobel Prize was awarded to Kohn and Hohenberg for their invention and development of DFT.

In the Hohenberg-Kohn theorem, there are two approximations made:

1. If two systems of electrons, one trapped in a potential $v_1(r)$ and the other in $v_2(r)$, have the same ground-state density $n(r)$ then necessarily have:

$$v_1(r) - v_2(r) = \text{constant}$$

   Corollary: the ground state density uniquely determines the potential and thus all properties of the system, including the many-body wave function. In particular, the "HK" functional, defined as $F[n] = T[n] + U[n]$ is a universal functional of the density (not depending explicitly on the external potential).

2. For any positive integer $N$ and potential $v(r)$, a density functional $F[n]$ exists such that equation (2.12)

$$E_{(v,N)}[n] = F[n] + \int v(\vec{r})n(\vec{r})d^3r \quad (2.12)$$

   obtains its minimal value at the ground-state density of $N$ electrons in the potential $v(r)$. The minimal value of $E$ is then the ground state energy of this system.

As a result, it can be derived that: (1) the energy and all observables of the ground state of a many-body system are functional of the electron density, (2) the minimization of the energy with respect to the electron density yields the actual ground state energy of the system, (3) the Schrödinger equation of a fictitious system (Kohn-Sham system) of non-interacting electrons generates the same density as any given system of interacting electrons. As the electrons in the Kohn-Sham system are non-interacting fermions, the Kohn-Sham wave function is a single Slater determinant, similar to a single electron

Schrödinger equation [104]. In a detail, the wave function for particle i can be written as equation (2.13):

$$\left(-\frac{2m}{\hbar^2}\nabla^2 + v_{eff}(r)\right)\Phi_i(r) = \varepsilon_i \Phi_i(r) \tag{2.13}$$

Since $\Phi_i(r)$ represents the wave function of each particle, the total density of states of a many-body system can be written as equation (2.14):

$$n(r) = \sum_{j=1}^{N} \Phi_i(r)^2 \tag{2.14}$$

Then, one can express the effective potential $V_{eff}$ through the electron density n(r) as equation (2.15):

$$v_{eff}(r) = V(r) + e^2 \int \frac{n(r')}{|r-r'|} dr' + V_{xc}(n(r)) \tag{2.15}$$

Where V(r) denotes the external potential given by nuclei; the second term describes the Coulomb interactions between electrons; and the last term is the so called exchange-correlation term. In the LDA approximation:

$$v_{xc}(r) = \int V_{xc}(n(r)) n(r) d^3r \tag{2.16}$$

And in the GGA approximation:

$$v_{xc}(r) = \int V_{xc}(n(r), \nabla n(r)) n(r) d^3r \tag{2.17}$$

## Appendix 2. Detailed information about Gaussian09.

There are many useful functions provided by Gaussian 09, for example, molecular structural optimization and molecular orbitals calculation, as shown in figure 2.8 a and b.

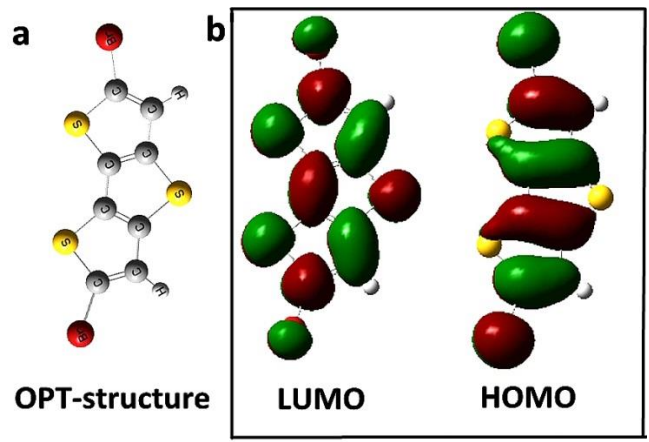

**Figure 2.8 (a) Optimized structure of Br2-SSS molecule (b) Molecular orbitals: HOMO and LUMO.**

On the other hands, most methods require a basis set be specified. A basis set in theoretical and computational chemistry is a set of functions (called basis functions) which are combined in linear combinations (generally as part of a quantum chemical calculation) to create molecular orbitals. For convenience these functions are typically atomic orbitals centered on atoms, but can theoretically be any function; plane waves are frequently used in materials calculations [108].

1. The detailed functions of Gaussian 09 are summarized as below [105]:
2. Fundamental Algorithms
3. Model Chemistries
4. Molecular Mechanics
5. Ground State Semi-Empirical
6. Self-Consistent Field (SCF)
7. Density Functional Theory
8. Electron Correlation:
9. Automated High Accuracy Energies
10. Basis Sets and DFT Fitting Sets
11. Geometry Optimizations and Reaction Modeling
12. Vibrational Analysis
13. Molecular Properties
14. ONIOM Calculations
15. Excited States
16. Self-Consistent Reaction Field Solvation Models
17. Ease-of-Use Features

During the calculation of Gaussian software, there are two key parameters determining the balance of computational accuracy and costs: calculation functional and basis set. In Hartree-Fock theory, the energy has the form:

$$E_{HF} = V + <hP> + 1/2<PJ(P)> - 1/2<PK(P)> \qquad (2.18)$$

Where the terms have the following meanings:

V- The nuclear repulsion energy; P- The density matrix; $<hP>$- The one-electron (kinetic plus potential) energy; $1/2<PJ(P)>$- The classical coulomb repulsion of the electrons; $-1/2<PK(P)>$- The exchange energy resulting from the quantum (fermion) nature of electrons. In the Kohn-Sham formulation of density functional theory, the exact exchange (HF) for a single determinant is replaced by a more general expression, the exchange-correlation functional, which can include terms accounting for both the exchange and the electron correlation energies, the latter not being present in Hartree-Fock theory:

$$E_{KS} = V + <hP> + 1/2<PJ(P)> + E_X[P] + E_C[P] \qquad (2.19)$$

Where $E_X[P]$ is the exchange functional, and $E_C[P]$ is the correlation functional. Within the Kohn-Sham formulation, Hartree-Fock theory can be regarded as a special case of density functional theory, with $E_X[P]$ given by the exchange integral $-1/2<PK(P)>$ and $E_C=0$. The functionals normally used in density functional theory are integrals of some function of the density and possibly the density gradient:

$$E_X[P] = \int f(\rho_\alpha(r), \rho_\beta(r), \nabla\rho_\alpha(r), \nabla\rho_\beta(r))dr \qquad (2.20)$$

Where the methods differ in which function f is used for $E_X$ and which (if any) f is used for $E_C$. In addition to pure DFT methods, Gaussian supports hybrid methods in which the exchange functional is a linear combination of the Hartree-Fock exchange and a functional integral of the above form. Proposed functionals lead to integrals which cannot be evaluated in closed form and are solved by numerical quadrature.

# Appendix 3. Detailed information and a computational example of VASP

Compared with other methods, there are several advantages of VASP:

1. VASP can solve larger system, which have at least hundreds of atoms in a unit cell. As a result, the systems like organic molecules adsorbed metallic substrates can be calculated with relative good accuracy.
2. Since the basic model of VASP calculation is usually a unit cell of a periodic structure, many molecular structures and metal-organic coordination can be calculated with substrates.
3. The interactions of molecules-substrates, atom-atom or dipoles systems are usually including covalent bonds, dipole interaction, ionic bonds and van der Waals forces. In VASP, these types of interactions can be accurately described by plenty of GGA functions with correlation.
4. A large number of electronic and magnetic properties can be obtained from VASP calculations, such as: molecular structure, molecular orbitals, density of states, charge distribution, magnetic moment, simulated STM image, band structure and electron localization function.
5. Calculations with Green function, such as GW, can be conducted under VASP software, which can give more accurate of magnetic properties and density of states. Bedsides, calculation with spin orbital coupling is also available for revised version of VASP.
6. For magnetic system, the magnetic properties can be calculated more accurately by the DFT+U method or DFT with dispersion.

**In the following contents, I will briefly go through an example of VASP calculations in order to introduce the functions of VASP in a comprehensive way.** *<u>(Note: the following models are only for demonstration purpose.)</u>*

a. The system was about TPyB molecules coordinated with Bi clusters. After the model was established with periodicity, the whole adsorption system was structural relaxed with van der Walls force and the binding energy between each Bi-cluster is 0.39eV. The optimized structure and structural related information are presented in figure 2.9.

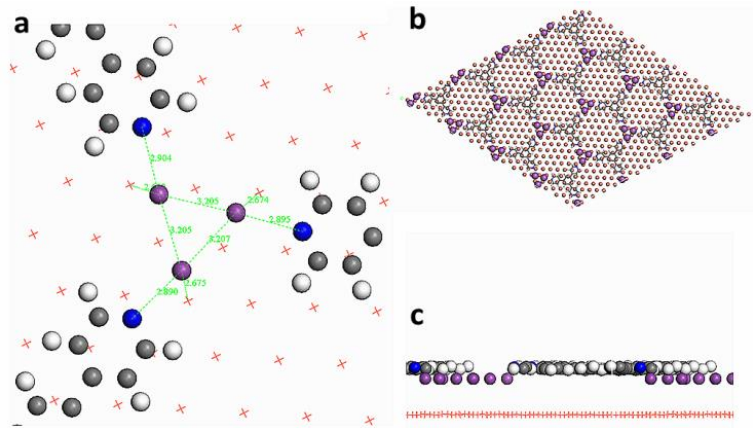

**Figure 2.9 (a) Distance between each Bi (purple) atom and N (blue) atom in the unit of angstrom (b) Optimized periodic structure (c) Side-view of the optimized structure and Bi atoms are attracted into the substrate due to the van der Waals force.**

b. Following structural optimization, the projected density of stated on the molecule and Bi atom were calculated, as shown in figure 2.10a. Based on the basic electronic properties obtained, the simulated STM images (molecular orbitals spatial distribution) were shown in figure 2.10b and c in a constant current mode. Based on the density functional theory, the distribution of electron in many-body electron system can be determined by calculating the wave function of the whole system. Firstly, the total wave function ($\psi n$) of the all the electrons inside the system can be exactly obtained by conducting DFT calculation. Then, the density of states, which can be regarded as the possibility of electrons located at specific location under certain energy value, is defined by $\rho(\varepsilon)= \sum n \langle \psi n | \psi n \rangle \delta(\varepsilon - \varepsilon n)$, where $\varepsilon n$ is the eigenvalue of the eigenstate $|\psi n\rangle |\psi n\rangle$. Secondly, if we want to calculate the density of state of a certain atom, one chooses to project the wave function of the entire system, $\psi n$, onto the each electron partial wave (i.e. the wave function of each isolated atom) $\phi^a_i$. For the DFT calculation performed by VASP, the wave function of each type of isolated atom is provided in the projector-augmented wave (PAW) input files. The projectors and pseudo partial waves form a complete basis within the augmentation spheres, this projection can be expressed as

$$\langle \phi_i^a | \psi_n \rangle = P_{ni}^a + \sum_{a' \neq a} \sum_{i_1 i_2} \langle \tilde{\phi}_i^a | \tilde{p}_{i_1}^{a'} \rangle \Delta S_{i_1 i_2}^{a'} P_{ni_2}^{a'}$$

If the chosen orbital index i correspond to a bound state, the overlaps $\langle \tilde{\phi}_i^a | \tilde{p}_{i_1}^{a'} \rangle$, a' ≠a will be small, we thus define an atomic orbital PDOS by

$$\rho_i^a(\varepsilon) = \sum_n |\langle \tilde{p}_i^a | \tilde{\psi}_n \rangle|^2 \delta(\varepsilon - \varepsilon_n) \approx \sum_n |\langle \phi_i^a | \psi_n \rangle|^2 \delta(\varepsilon - \varepsilon_n)$$

As a result, the density of state in each atom can be calculated by this projection method [109].

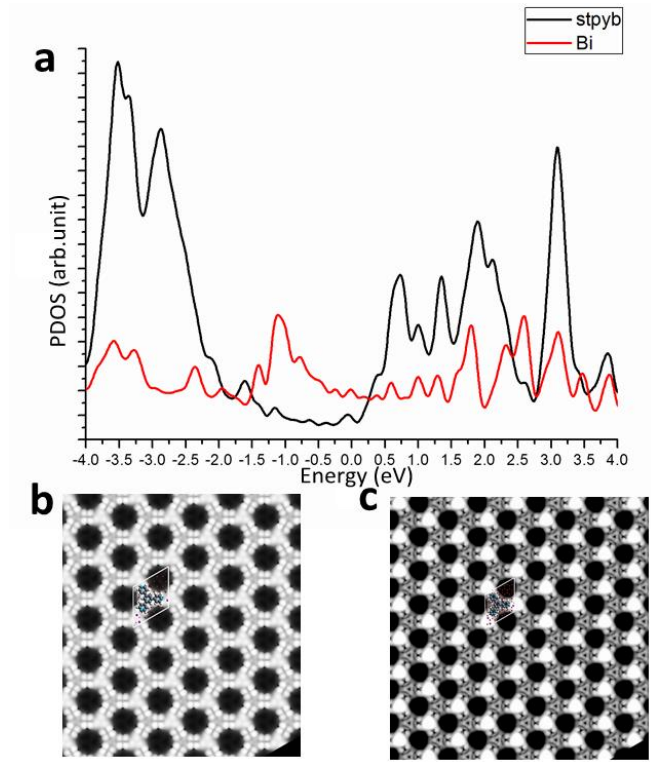

**Figure 2.10 (a) PDOS of TPyB molecules (black) and Bi atoms (red) (b) STM simulated image of 1.5~2.2eV in a constant current mode (c) STM simulated image of 2.8~3.5 eV in a constant current mode.**

c. Based on the wavefunctions and charges distribution results obtained in previous step of self-consistent calculation, the Barder charges transferred and related electrons localization functions were calculated, as shown in figure 2.11.

The Barder charge analysis is an intuitive way of dividing molecules into atoms with the definition of an atom is based purely on the electronic charge density. Bader uses what are called zero flux surfaces to divide atoms. A zero flux surface is a 2-D surface on which the charge density is a minimum perpendicular to the surface. Typically in molecular systems, the charge density reaches a minimum between atoms and this is a natural place to separate atoms from each other. Besides being an intuitive scheme for visualizing atoms in molecules, Bader's definition is often useful for charge analysis. For example, the charge enclosed within the Bader volume is a good approximation to the total electronic charge of an atom. The charge distribution can be used to determine multipole moments of interacting atoms or molecules. Bader's analysis has also been used to define the hardness of atoms, which can be used to quantify the cost of removing charge from an atom. In quantum chemistry, the electron localization function (ELF) is a measure of the likelihood of finding an electron in the neighborhood space of a reference electron located at a given point and with the same spin. Physically, this measures the extent of spatial localization of the reference electron and provides a method for the mapping of electron pair probability in multielectronic systems [116, 117].

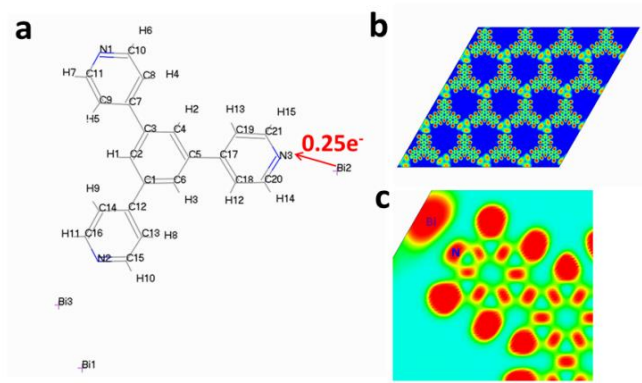

**Figure 2.11 (a) Based on the Barder charge analysis, the charge of 0.25e is transferred from each Bi atom to N atom (b) The ELF image of the whole system, red is strongest (c) Zoom-in ELF image of the interaction between N and Bi atom.**

**Appendix 4. Simulation steps and force fields in MD simulation.**

Figure 2.12 illustrates the flowchart of MD simulation process. At first, the initial coordinates of each particle inside the system are given, and then the strain of whole structure is totally relaxed usually by some DFT methods. Secondly, the force fields for all the particles will be calculated by potential. Then, each particle is initialized with a starting velocity, which is starting state of this dynamics process. Lastly, the simulation loop of all the trajectories are conducted according to the equilibration dynamics for a certain period of time [124]. Practically, there are several types of software applied in MD simulation works, such as CHARMM, Amber, Gromacs and NAMD. In chapter 4 of this thesis, I will introduce some simulation results by MD methods in detail.

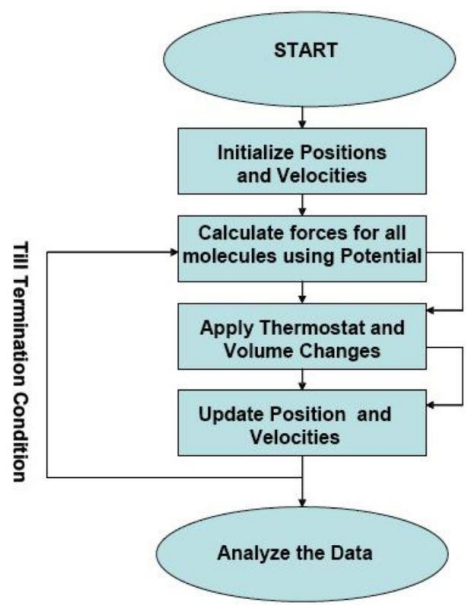

**Figure 2.12 Simulation flowchart of MD methods [124].**

**Force fields in MD:** In molecular dynamics simulation, a molecular system is usually described as a series of charged points (atoms) linked by springs (bonds). To describe the time evolution of each property in the model, such as bond lengths, bond angles and torsions, non-bonding van der Waals and elecrostatic interactions, we have to apply a force field. A force field refers to both the functional form and parameter sets used to calculate the potential energy of a model used in molecular dynamics simulations. The force field is thus a collection of equations and associated constants designed to reproduce molecular geometry and selected properties of tested structures.

Firstly, the basic functional form of potential energy in molecular mechanics includes bonded terms for interactions of atoms that are linked by covalent bonds, and nonbonded (also termed noncovalent) terms that describe the combination of long-range electrostatic and van der Waals forces. The general form for the total energy in an additive force field can be written as: $E_{total} = E_{bonded} + E_{nonbonded}$ ; $E_{bonded} = E_{bond} + E_{angle} + E_{dihedral}$ and $E_{nonbonded} = E_{electronstatic} + E_{vdw}$. A more specific form of the functional form is listed as below, in which $U_{bond}$ = oscillations about the equilibrium bond length; $U_{angle}$ = oscillations of 3 atoms about an equilibrium bond angle; $U_{dihedral}$ = torsional rotation of 4 atoms about a central bond; $U_{nonbond}$ = non-bonded energy terms (electrostatics and Lenard-Jones). [274, 275]

$$U(\vec{R}) = \underbrace{\sum_{bonds} k_i^{bond}(r_i - r_0)^2}_{U_{bond}} + \underbrace{\sum_{angles} k_i^{angle}(\theta_i - \theta_0)^2}_{U_{angle}} +$$

$$\underbrace{\sum_{dihedrals} k_i^{dihe}[1 + \cos(n_i\phi_i + \delta_i)]}_{U_{dihedral}} +$$

$$\underbrace{\sum_i \sum_{j \neq i} 4\epsilon_{ij}\left[\left(\frac{\sigma_{ij}}{r_{ij}}\right)^{12} - \left(\frac{\sigma_{ij}}{r_{ij}}\right)^{6}\right] + \sum_i \sum_{j \neq i} \frac{q_i q_j}{\epsilon r_{ij}}}_{U_{nonbond}}$$

The bond and angle terms in this form are usually modeled by quadratic energy functions that do not allow bond breaking. The functional form for dihedral energy is highly variable. Additional, "improper torsional" terms may be added to enforce the planarity of aromatic rings and other conjugated systems, and "cross-terms" that describe coupling of different internal variables, such as angles and bond lengths. Some force fields also include explicit terms for hydrogen bonds. The nonbonded terms are most computationally intensive. The van der Waals term is usually computed with a Lennard-Jones potential and the electrostatic term with Coulomb's law.

Secondly, the force fields define a set of parameters for different types of atoms, chemical bonds, dihedral angles and so on. For instance, all-atom force fields provide parameters for every type of atom in a system, including hydrogen, while united-atom interatomic potentials treat the hydrogen and carbon atoms in each methyl group (terminal methyl) and each methylene bridge as one interaction center. The parameter sets are usually empirical, which are defined by interatomic potentials developers to be self-consistent. The parameters need to determinate in a force consist of some typical parameter set values for atomic mass, van der Waals radius, and partial charge for individual atoms, and equilibrium values of bond lengths, bond angles, and dihedral angles for pairs, triplets, and quadruplets of bonded atoms, and values corresponding to the effective spring constant for each potential [275].

As the works discussed in chapter 4, all-atom MD simulations is conducted by using GROMACS (version 4.5.4) software. The atom types and force field parameters of Br$_2$TPP were taken from the General Amber Force Field (GAFF) [276]; GAFF is an extension of the AMBER force fields, and it was parameterized for most of the organic molecules. It is a complete force field, and all parameters are available for the basic atom types. Moreover, it is designed to be compatible with existing Amber force fields for proteins and nucleic acids, and has parameters for most organic and pharmaceutical molecules that are composed of H, C, N, O, S, P, and halogens. It uses a simple functional form and a limited number of atom types, but incorporates both empirical and heuristic models to estimate force constants and partial atomic charges [276]. Bedsides, the partial charge of each atom is obtained by the optimization of geometry, calculation of the electrostatic potential, and finally application of the Restrained Electrostatic Potential (RESP) method [277, 278]. To note, all the quantum mechanics calculations were performed by the Density Functional Theory (DFT) at B3LYP/6-31G* level using the Gaussian 09 software [279]. The charge for Au was set to zero. The Lennard-Jones parameters for Au were assigned as $\sigma_{AuAu}$ = 0.32 nm and $\varepsilon_{AuAu}$ = 0.65 kJ/mol, respectively [280].

## Appendix 5. The Tight-binding Simulation

The tight-binding model (TB model) is a well-known computational approach in the field of solid-state physics to simulate the electronic band behaviors. Usually, the 2-dimension wave functions sets are used based on the superposition of wave functions in order to study isolated atoms/molecules located at all sites in the materials. The forerunner of tight-binding theory, bond-orbital model, was firstly investigated by Salim Ciraci in1970s. This work was originated from the idea of Felix Bloch, which had not drawn any attention until the development of pseudopotential theory. According to Salim Ciraci's works, TB theory can provide a comprehensive understanding of the electronic structure for tetrahedral coordinated semiconductors. Later on, the subsequent developments in this theory have utilized more accurate and complete TB-calculations to access the electronic properties of all the solid-based materials, especially for 2D semiconductors.

**Discretization of the Hamiltonian**

In general, the Hamiltonian for a 2D system can be written as

$$H = -\frac{\hbar^2}{2m}(\partial_x^2 + \partial_y^2) + U(x,y) \tag{2.21}$$

where the 1$^{st}$ and 2$^{nd}$ terms are the kinetic and potential term respectively.

For each site, the spatial coordinates are just $(x,y) = (ai, aj)$, where $i, j$ are integers and $a$ is lattice constant. The position state at each site is therefore

$$|i,j\rangle \equiv |ai, aj\rangle \tag{2.22}$$

The (continuum) Hamiltonian can be discretized by the method of finite difference as follows:

$$\partial_x = \frac{1}{a}(|i+1,j\rangle\langle i,j| - |i,j\rangle\langle i,j|) \tag{2.23}$$

$$\Rightarrow \partial_x^2 = \frac{1}{a^2}(|i+1,j\rangle\langle i,j| + |i,j\rangle\langle i+1,j| - 2|i,j\rangle\langle i,j|) \text{ and similar for } \partial_y^2$$

$$U(x,y) = U(ai, aj) \tag{2.24}$$

The resulting Hamiltonian is

$$H = \sum_{i,j}[4t|i,j\rangle\langle i,j| - t(|i+1,j\rangle\langle i,j| + |i,j\rangle\langle i+1,j| + |i,j+1\rangle\langle i,j| +$$
$$|i,j\rangle\langle i,j+1| + U(ai, aj)] \tag{2.25}$$

where $t = \hbar^2/2ma^2$ is the n.n. hopping strength. The diagonal term $4t|i,j\rangle\langle i,j|$ is unimportant and can be dropped.

For example, in the case of a 1D chain with 50 atoms (which has only one orbital) and *periodic boundary condition*:

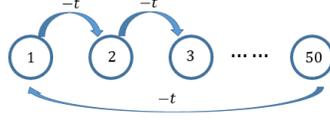

The lattice constant $a$ is assumed to be 1 for convenience. The Hamiltonian is just (when no external potential presents and note that the notation is changed here from $|i\rangle$ to $|n\rangle$ to avoid confusion with the imaginary part i)

$$H = -t \sum_n (|n\rangle\langle n+1| + |n+1\rangle\langle n|) \qquad (2.26)$$

One can see that hopping only exists between the nearest neighbor $|n\rangle$ and $|n+1\rangle$. In matrix form

$$H = t \begin{bmatrix} 0 & -1 & 0 & \cdots & -1 \\ -1 & 0 & -1 & \cdots & 0 \\ 0 & -1 & 0 & \cdots & 0 \\ \cdots & \cdots & \cdots & \cdots & \cdots \\ -1 & 0 & 0 & -1 & 0 \end{bmatrix} \begin{matrix} 1 \\ 2 \\ 3 \\ \\ 50 \end{matrix}$$

The eigenvalues can be readily obtained through diagonalization of H. The result agrees with the analytical expression $E_k = -2t\cos(k)$, where $k = \frac{2p\pi}{N}$ and $p$ goes from 1 to N (N = 50 here).

To obtain the band structure, one can Fourier transform the Hamiltonian. Introducing the momentum state

$$|k\rangle = \frac{1}{\sqrt{N}} \sum_n e^{-ikn}|n\rangle \ \text{ or } \ |n\rangle = \frac{1}{\sqrt{N}} \sum_k e^{ikn}|k\rangle \qquad (2.27)$$

Plugging the above expression in H yields

$$H = -t \sum_n \frac{1}{N} \sum_{k,k'} [e^{ikn} e^{-ik'(n+1)} + e^{ik(n+1)} e^{-ik'n}] \qquad (2.28)$$

Using $\sum_n \exp(i(k-k')n) = N\delta_{k,k'}$ we have

$$H = -t\sum_k (e^{-ik} + e^{ik}) = -2t\cos k \qquad (2.29)$$

Numerically, we introduce the discrete Fourier transform matrix in order to change the basis from $|n\rangle$ to $|k\rangle$

$$W = \frac{1}{\sqrt{N}}\begin{bmatrix} 1 & 1 & 1 & 1 & \cdots & 1 \\ 1 & \omega & \omega^2 & \omega^3 & \cdots & \omega^{N-1} \\ 1 & \omega^2 & \omega^4 & \omega^6 & \cdots & \omega^{2(N-1)} \\ 1 & \omega^3 & \omega^6 & \omega^9 & \cdots & \omega^{3(N-1)} \\ \vdots & \vdots & \vdots & \vdots & \ddots & \vdots \\ 1 & \omega^{N-1} & \omega^{2(N-1)} & \omega^{3(N-1)} & \cdots & \omega^{(N-1)(N-1)} \end{bmatrix}$$

where $\omega = e^{-\frac{2\pi i}{N}}$. The Hamiltonian in the k space is then

$$\mathcal{H} = W^+ H W \qquad (2.30)$$

which is diagonalized (as checked numerically). This technique can be generalized to 2D or 3D by amending the exponential factor $e^{ikn}$ to $e^{i\vec{k}\cdot\hat{r}}$.

**LDOS**

The local density of state defined in the program is

$$\rho(r) = \sum_n |\Psi_n(r)|^2 \delta_{E_n,E} \approx \frac{1}{c\sqrt{2\pi}} \sum_n |\Psi_n(r)|^2 e^{-\frac{(E_n-E)^2}{2c^2}} \qquad (2.31)$$

Since locating the exact eigenvalue $E_n$ is not convenient, a Gaussian braodening is used to approximate the delta function. The width c used in my computation is 0.1 or 0.2 eV, depending on the resolution required. While the LDOS can be obtained through the above equation, the method of Green function is employed in the package.

In the diagonal basis, the Green function is

$$G(\omega) = \sum_k \frac{|k\rangle\langle k|}{\omega - E_k + i\delta} \qquad (2.32)$$

where $\delta$ is a infinisimal number ($\hbar$ taken to be 1)

The matrix elements can be obtained by the projection of G on site n

$$G_{nn}(\omega) = \sum_k \frac{|\Psi_n(k)|^2}{\omega - E_k + i\delta} \tag{2.33}$$

where $\Psi_n(k) = \langle k|n\rangle$ is the amplitude of the wavefunction at site n. The local density of state is just the spectral function, which is defined as

$$A_{nn}(\omega, r) = -\frac{1}{\pi} Im(G_{nn}(\omega)) = \sum_k |\Psi_n(k)|^2 \delta(\omega - E_k) \tag{2.34}$$

$$\rho_n(\omega) = \sum_k |\Psi_n(k)|^2 \delta(\omega - E_k) \tag{2.35}$$

**DOS**

There is no tricks in the computation of the DOS. It can be easily obtained from the eigenvalues of the Hamiltonian by using the definition

$$D(E) = \sum_n \delta_{E,E_n} \approx \frac{1}{c\sqrt{2\pi}} \sum_n e^{-\frac{(E_n - E)^2}{2c^2}} \tag{2.36}$$

where, again, a Gaussian broadening is used for computational convenience

# Appendix 6. Plane wave expansion method for calculation of electronic structures of 2DEG systems

Plane wave expansion method is a widely used simulation method for calculating the electronic properties of photonic crystals [125]. This method is entirely based on the numerical solution of Schrodinger equation. It is featured as fast calculation on the photo intensity map and spectra, band structure and local density of states of photonic crystals. This method is not only applicable in 3D cases, but also can be transferred to simulate the electronic behaviors of 2 dimension electron gas (2DEG) systems. The Bloch's theorem gives a strong theoretical basic of the plane wave expansion method in the band structure calculation. Based on the Bloch's theorem, in the periodical potential fields, the wave function ($\psi_k$) of each electron can be written as $\psi_k(r + R) = e^{ik \cdot R}\psi_k(r)$, where r is one arbitrary vector inside the an unit cell, R is the cell translation vectors. The Schrödinger equation of this many-body system can then be simplified as the case of one electron inside the periodical potential:

$$H\psi(r) = \left(-\frac{\hbar^2}{2m}\nabla^2 + U(r)\right)\psi(r) = \varepsilon\psi(r) \tag{2.37}$$

where the wavefunction $\psi(r)$ is expanded by plane waves: $\psi(r) = \sum_q c_q e^{iq \cdot r}$.

Therefore, the Schrödinger equation as:

$$\left[-\frac{\hbar^2}{2m}q^2 - \varepsilon\right]c_q + \sum_{K'} U_{K'-K}c_{q-K'} = 0 \tag{2.38}$$

where $U_K$ are the Fourier coefficients in plane wave expansion of a potential:

$$U_K = \frac{1}{\upsilon}\int_{cell} e^{-iK \cdot r}U(r)dr \tag{2.39}$$

where $\upsilon$ is a unit volume.

After setting that q=k − K, and $\varepsilon_{k-K}^0 = \frac{\hbar^2}{2m}(k-K)^2$, the Schrödinger equation is reformulated in the form of Fourier coefficient in the reciprocal space:

$$\left(\varepsilon - \varepsilon_{k-K}^0\right)c_{k-K} = \sum_{K'} U_{K'-K}\, c_{k-K'} \tag{2.41}$$

where k and K are both the reciprocal lattice vectors, with K chosen in a way such that k located at the first Brillouin Zone. This set of equations can then be expressed as a matrix:

$$\begin{pmatrix} \varepsilon^0_{k-K_1} & U_{K_2-K_1} & U_{K_3-K_1} & & U_{K_n-K_1} \\ U_{K_1-K_2} & \varepsilon^0_{k-K_2} & U_{K_3-K_2} & \cdots & U_{K_n-K_2} \\ U_{K_1-K_3} & U_{K_2-K_3} & \varepsilon^0_{k-K_3} & & U_{K_n-K_3} \\ \vdots & & & \ddots & \vdots \\ U_{K_1-K_n} & U_{K_2-K_n} & U_{K_3-K_n} & \cdots & \varepsilon^0_{k-K_n} \end{pmatrix} \begin{pmatrix} c_{k-K_1} \\ c_{k-K_2} \\ c_{k-K_3} \\ \vdots \\ c_{k-K_n} \end{pmatrix} = \begin{pmatrix} \varepsilon_1 \\ \varepsilon_2 \\ \varepsilon_3 \\ \vdots \\ \varepsilon_n \end{pmatrix} \begin{pmatrix} c_{k-K_1} \\ c_{k-K_2} \\ c_{k-K_3} \\ \vdots \\ c_{k-K_n} \end{pmatrix} \quad (2.42)$$

Once we obtained the electronic band structures, the LDOS for and location r and an energy E can be simplified as the formula:

$$\text{LDOS}(E, r) = \sum_{k'} \psi_E(k', r)^* \times \psi_E(k', r) \quad (2.43)$$

where $\psi_E(k, r)$ is a summation of wave functions of each plane-wave expanded K components:

$$\psi_E(k', r) = \sum_K \psi_E(K, r) \quad (2.44)$$

As discussed in Chapter 5, the detailed analysis of using plane wave expansion to simulate a 2DEG systems (artificial graphene nano flakes) according to experimental measurements will be presented.

# Bibliography

(1) Kuang, G.; Chen, S.-Z.; Chen, K.; Zhang, Q.; Lin, N. Charge Transport through On-Surface Synthesized Oligomers. 1.

(2) Miao, Q.; Wang, J.; Hu, M.; Zhang, F.; Zhang, Q.; Xia, C. Promotion of Cooperation Induced by a Self-Questioning Update Rule in the Spatial Traveler's Dilemma Game. *Eur. Phys. J. Plus* **2014**, *129* (1), 8. https://doi.org/10.1140/epjp/i2014-14008-4.

(3) Kuang, G.; Zhang, Q.; Li, D. Y.; Shang, X. S.; Lin, T.; Liu, P. N.; Lin, N. Cross-Coupling of Aryl-Bromide and Porphyrin-Bromide on an Au(111) Surface. *Chem. Eur. J.* **2015**, *21* (22), 8028–8032. https://doi.org/10.1002/chem.201501095.

(4) Zhang, Q.; Kuang, G.; Pang, R.; Shi, X.; Lin, N. Switching Molecular Kondo Effect via Supramolecular Interaction. *ACS Nano* **2015**, *9* (12), 12521–12528. https://doi.org/10.1021/acsnano.5b06120.

(5) Lyu, G.; Zhang, Q.; I. Urgel, J.; Kuang, G.; Auwärter, W.; Ecija, D.; V. Barth, J.; Lin, N. Tunable Lanthanide-Directed Metallosupramolecular Networks by Exploiting Coordinative Flexibility through Ligand Stoichiometry. *Chemical Communications* **2016**, *52* (8), 1618–1621. https://doi.org/10.1039/C5CC08526H.

(6) Zhang, Q. P. Characterization of the magnetism and conformation of single porphyrin molecules adsorbed on surfaces, and artificial graphene nanoflakes.

(7) Zhang, Q.; Zheng, X.; Kuang, G.; Wang, W.; Zhu, L.; Pang, R.; Shi, X.; Shang, X.; Huang, X.; Liu, P. N.; Lin, N. Single-Molecule Investigations of Conformation Adaptation of Porphyrins on Surfaces. *J. Phys. Chem. Lett.* **2017**, *8* (6), 1241–1247. https://doi.org/10.1021/acs.jpclett.7b00007.

(8) Kuang, G.; Zhang, Q.; Lin, T.; Pang, R.; Shi, X.; Xu, H.; Lin, N. Mechanically-Controlled Reversible Spin Crossover of Single Fe-Porphyrin Molecules. *ACS Nano* **2017**, *11* (6), 6295–6300. https://doi.org/10.1021/acsnano.7b02567.

(9) Yan, L.; Kuang, G.; Zhang, Q.; Shang, X.; Liu, P. N.; Lin, N. Self-Assembly of a Binodal Metal–Organic Framework Exhibiting a Demi-Regular Lattice. *Faraday Discuss.* **2017**, *204* (0), 111–121. https://doi.org/10.1039/C7FD00088J.

(10) Yan, L.; Xia, B.; Zhang, Q.; Kuang, G.; Xu, H.; Liu, J.; Liu, P. N.; Lin, N. Stabilizing and Organizing Bi3Cu4 and Bi7Cu12 Nanoclusters in Two-Dimensional Metal–Organic Networks. *Angewandte Chemie* **2018**, *130* (17), 4707–4711. https://doi.org/10.1002/ange.201800906.

(11) Yan, L.; Hua, M.; Zhang, Q.; Ngai, T. U.; Guo, Z.; Wu, T. C.; Wang, T.; Lin, N. Symmetry Breaking in Molecular Artificial Graphene. *New J. Phys.* **2019**, *21* (8), 083005. https://doi.org/10.1088/1367-2630/ab34a6.

(12) Zhang, Q.; Pang, R.; Luo, T.; Van Hove, M. A. Controlling the Rotational Barrier of Single Porphyrin Rotors on Surfaces. *J. Phys. Chem. B* **2020**, *124* (6), 953–960. https://doi.org/10.1021/acs.jpcb.9b09986.

(13) Zhang, Q.; Wu, T. C.; Kuang, G.; Xie, A.; Lin, N. Investigation of Edge States in Artificial Graphene Nano-Flakes. *J. Phys.: Condens. Matter* **2021**, *33* (22), 225003. https://doi.org/10.1088/1361-648X/abe819.